\documentclass{article}
\usepackage[utf8]{inputenc}
\usepackage[letterpaper, margin=1in]{geometry}
\usepackage[sorting=none]{biblatex}
\usepackage{graphicx}
\usepackage{amsmath}
\usepackage[]{hyperref}
\usepackage{xfrac}
\hypersetup{
	colorlinks=true,       
	linkcolor=blue,        
	citecolor=blue,        
	filecolor=magenta,     
	urlcolor=blue         
}

\def\MC{MuCol}
\def\mm{$\mu^+\mu^-$}

\def\ten34{10$^{34}$ cm$^{-2}$ s$^{-1}$~}

\title{
\bf Prospects for the Measurement of the Standard Model\\
Higgs Pair Production at the Muon Colliders 
}
\author{
K. Black, T. Bose, S. Dasu, H. Jia, S. Lomte, V. Sharma, C. Vuosalo
\\{\bf University of Wisconsin - Madison}\\ \\
I. Ojalvo
\\{\bf Princeton University}\\ \\
T. Holmes, L. Lee
\\{\bf University of Tennessee}\\ \\
M. Swiatlowski, M. Valente
\\{\bf TRIUMF} \\ \\
J. Oliver
\\{\bf University of California - Irvine}
}

\date{December 2022}

\begin{document}

\maketitle

\begin{abstract}
We study the Higgs pair production process at a muon collider using $b$-pair decays of the Higgs bosons. Efficient identification and good measurement resolution for the $b$-jet pair invariant mass are crucial for unearthing the di-Higgs signal. However, the beam-induced background has potential to drastically degrade the performance. We report on the full simulation studies of the degradation of the reconstructed $b$-jet pair invariant mass in di-Higgs events, considering only the beam-induced background in the calorimeter. Mitigation strategies for the suppression of the beam-induced background are underway. We also report prospects for the measurement of the Standard Model Higgs pair production at the Muon Colliders at various benchmarks of the collider center of mass energy and integrated luminosity using a fast simulation program.
\end{abstract}

\section{Introduction}

A future high-energy Muon Collider (\MC) collider~\cite{mucol} could greatly improve our understanding of the Higgs self-coupling, extending our understanding of the Higgs boson acquired at the LHC. At \mm~center of mass energies above 3 TeV, with luminosities above few $\times$ \ten34, \mm$-$ collisions provide significantly cleaner final states than those produced at the HL-LHC. Even though the idea of a muon collider is not new, as it was first proposed in the late ‘60s~\cite{tikhonin1968effects, budker1982accelerators} and studied in detail by the Muon Accelerator Program (MAP), it is now receiving renewed interest because of its potential to overcome key limitations of other proposed collider concepts. The enormous physics potential of colliding \mm beams has sparked a wave of studies aimed at quantifying a possible physics program. Circular \mm~colliders could reach multi-TeV regime, within a limited spatial footprint and power budget because synchrotron radiation from muons is significantly lower than that of electrons. Furthermore, \mm~collisions are expected to take place with a small energy spread and lead to an improved energy resolution for physics measurements. There has been much attention paid to the Higgs physics prospects at \MC~recently~\cite{AlAli:2021let}. In this paper we explore the prospects for measuring di-Higgs production at the muon collider.

The International Muon Collider Consortium~\cite{imcc} (IMCC) has taken the lead in making muon colliders studies and has established benchmark concepts for the collider and a detector~\cite{Bartosik:DetPhys}, based on the CLIC detector. Detector simulations at the muon collider~\cite{Bartosik:Simulation} are extremely important in understanding the physics reach of a \MC. Necessary detector simulation software~\cite{mcsoft:confluence} has been published~\cite{mcsoft:git} including both GEANT-based detector simulation and reconstruction~\cite{Marlin}.

Because muons are fundamental particles, the \MC-events do not have the underlying event contributions as in the LHC-events. Further, their lack of strong interactions also eliminates the large pileup problem for the proton collisions at the LHC. Nevertheless, one of the major challenges to detector performance is the beam-induced background (BIB)~\cite{Bartosik:BIB,Collamati_2021}, which comes from muon decays along the beam line. The upstream and downstream electromagnetic showers blanket the detector with low energy photons, electrons and neutrons. Although individually the particles are low energy there are enough of them expected that several TeV of energy is integrated over the detector at each beam crossing. Therefore, powerful BIB mitigation strategies have to be employed to study physics potential at a muon collider environment. In the first section of the report, we focus on the BIB appearing in the calorimeter and study the performance of jet reconstruction algorithm prior to employing any mitigation strategy. As an example signal process, we use di-Higgs production at 3 TeV in presence of full beam-induced background. 

In the second section of the detector we use fast simulation, assuming that BIB problem is mitigated fully, to estimate di-Higgs signal significance at various benchmark centers of mass energy for the collider and attainable integrated luminosity.

\section{Calorimeter BIB Mitigation Study - S. Lomte}

We are studying jet reconstruction performance in \mm $\to \nu \bar{\nu} H H \to \nu \bar{\nu} b\bar{b}b\bar{b}$ process at 3 TeV in presence of full beam-induced background (BIB). To attain a reasonable simulation run time, only the calorimeter hits from BIB are overlaid. However, both the tracks reconstructed from the main signal event and the calorimeter hits are included in the jet finding using Pandora particle-flow techniques~\cite{Marlin}. We tested jet performance by optimizing the calorimeter hit energy threshold and timing cuts. 

Muons in the beam pipe can decay and produce energetic electrons and neutrinos, which interact with the detector material by producing electromagnetic showers and deposit large amounts of energy which affects the signal reconstruction. To reduce this background, we need to study the full simulation of the detector with the effect of beam-induced background. This is computationally demanding in terms of CPU and memory consumption. Thus, the strategy used is to simulate one bunch crossing of $\mu^{+} \mu^{-}$ decay at 1.5~TeV which is overlaid to each hard scatter event. It is also shown~\cite{Bartosik:BIB} that the BIB effects at higher energy like 3~TeV are compensated by the increase in muon `lifetime' at higher center of mass energy. For the purpose of this study, we have used 1.5~TeV BIB samples under the above assumption. The simulated BIB samples used in this study are provided by the IMCC community. 

\begin{figure}[ht!]
    \centering
    \includegraphics[scale=0.35]{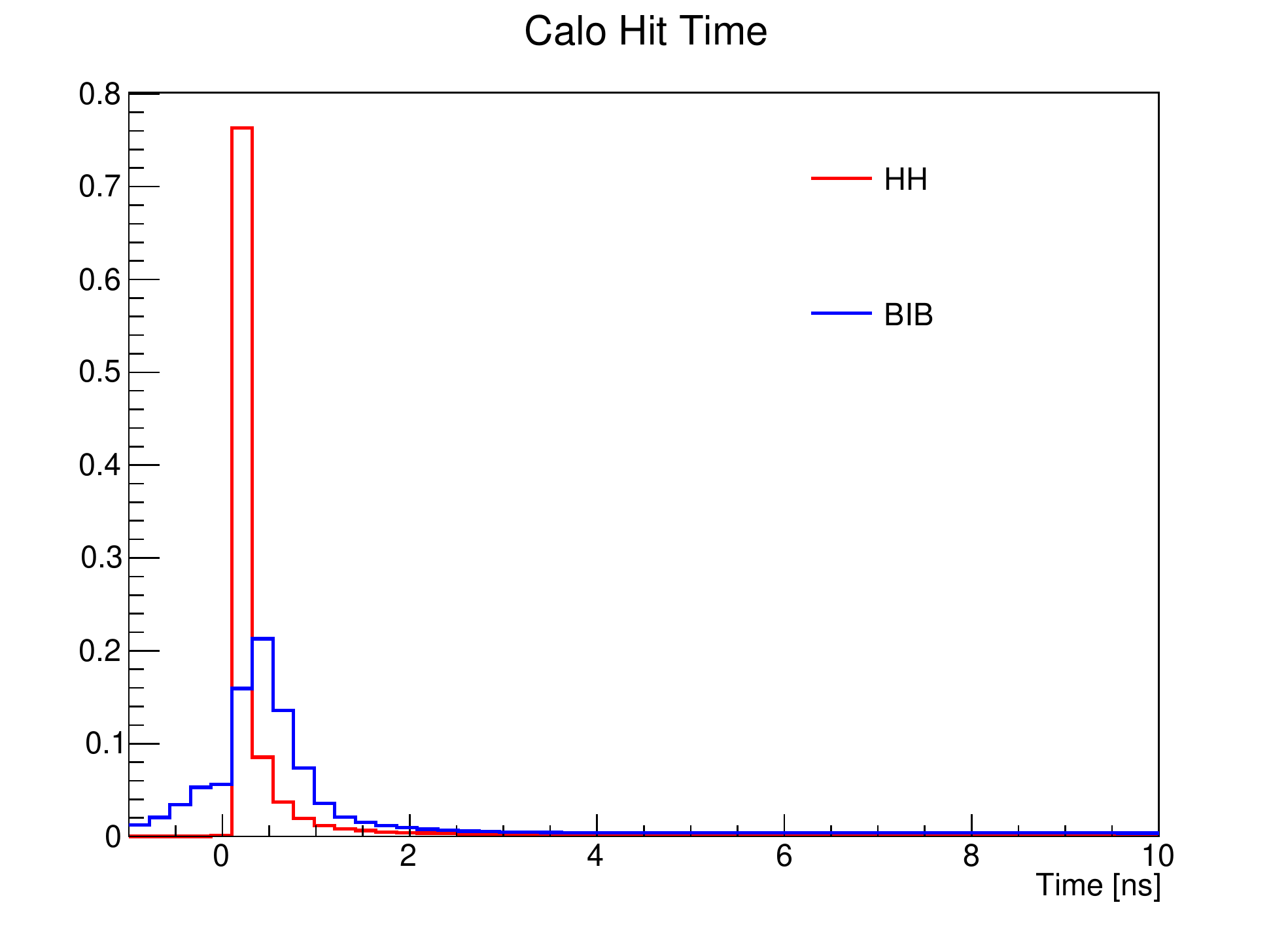}
    \includegraphics[scale=0.35]{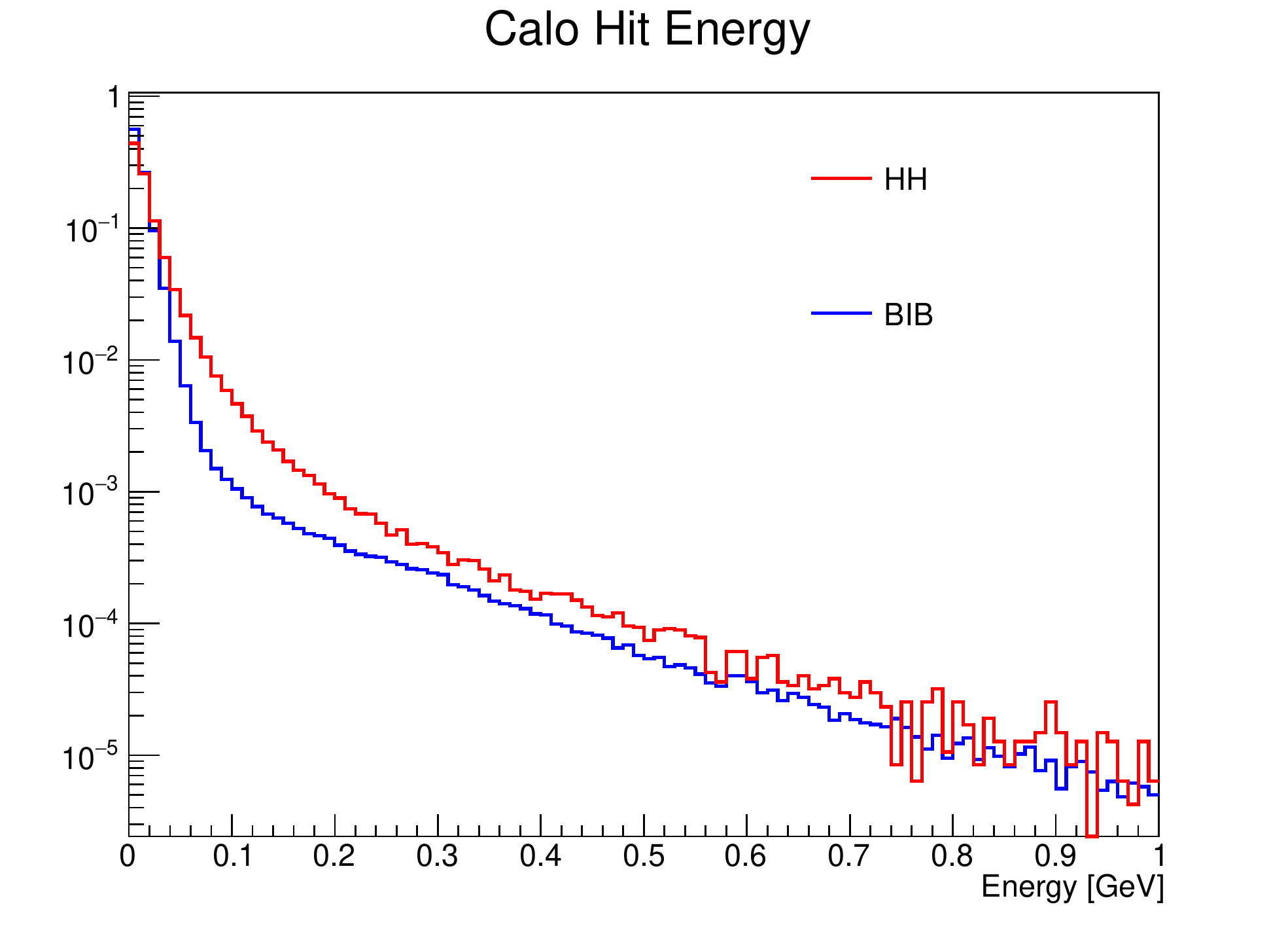}
    \caption{Calorimeter hit time (left) and energy (right) distribution for HH and BIB event.}
    \label{calohits}
\end{figure}

Figure \ref{calohits} shows calorimeter hit time and energy distributions for hard scatter event (HH) and BIB. The time of flight of BIB particles is spread out in comparison to the hard scatter event with respect to the bunch crossing and this feature is exploited to reduce BIB contribution to jet clustering. Another distinguishing feature is the energy deposit of hits in the calorimeter. BIB particles have a lower energy deposit and a more uniform spread in the calorimeter which is also used to reduce some contribution of BIB. 

\subsection{Jet reconstruction}

Jets are crucial to study the hadronic decay of HH process which has a high branching fraction to $b$-jets. The presence of beam-induced background poses a major challenge to reconstructing jets. Pandora Particle Flow Algorithm~\cite{Marshall:2013bda} (PandoraPFA) is used to cluster calorimeter hits of hard event and beam-induced background, and tracker hits of only the hard event. The digitized calorimeter hits are given to PandoraPFA as input. The kt-algorithm with cone size of 0.7 is used to reconstruct jets as output. 

\begin{figure}[ht!]
    \centering
    \includegraphics[scale=0.35]{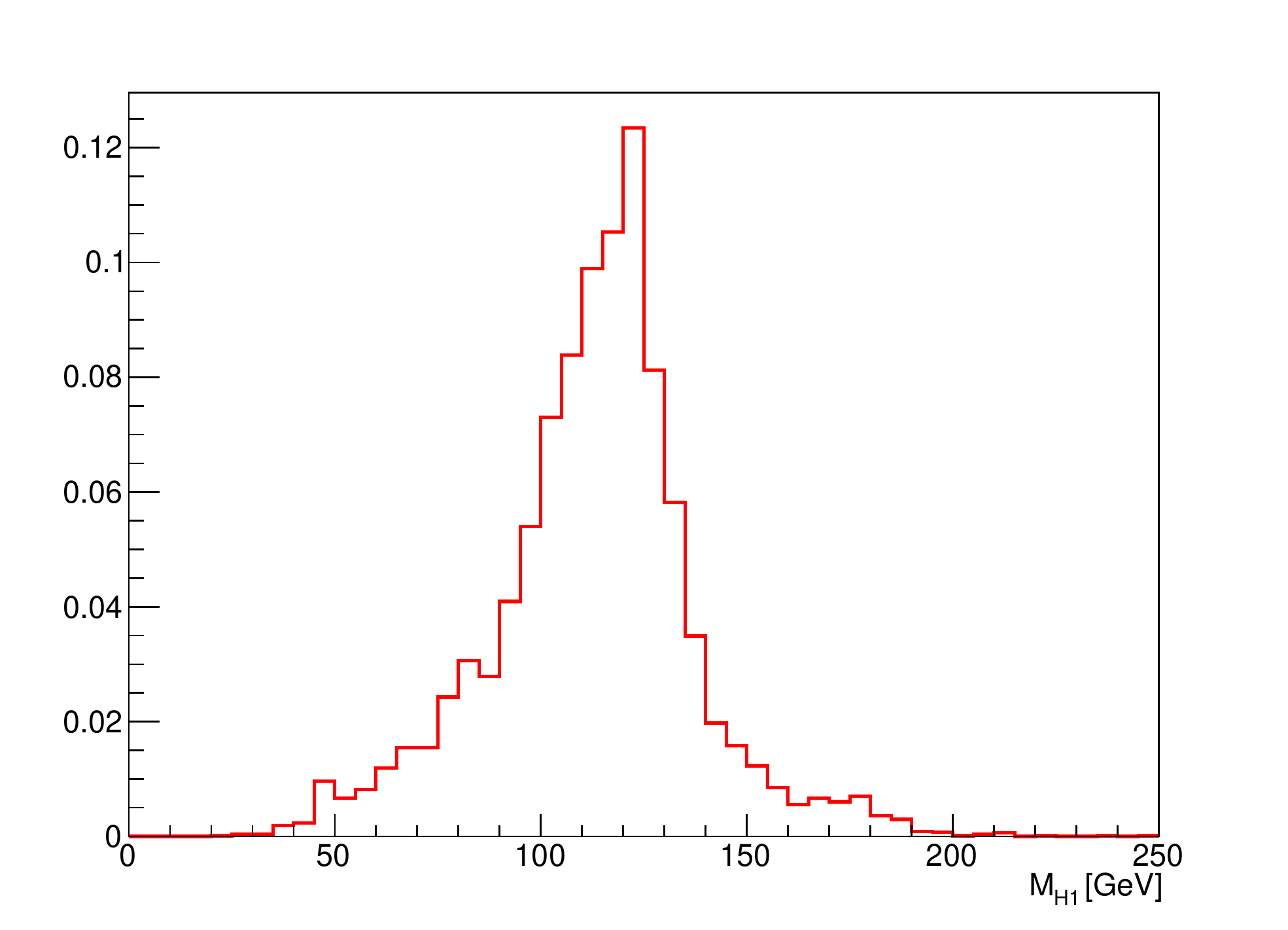}
    \includegraphics[scale=0.35]{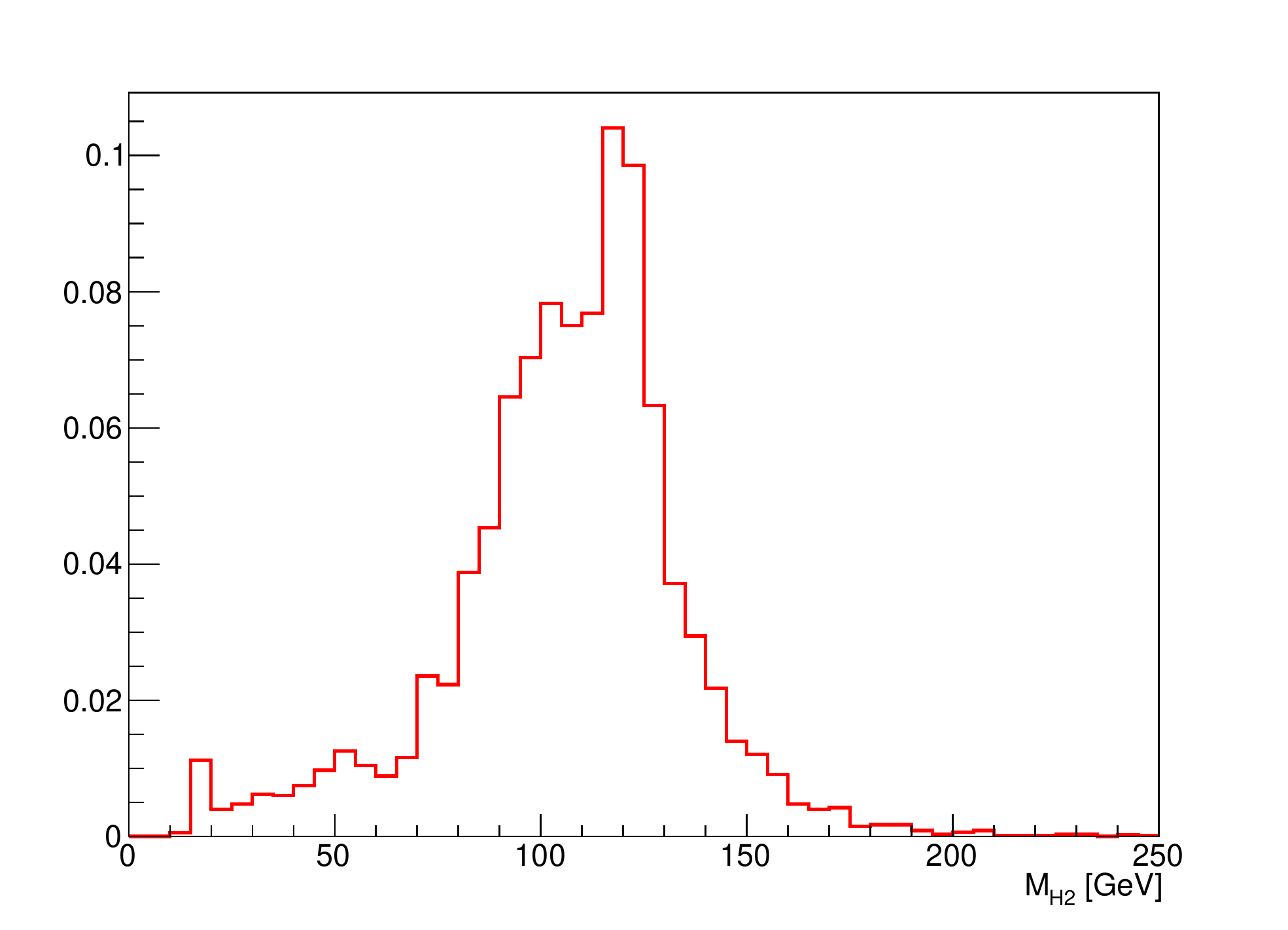}
    \caption{Invariant mass of leading (left) and subleading (right) di-jet pair at loose calorimeter energy threshold of 50 keV (ECal), 250 keV (HCal) and hit Timing window of [-1,10] ns without BIB.}
    \label{m1_s_default}
    \includegraphics[scale=0.35]{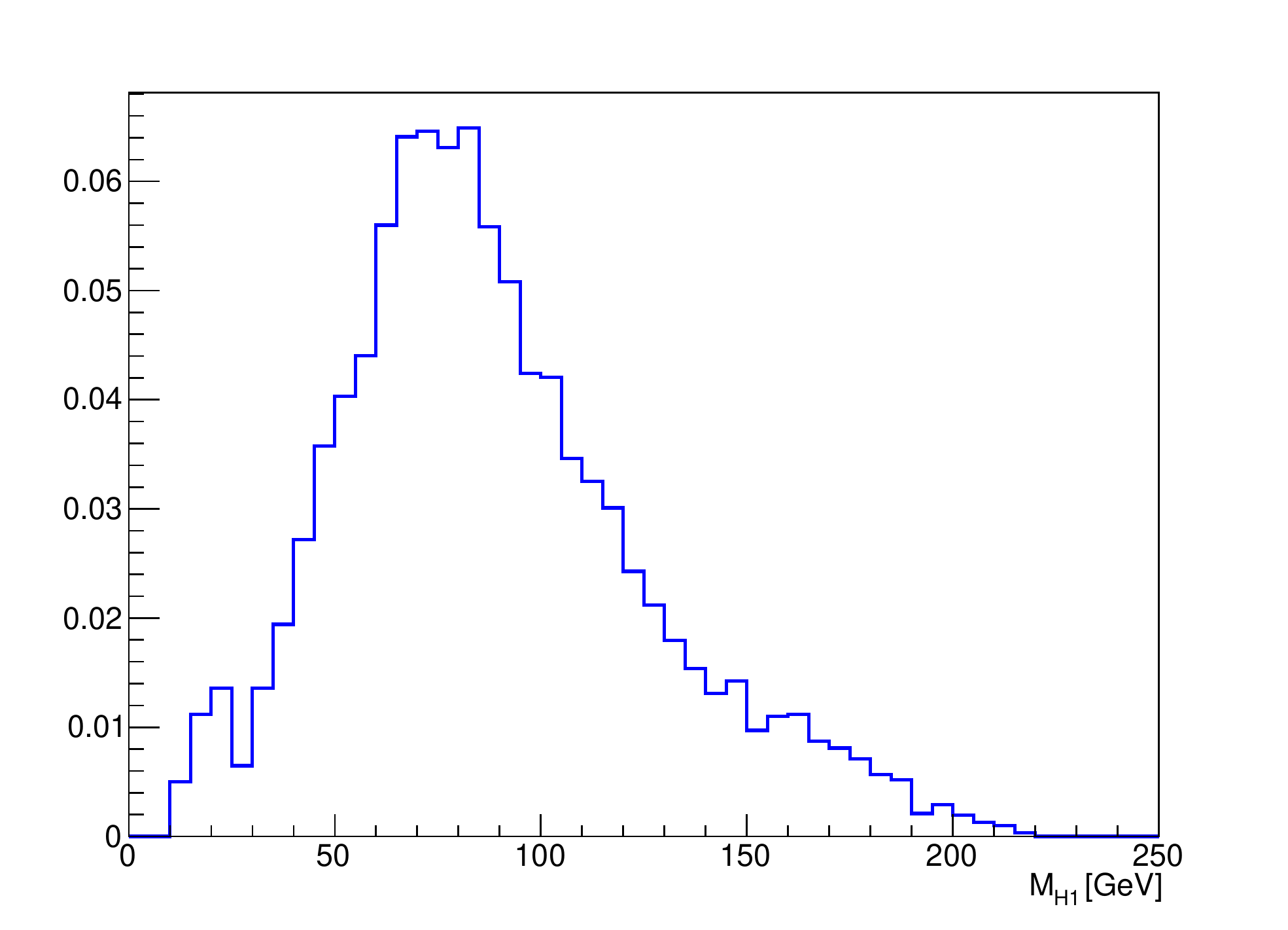}
    \includegraphics[scale=0.35]{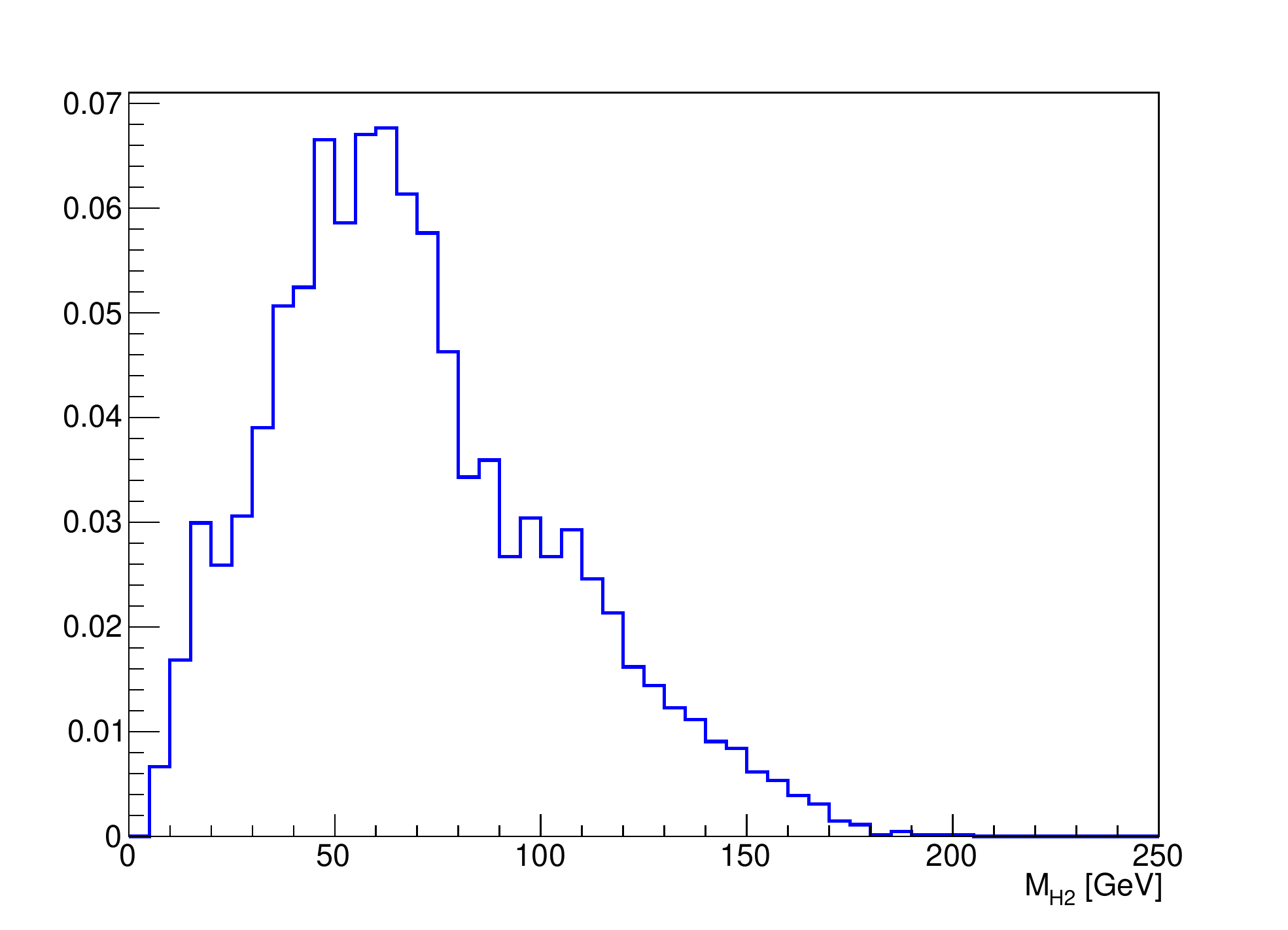}
    \caption{Invariant mass of leading (left) and subleading (right) di-jet pair at tight calorimeter energy threshold of 2 MeV (ECal), 2 MeV (HCal) and hit Timing window of [-1,2] ns without BIB.}
\label{m1_s_2mev2ns}
\end{figure}

The calorimeter hit energy and time selections affect the performance of jet reconstruction significantly. Figure \ref{m1_s_default} shows the invariant mass plot of di-jet pair without the presence of BIB using a loose selection of 50~keV for ECal energy and 250~keV for HCal along with a hit timing window of [-1, 10] ns. Figure \ref{m1_s_2mev2ns} compares that using a tighter selection of hit energy of 2 MeV for both ECal and HCal, and timing window of [-1, 2] ns.

The di-jet invariant mass pairs ($M_{H1}$ and $M_{H2}$) are found with minimizing $\Delta M$ condition, where $\Delta M = \Delta m_{1}^{2} + \Delta m_{2}^{2}$, and $\Delta m_{1} = M_{H} - M_{ij}$, $\Delta m_{2} = M_{H} - M_{kl}$ with $ij$ and $kl$ as the best two di-jet pairs, and $M_{H} = 125$~GeV. We observe that the mass peak degrades with tighter cuts. However, these selections allow for a reasonable run-time of the reconstruction code, i.e hours as opposed to days when including the beam-induced background. We plan to use the above tight selections (can be further optimized) and apply jet corrections to improve the mass peak.

\begin{figure}[ht!]
    \centering
    \includegraphics[scale=0.35]{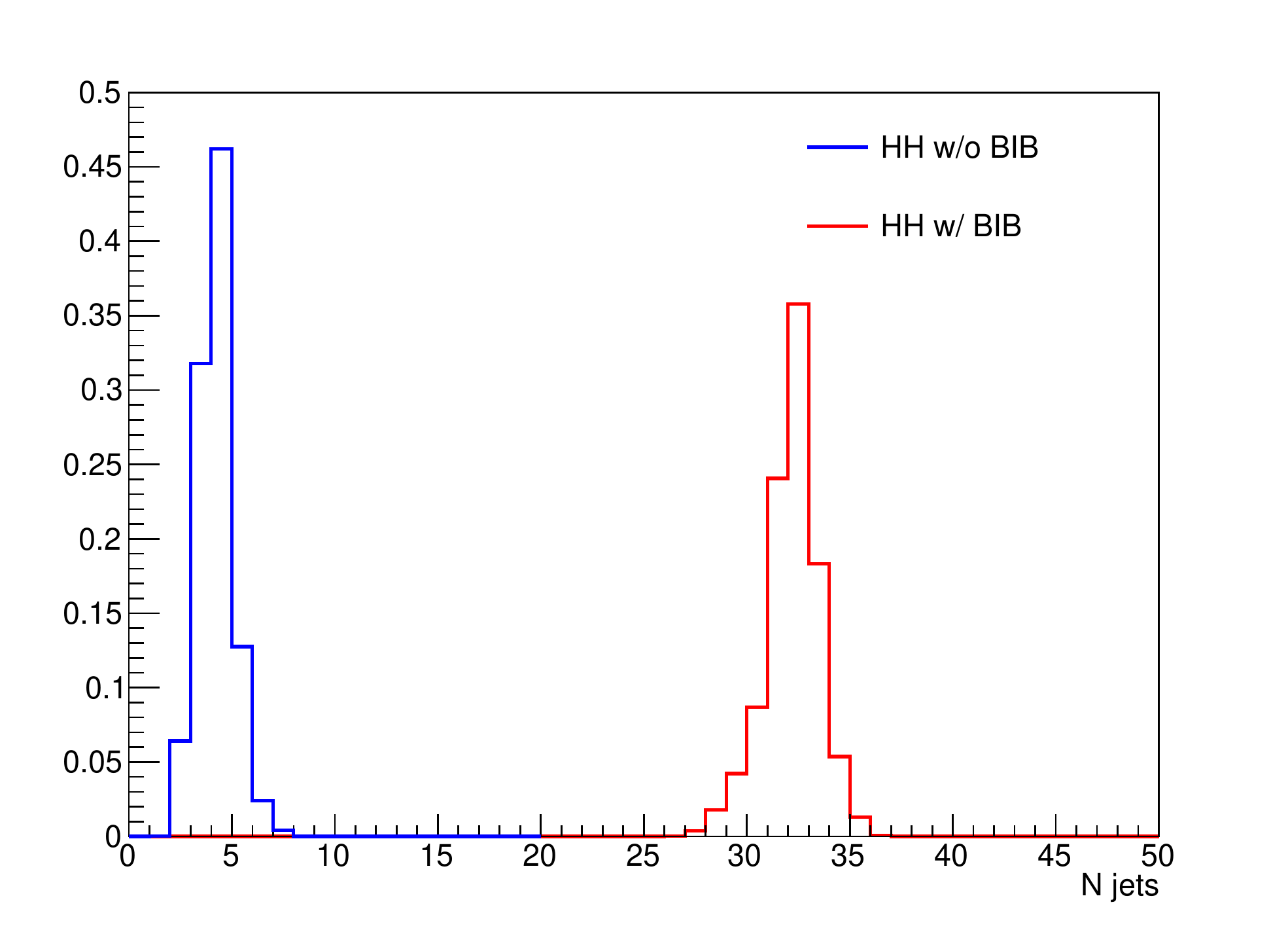}
    \includegraphics[scale=0.35]{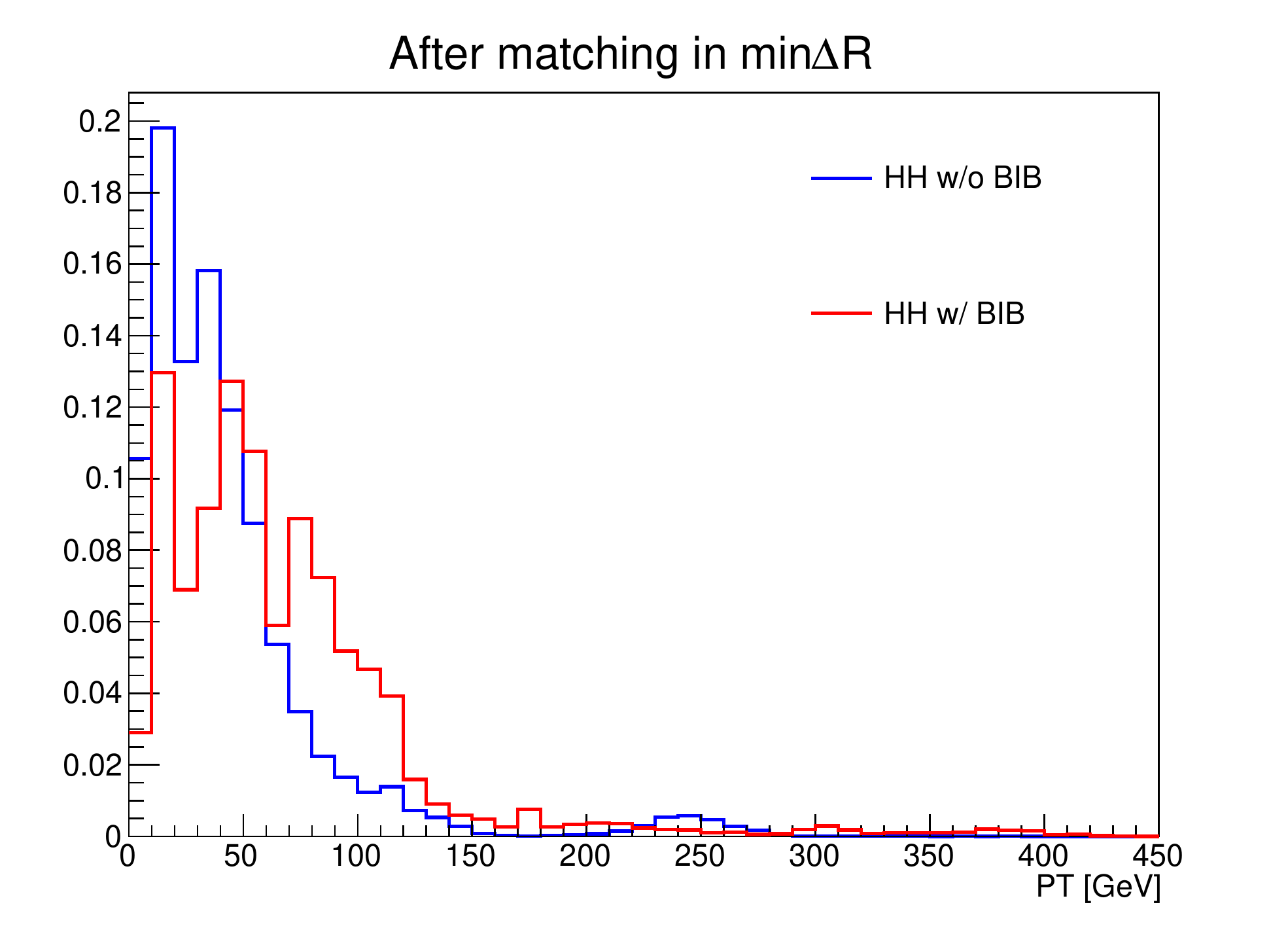}
    \caption{Number of reconstructed jets (left). $p_{T}$ of jets with BIB matched to jets without BIB (right).}
    \label{Nj_PT}
    \includegraphics[scale=0.35]{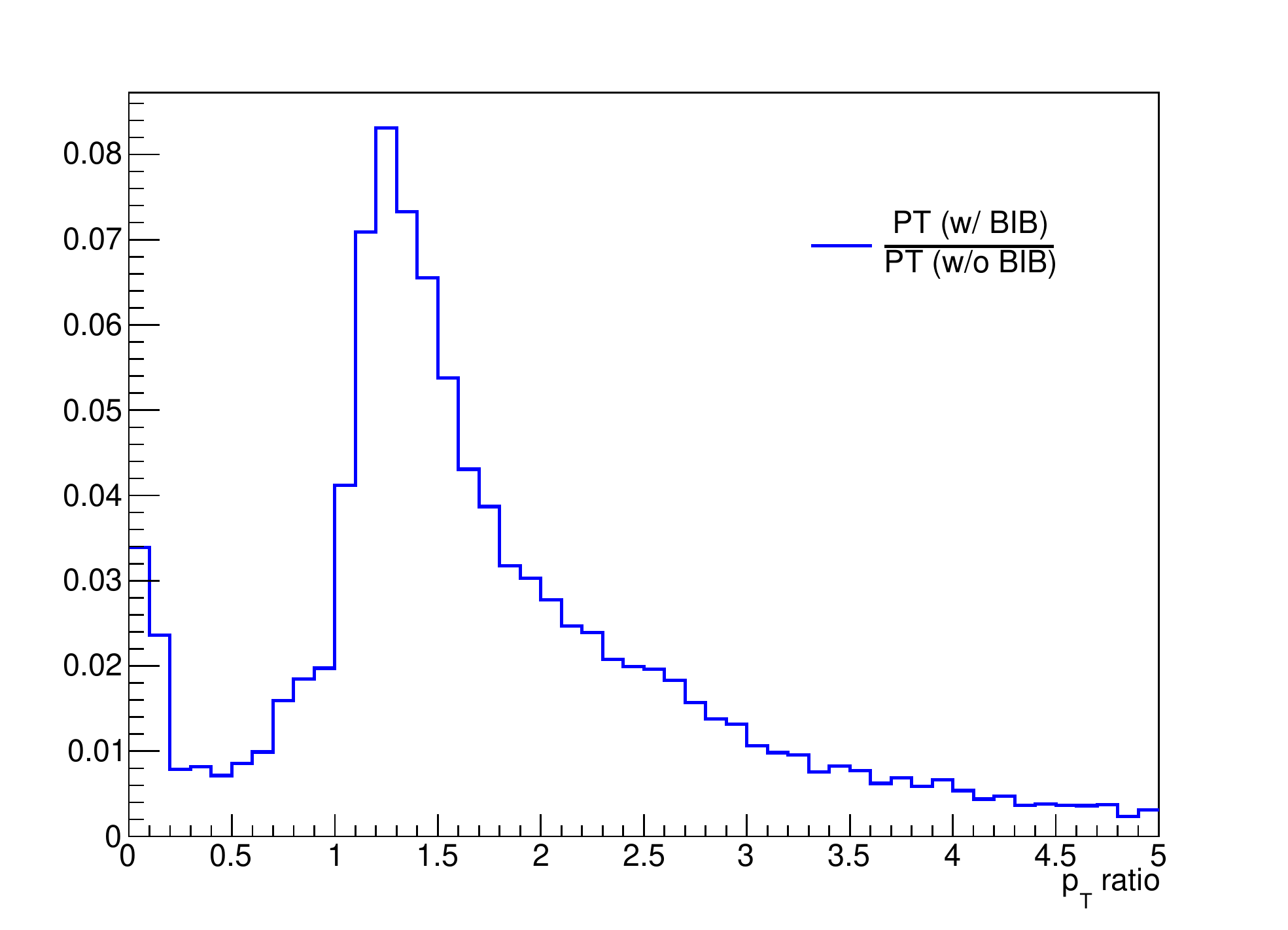}
    \includegraphics[scale=0.35]{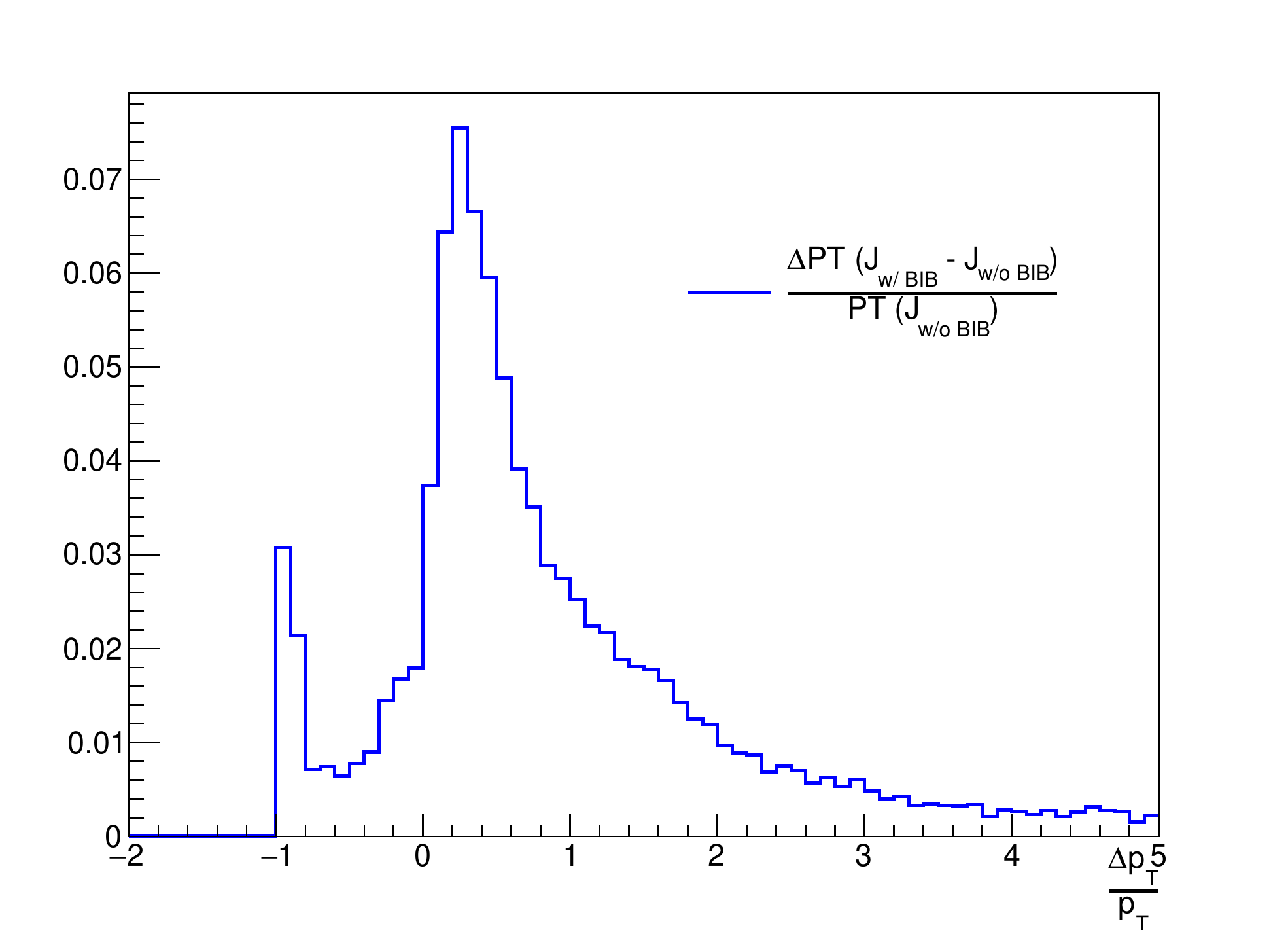}
    \caption{Ratio of jet $p_{T}$ with and without BIB after matching (left). Difference in $p_{T}$ of jets with and without BIB scaled by jet $p_{T}$ without BIB (right).}
    \label{PTratio_delPTbyPT}
    \includegraphics[scale=0.44]{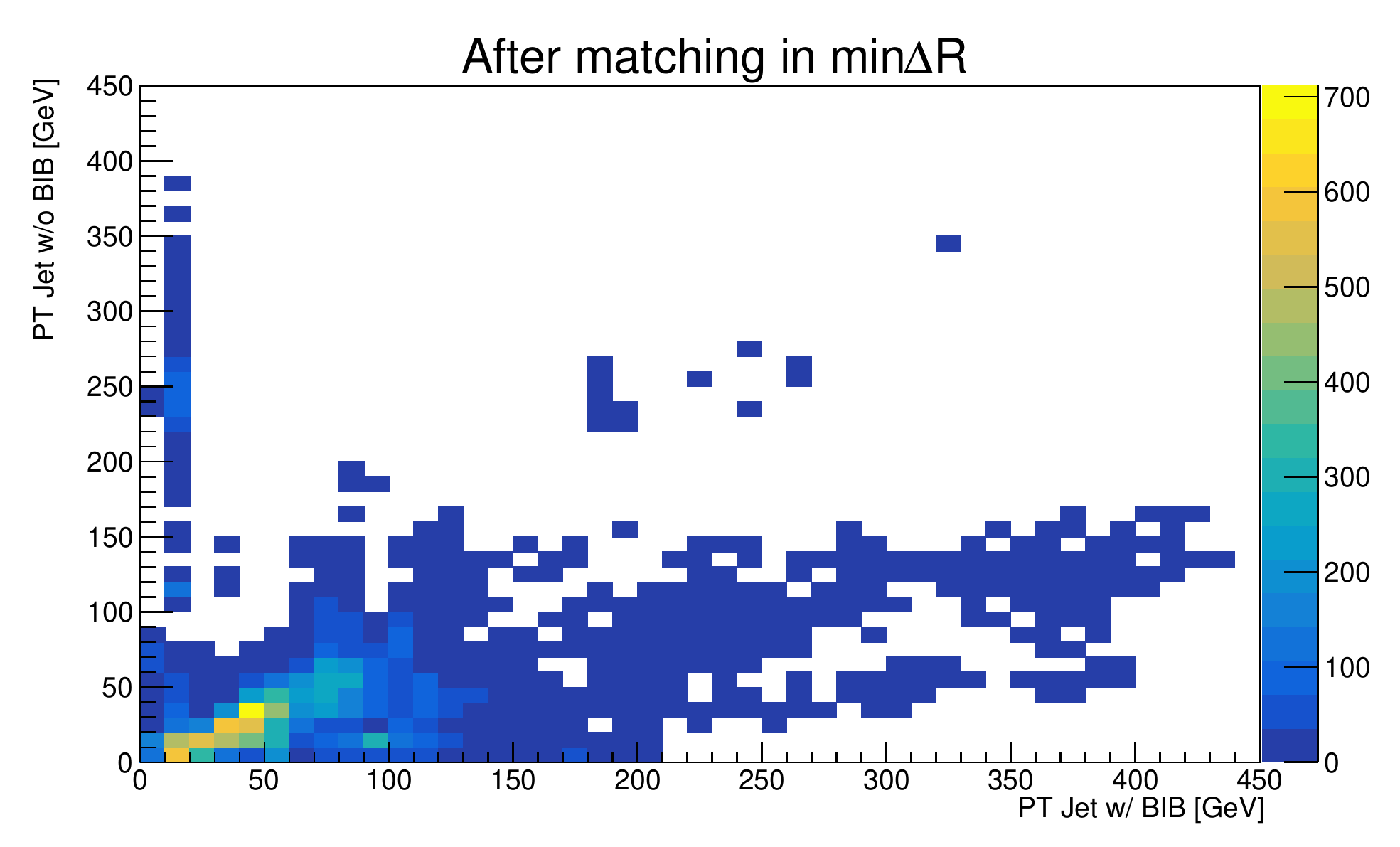}
    \caption{2D histogram of jet $p_{T}$ correlation after matching.}
    \label{2D_PT_Jsb_Js}
\end{figure}

\begin{figure}
    \centering
    \includegraphics[scale=0.35]{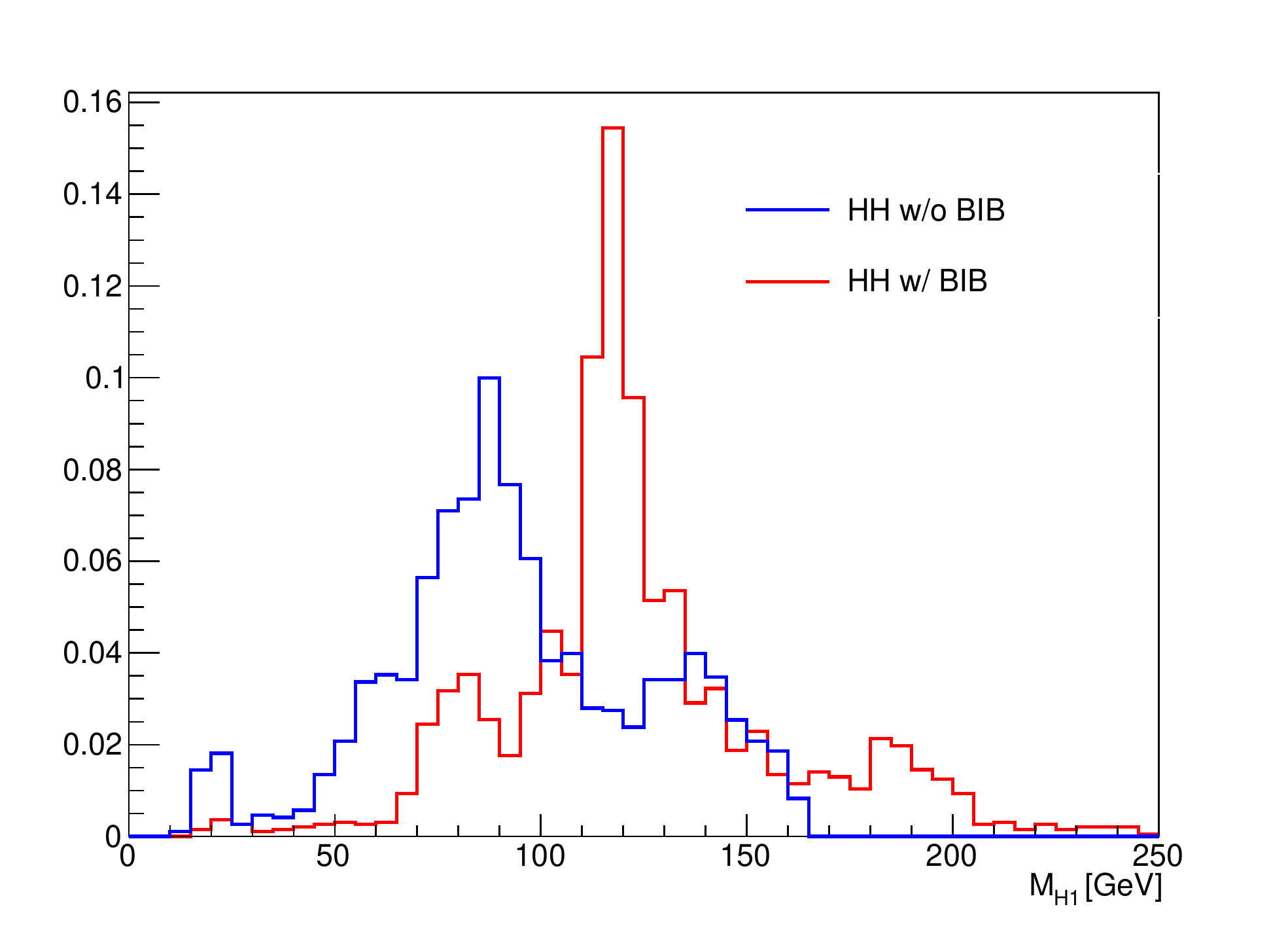}
    \includegraphics[scale=0.35]{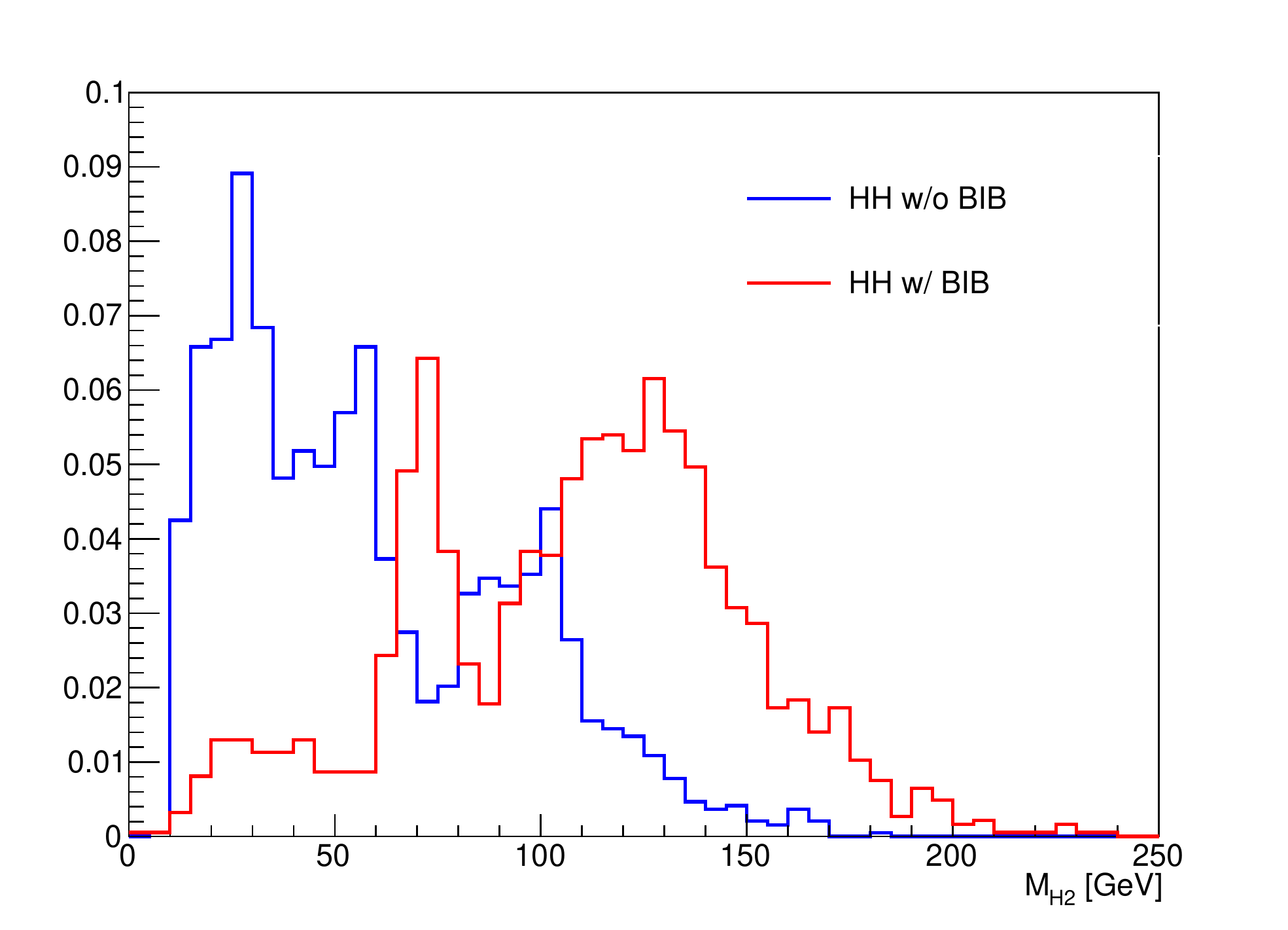}
    \caption{Invariant mass of leading (left) and sub-leading (right) di-jet pair after matching.}
    \label{m1_m2_matching}
\end{figure}

The presence of beam-induced background significantly increases the number of reconstructed jets and their momenta. Figure \ref{Nj_PT} (left) shows the number of jets before and after including BIB. Each jet without BIB is matched to its corresponding jet with BIB closest in $\Delta R$ (jet is only matched if $\Delta R < 0.7$). Figure \ref{Nj_PT} (right) shows the $p_{T}$ of matched jets. Figure \ref{PTratio_delPTbyPT} left plot shows the $p_{T}$ ratio of matched jets and the right plot shows $\frac{\Delta p_{T}}{p_{T,\mathrm{NoBIB}}}$, where $\Delta p_{T} = p_{T,\mathrm{BIB}} - p_{T,\mathrm{NoBIB}}$. Figure \ref{2D_PT_Jsb_Js} shows $p_{T}$ correlation between matched jets with and without BIB. These plots suggest that the presence of BIB increases the reconstructed momentum of jets and jet cleaning is required before proceeding with further studies. Figure \ref{m1_m2_matching} shows the invariant mass of leading (left) and sub-leading (right) di-jet pairs after employing the matching scheme. The mass peak near Higgs mass for HH with BIB is misleading as the BIB increases the energy contribution to the jets.



These are preliminary studies on jet reconstruction in presence of beam-induced background. We plan to employ a jet cleaning strategy to mitigate BIB and reduce the number of fake jets. The jet algorithm can be further improved to exploit calorimeter hit distribution features of BIB which our group is currently working on. 
\section{Di-Higgs Signal Significance Study - H. Jia}

Fast simulation using \texttt{Delphes} based description of the Muon Collider detector is used to study the performance at various center-of-mass energies and integrated luminosities~\cite{deFavereau2014}. Among all channels of di-Higgs decay, we have selected the $\nu \bar{\nu} b \bar{b} b \bar{b}$ channel with the largest branching ratio and other two channels with relatively low backgrounds: $\nu \bar{\nu} b \bar{b} \gamma \gamma$ and $\nu \bar{\nu} b \bar{b} \tau\tau$. For each channel, we neglect all backgrounds without neutrinos in the final states as we expect those backgrounds to be negligible with a simple cut on Missing Transverse Energy. All samples of each channel's signal and their respective dominant backgrounds are simulated using \texttt{MadGraph5\_aMC\@NLO} (MadGraph5) and hadronized using \texttt{Pythia8}~\cite{2015,2014}. \texttt{FastJet} anti-$k_T$ jet algorithm is used to reconstruct the jets using the Delphes physics objects, smeared charged tracks and calorimeter clusters~\cite{2012,2008}. The efficiency for very loose $b-$tagging configuration is modeled using a simple probability distribution with a mean efficiency of 90\%. The efficiency for $\tau-$tagging configuration is modeled with a mean efficiency of 80\% with a false rate of $0.1\%$ for electron and positron. For each channel, the signal significance is estimated by 
\begin{align*}
    \frac{S}{\sqrt{S+B}}.
\end{align*}
Comparing the result of full simulation without the BIB (Figure~\ref{m1_s_default}) and that of the Delphes study for $\nu \bar{\nu} b \bar{b} b \bar{b}$ channel, the resolution of the leading reconstructed Higgs invariant mass are comparable, with $\sigma\approx 20$ GeV. Improvements to $b$-tagging simulation, $\tau$-tagging simulation, and optimization of selection may be possible, but we anticipate that we have underestimated the performance that is achievable. However, we have neglected the all-important large beam-induced background in this Delphes study.

\subsection{The $\nu \bar{\nu} b \bar{b} b \bar{b}$ channel} 
For the $\nu \bar{\nu} b \bar{b} b \bar{b}$ channel, we have simulated both the signal, $\mu^+\mu^- \to \nu \bar{\nu} H H$, and background, 
$\mu^+\mu^- \to \nu \bar{\nu} H Z$ (Single Higgs), 
$\mu^+\mu^- \to \nu \bar{\nu} Z Z$ (ZZ), 
$\mu^+\mu^- \to \nu \bar{\nu} Z q \bar{q}$ (Z + qq), and 
$\mu^+\mu^- \to \nu \bar{\nu} b \bar{b} b \bar{b}$ (QCD) 
events using fast simulation. All $Z$ and $H$ boson decays are forced to be $H, Z \to b \bar{b}$. The analysis starts with the requirement of four jets with $p_T > 20$ GeV, which at least three of those are identified as $b$-tagged.

\begin{figure}[ht!]
    \centering
    \includegraphics[width=3in]{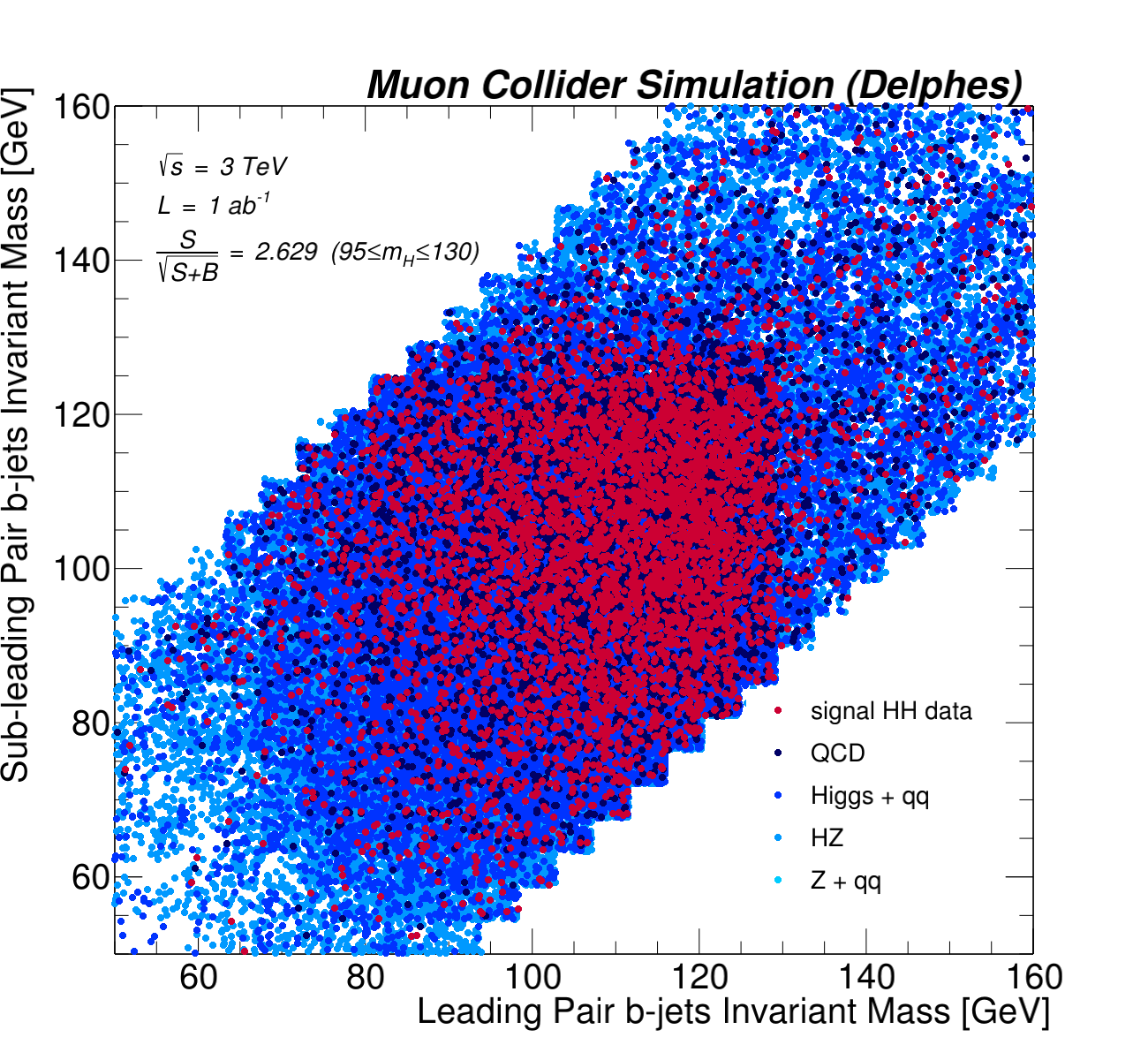}
    \includegraphics[width=3in]{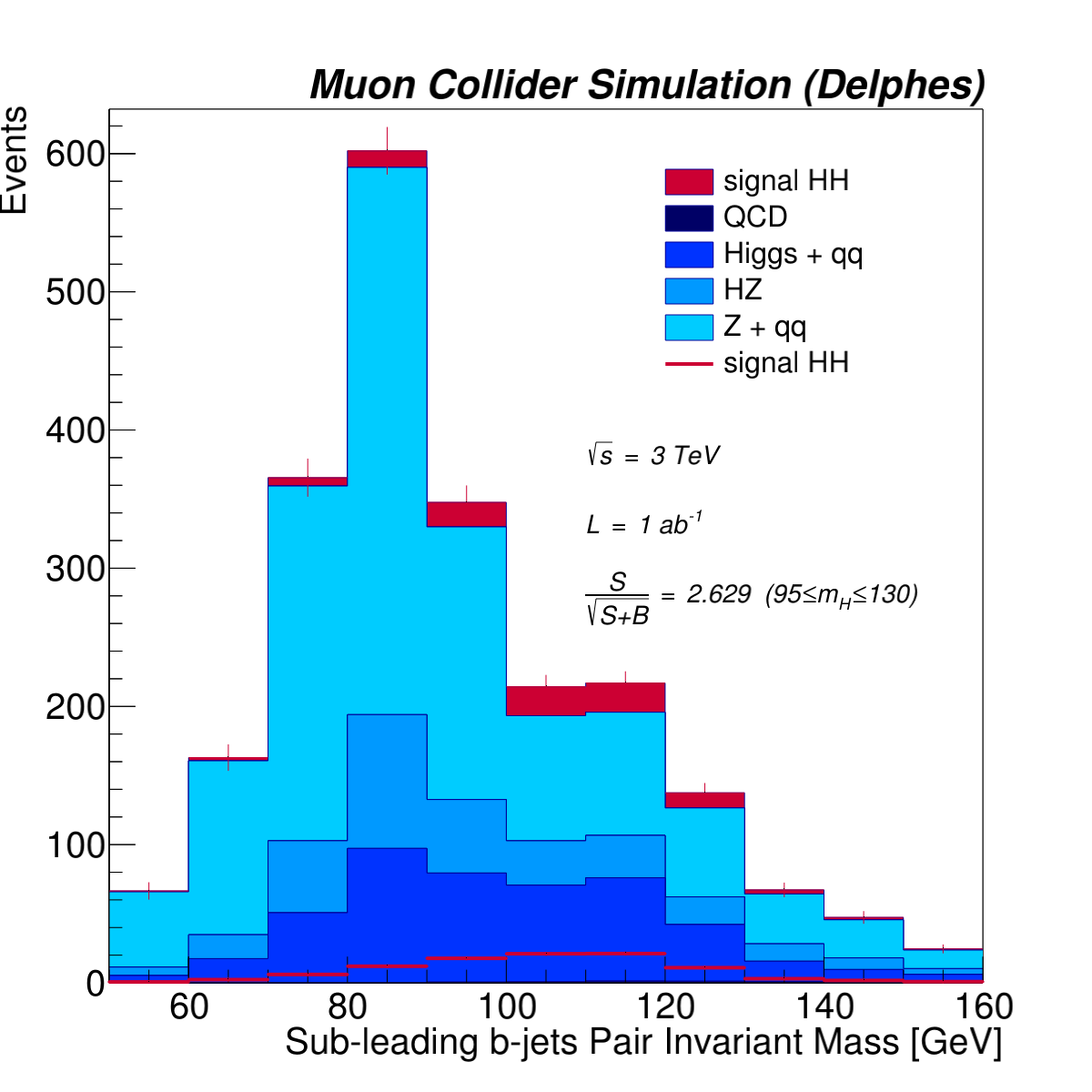}\\
    \includegraphics[width=3in]{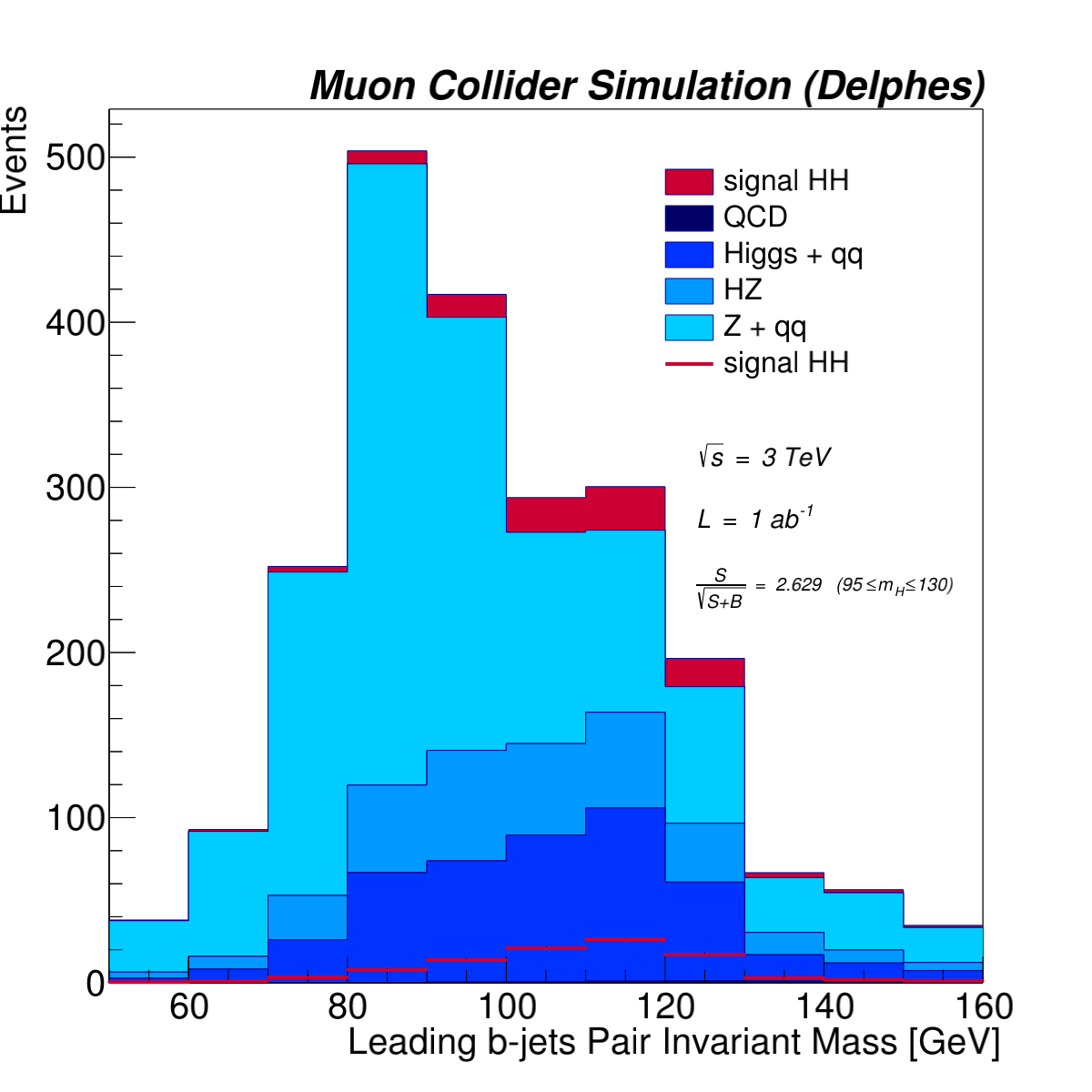}
    \includegraphics[width=3in]{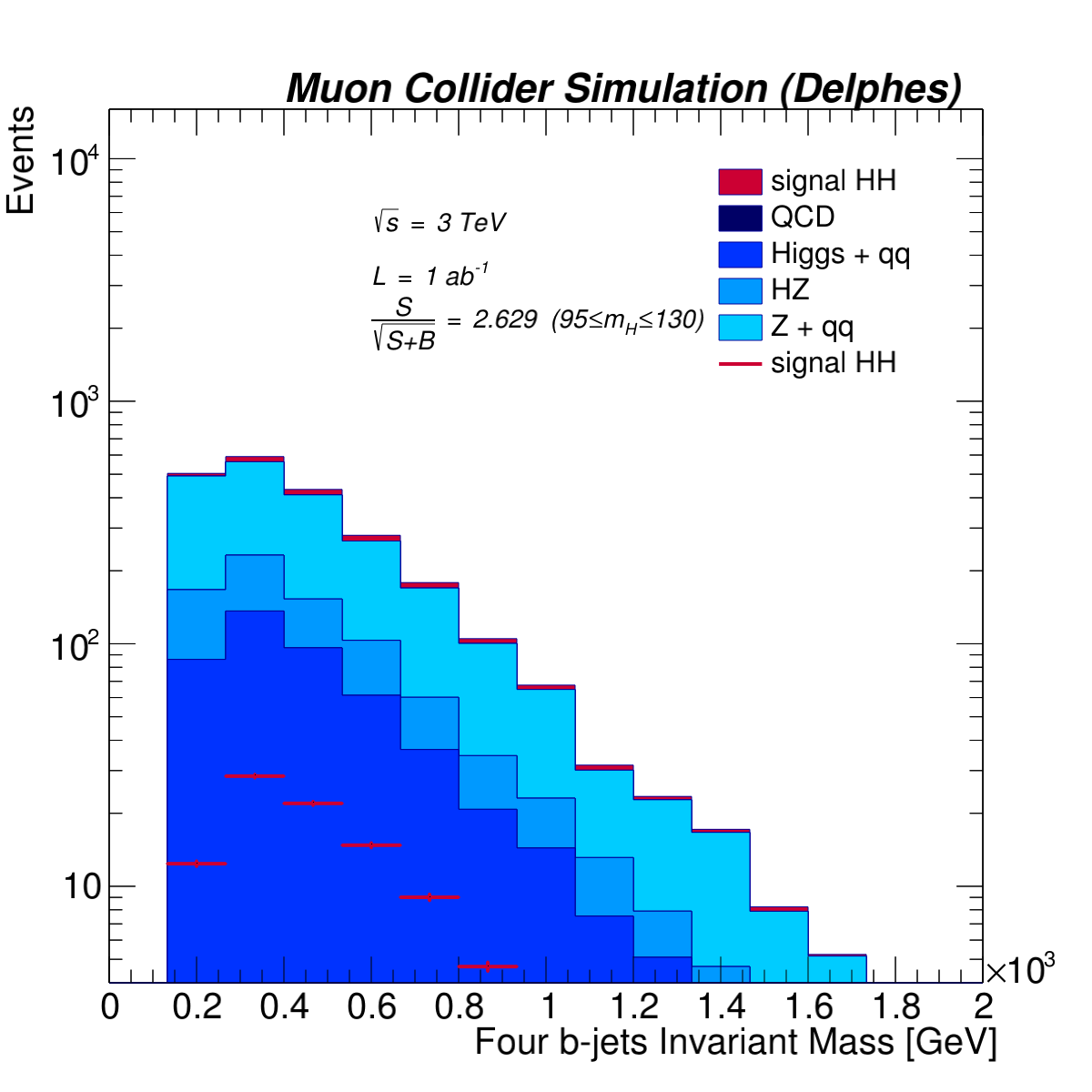}
    \caption{Event distribution of both signal and backgrounds channels is shown in the plane of the leading and sub-leading $b$-jets pair invariant mass for $\sqrt{s}$ = 3 TeV data. Signal is shown in red and the background in various shades of blue. Projections to the leading (bottom-left) and sub-leading (top-right) $b$-jets pair invariant mass are also shown. The four $b$-jet invariant mass is shown (bottom-right).}
    \label{JetPair_3TeV}
\end{figure}

\begin{figure}[ht!]
    \centering
    \includegraphics[width=3in]{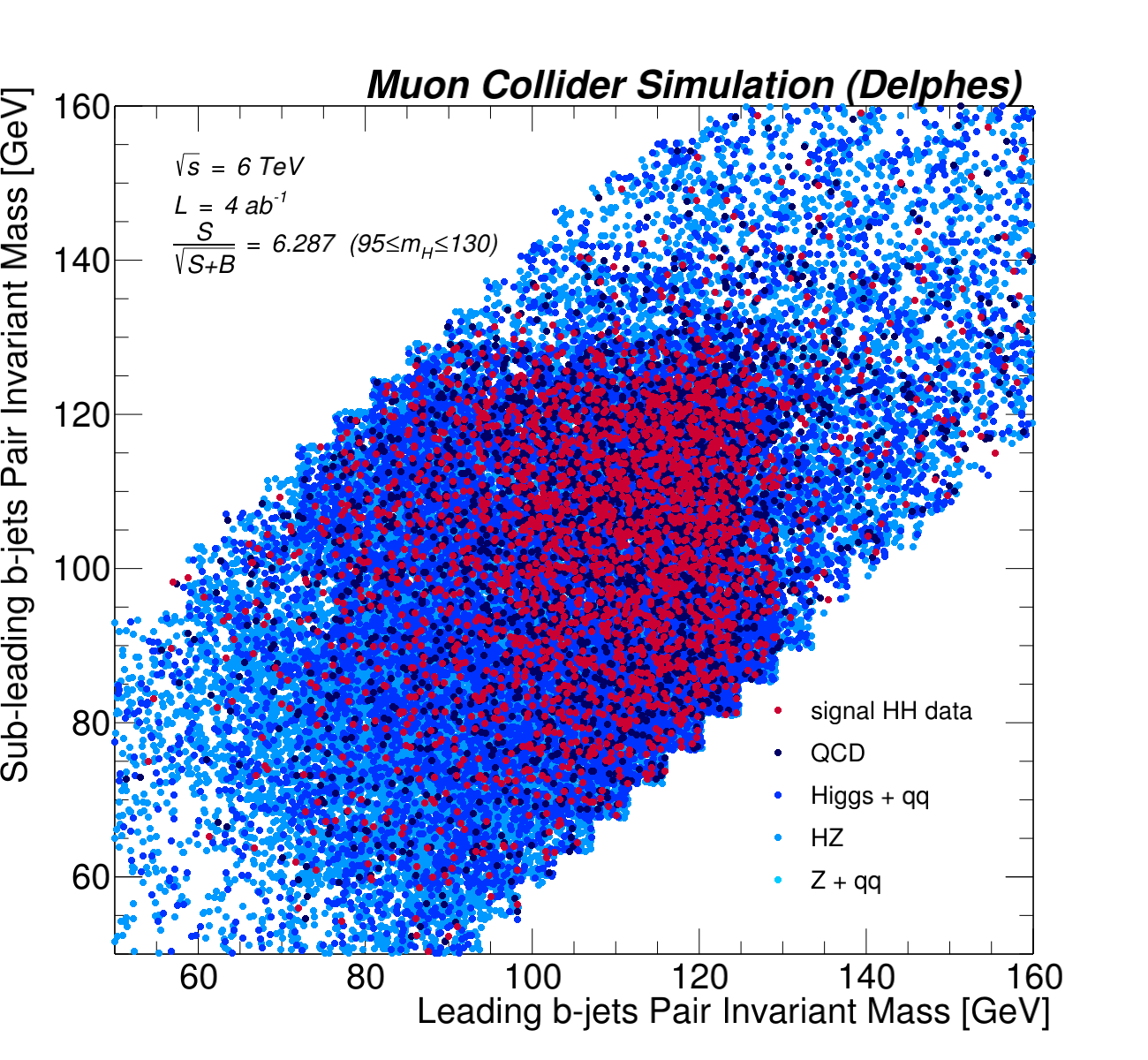}
    \includegraphics[width=3in]{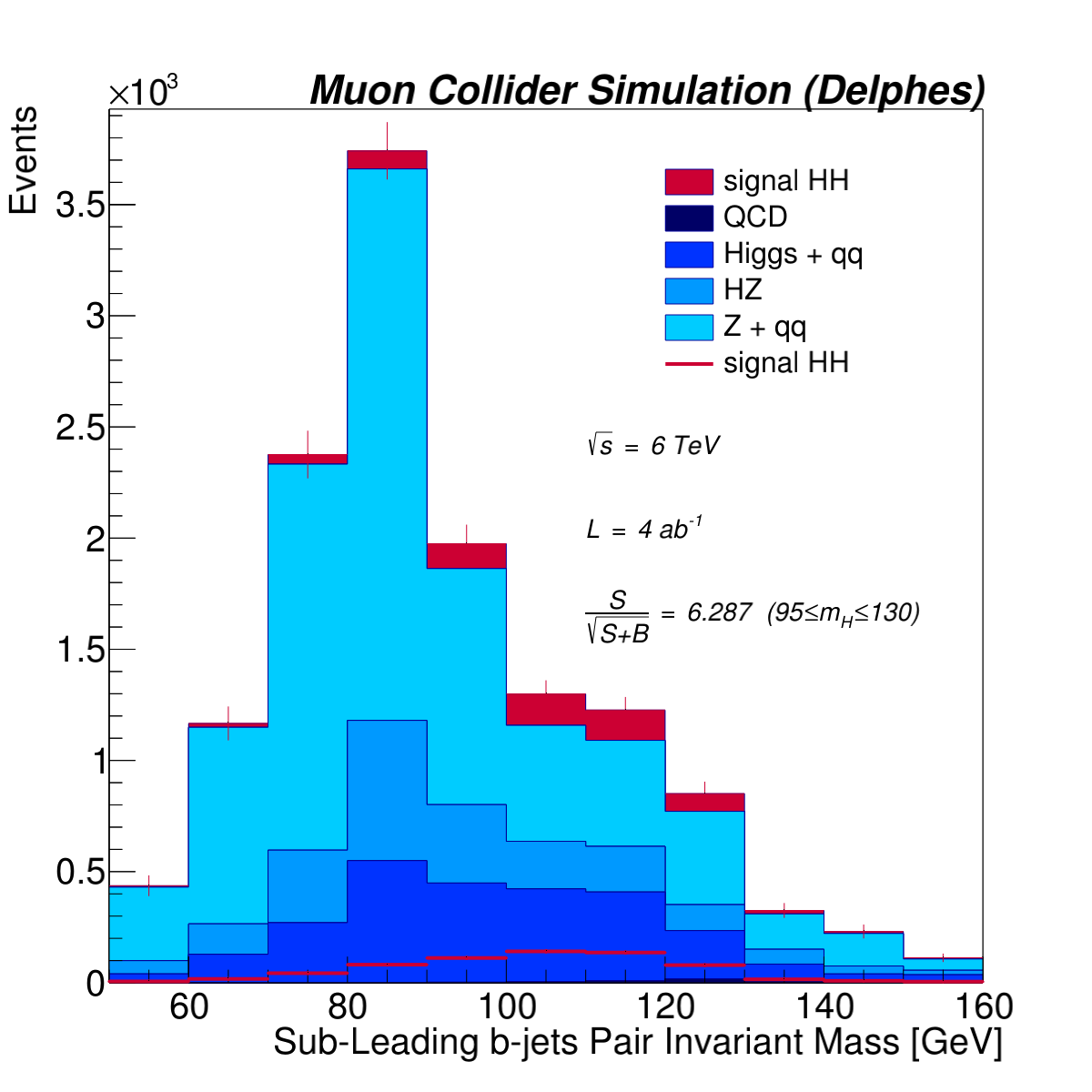}\\
    \includegraphics[width=3in]{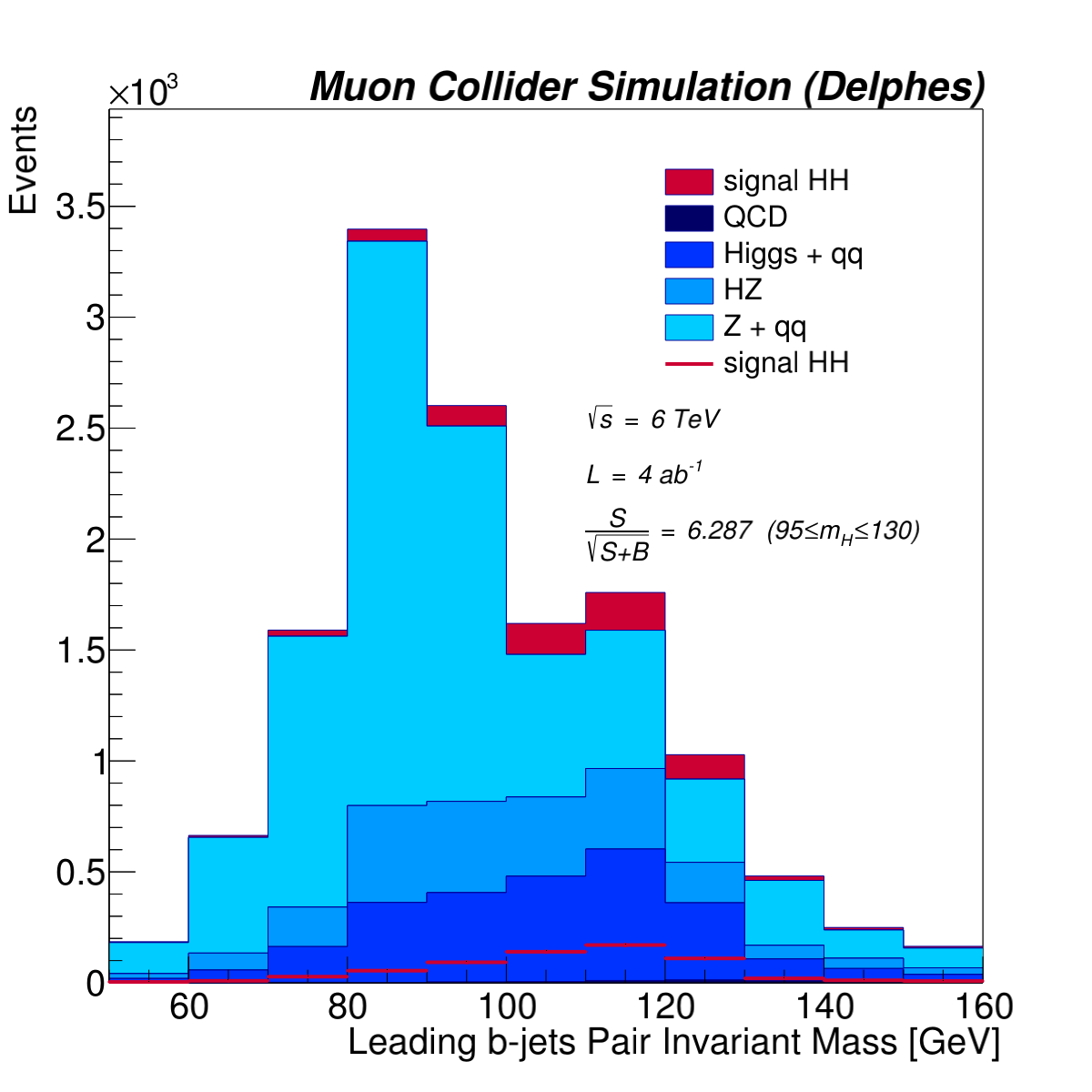}
    \includegraphics[width=3in]{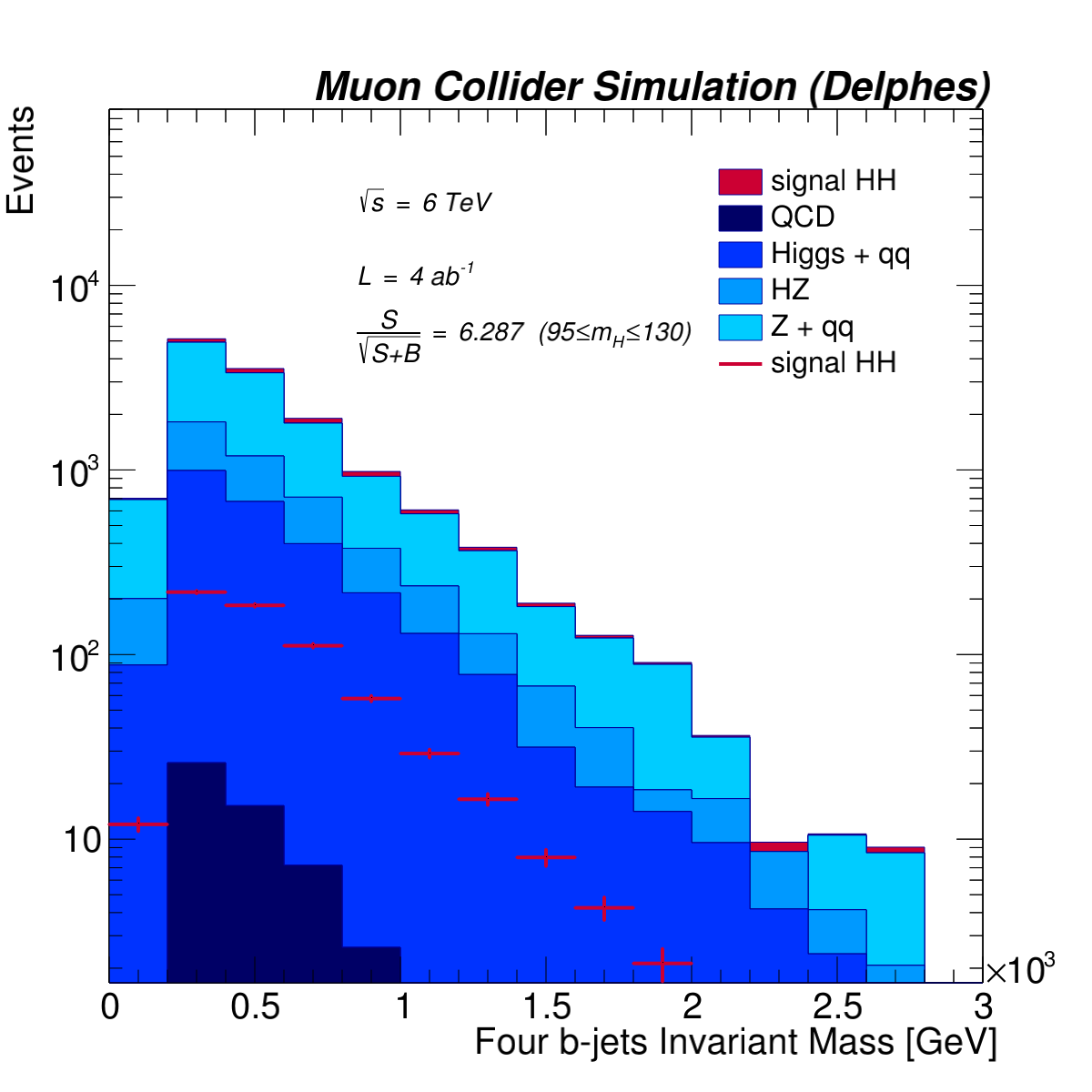}
    \caption{Event distribution of both signal and backgrounds channels is shown in the plane of the leading and sub-leading $b$-jets pair invariant mass for $\sqrt{s}$ = 6 TeV data. Signal is shown in red and the background in various shades of blue. Projections to the leading (bottom-left) and sub-leading (top-right) $b$-jets pair invariant mass are also shown. The four $b$-jet invariant mass is shown (bottom-right).}
    \label{JetPair_6TeV}
\end{figure}

\begin{figure}[ht!]
    \centering
    \includegraphics[width=3in]{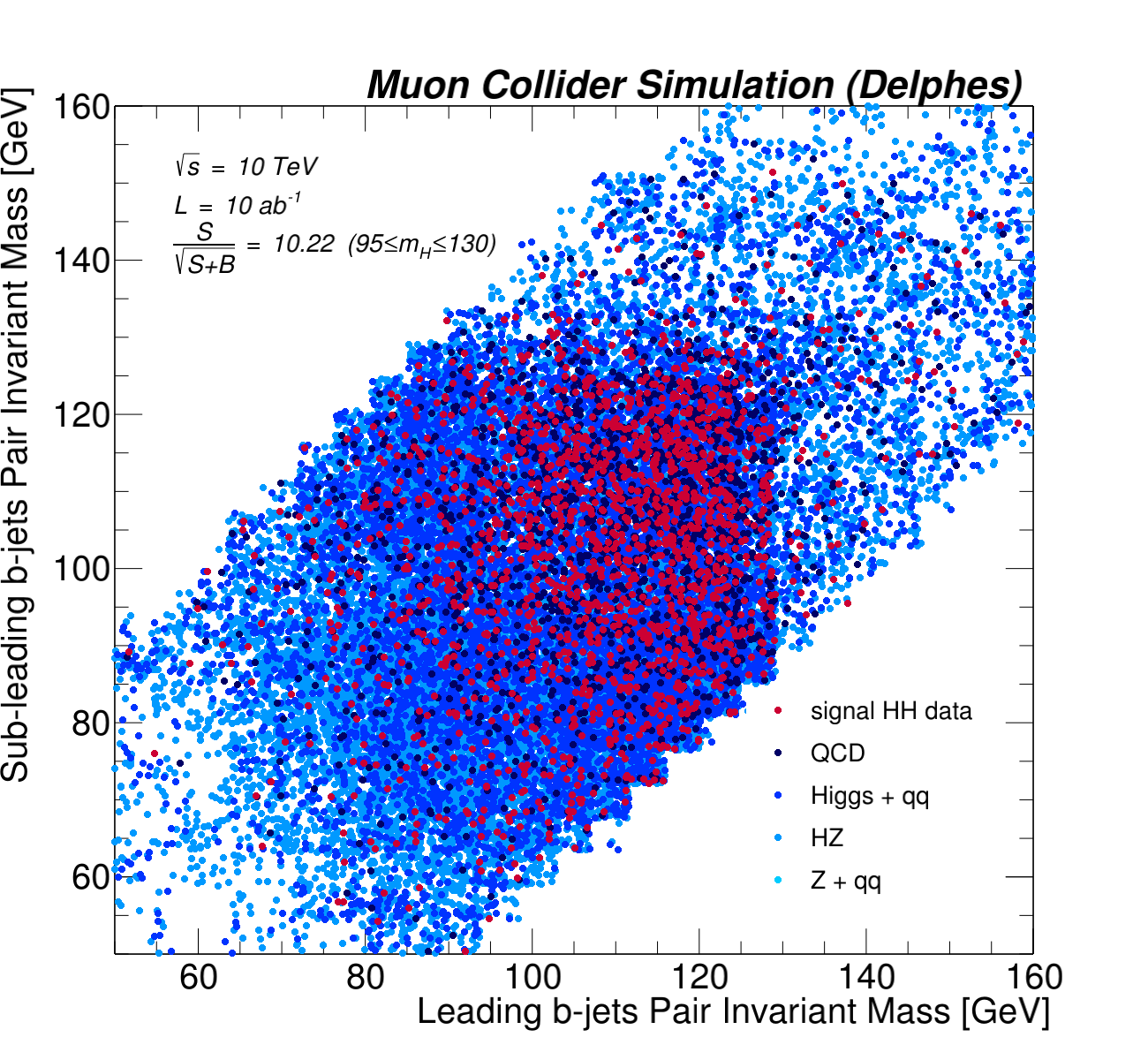}
    \includegraphics[width=3in]{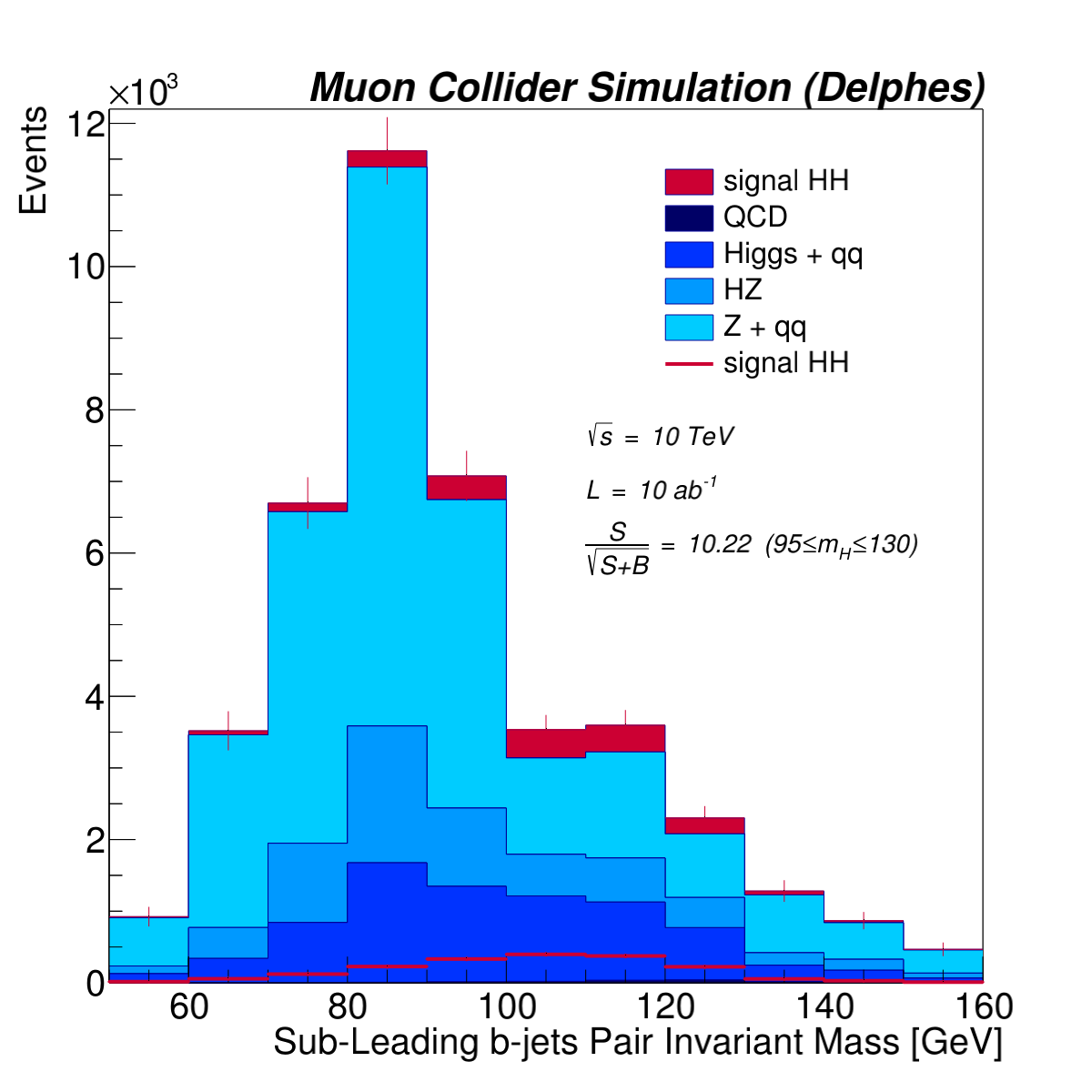}\\
    \includegraphics[width=3in]{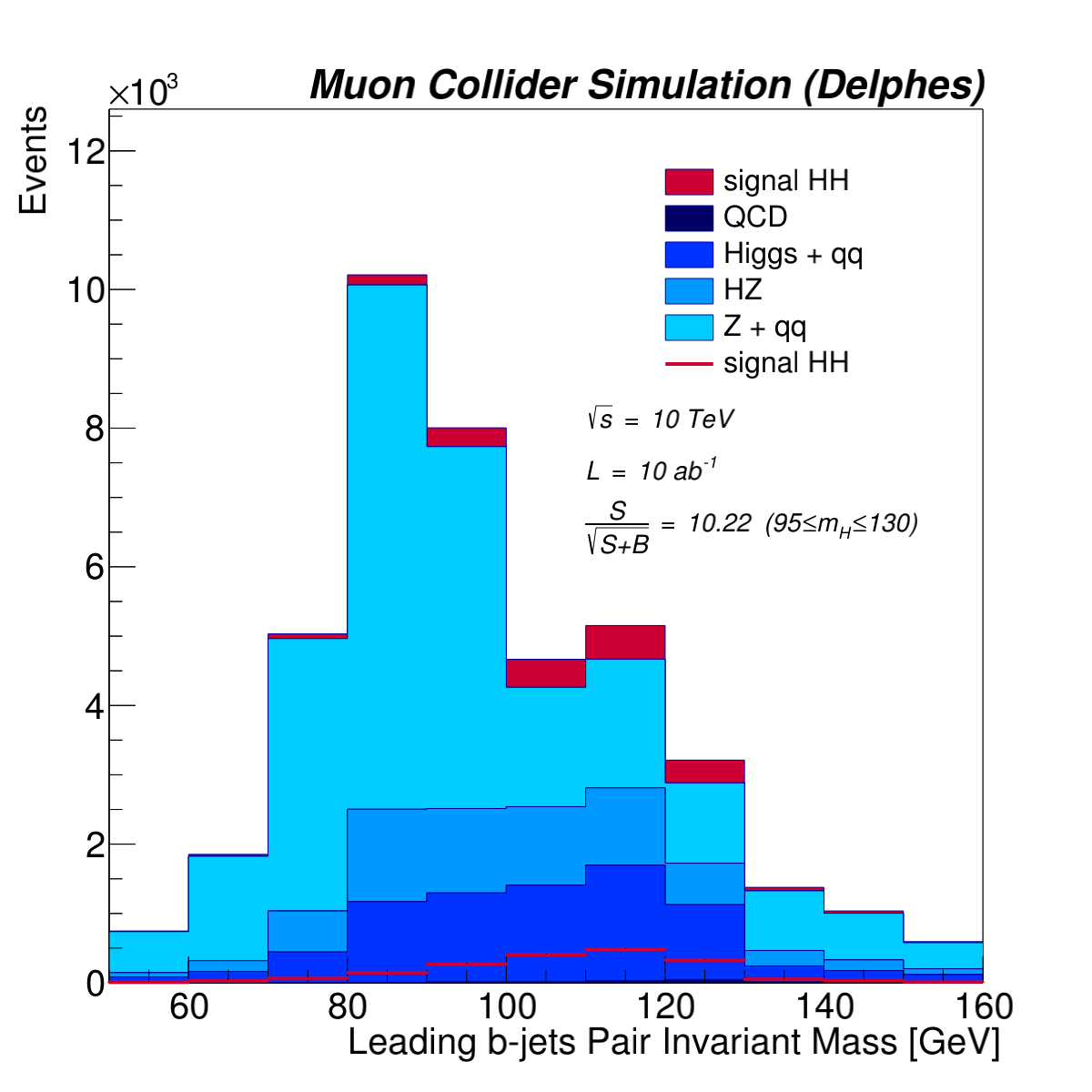}
    \includegraphics[width=3in]{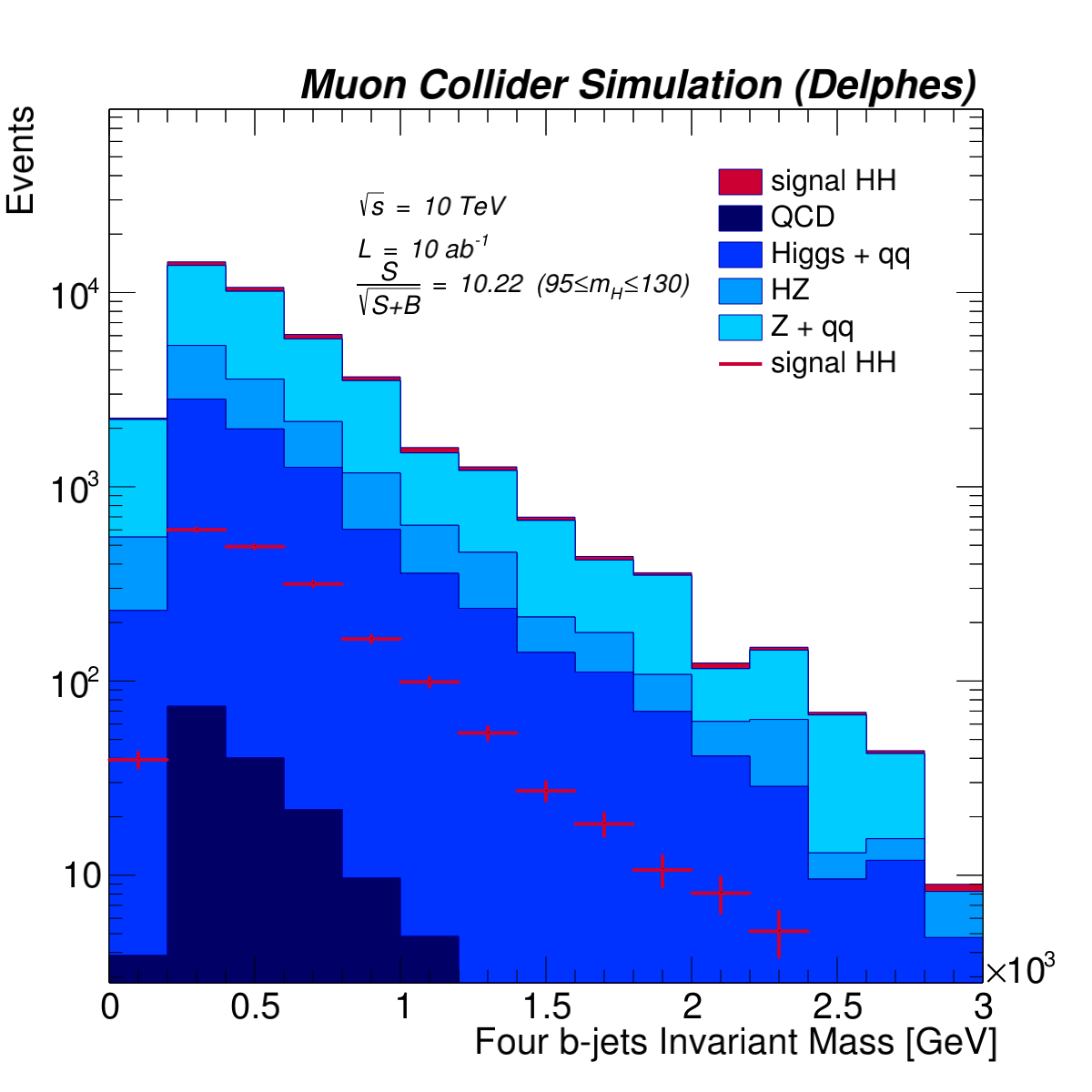}
    \caption{Event distribution of both signal and backgrounds channels is shown in the plane of the leading and sub-leading $b$-jets pair invariant mass for $\sqrt{s}$ = 10 TeV data. Signal is shown in red and the background in various shades of blue. Projections to the leading (bottom-left) and sub-leading (top-right) $b$-jets pair invariant mass are also shown. The four $b$-jet invariant mass is shown (bottom-right).}
    \label{JetPair_10TeV}
\end{figure}

\begin{figure}[ht!]
    \centering
    \includegraphics[width=3in]{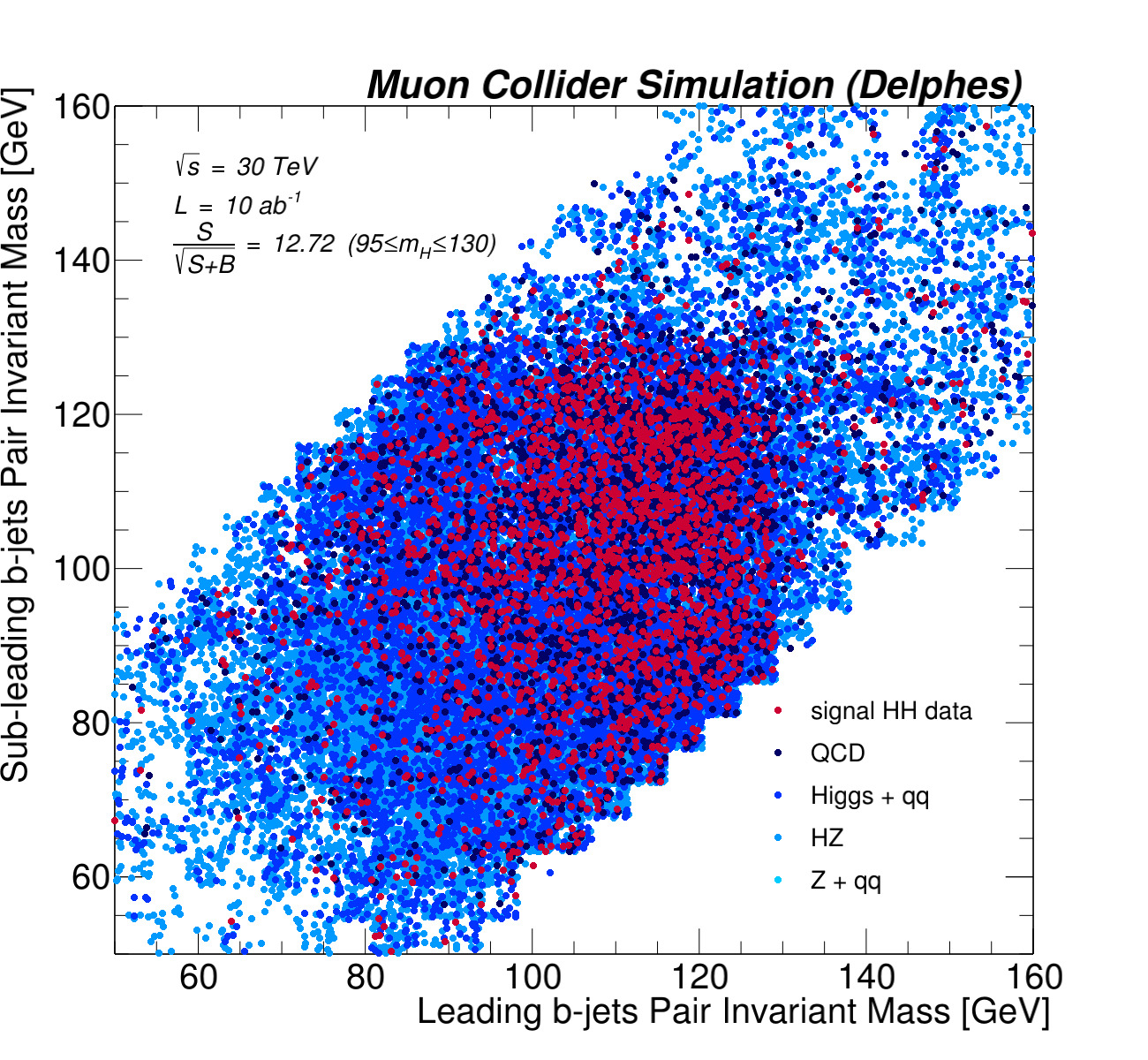}
    \includegraphics[width=3in]{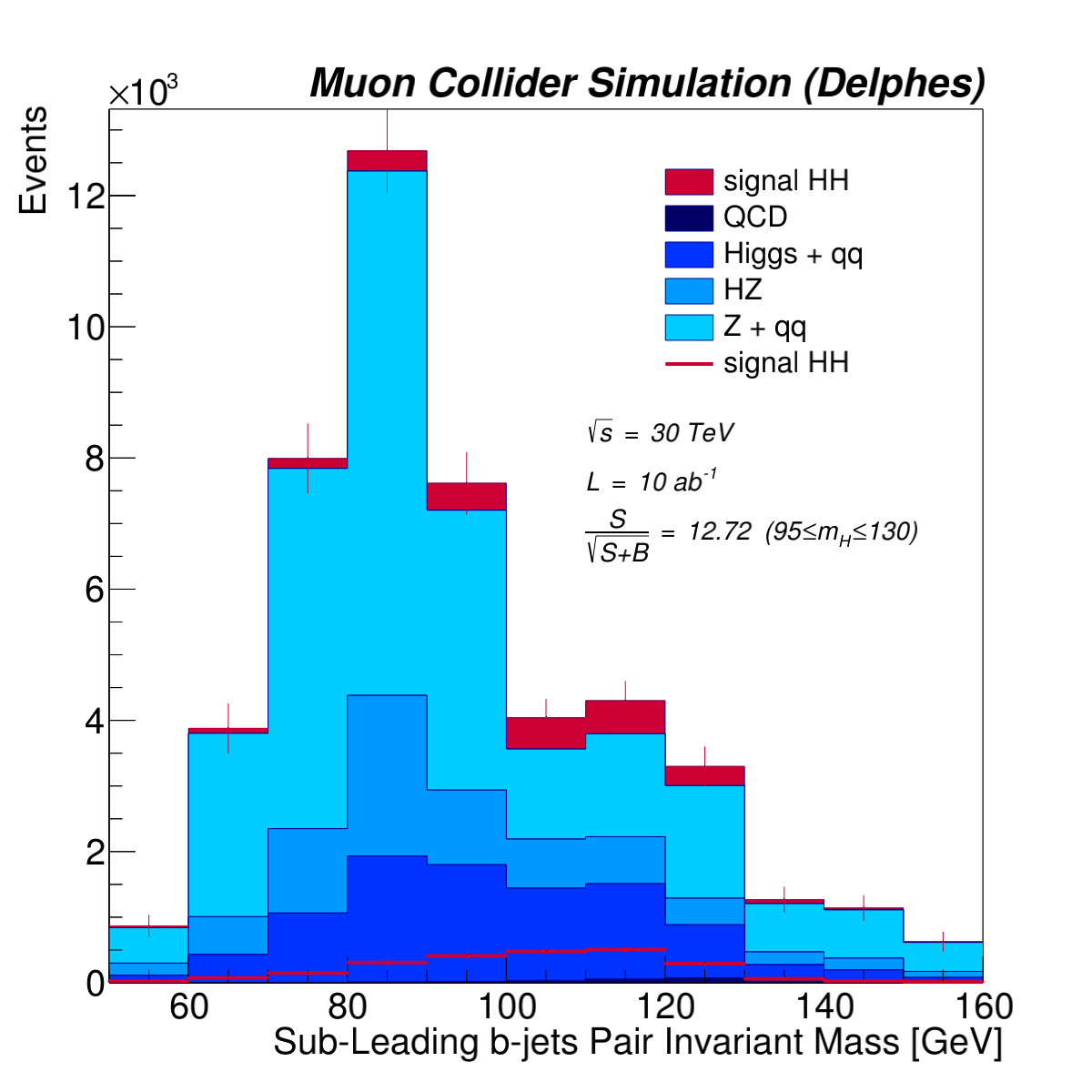}\\
    \includegraphics[width=3in]{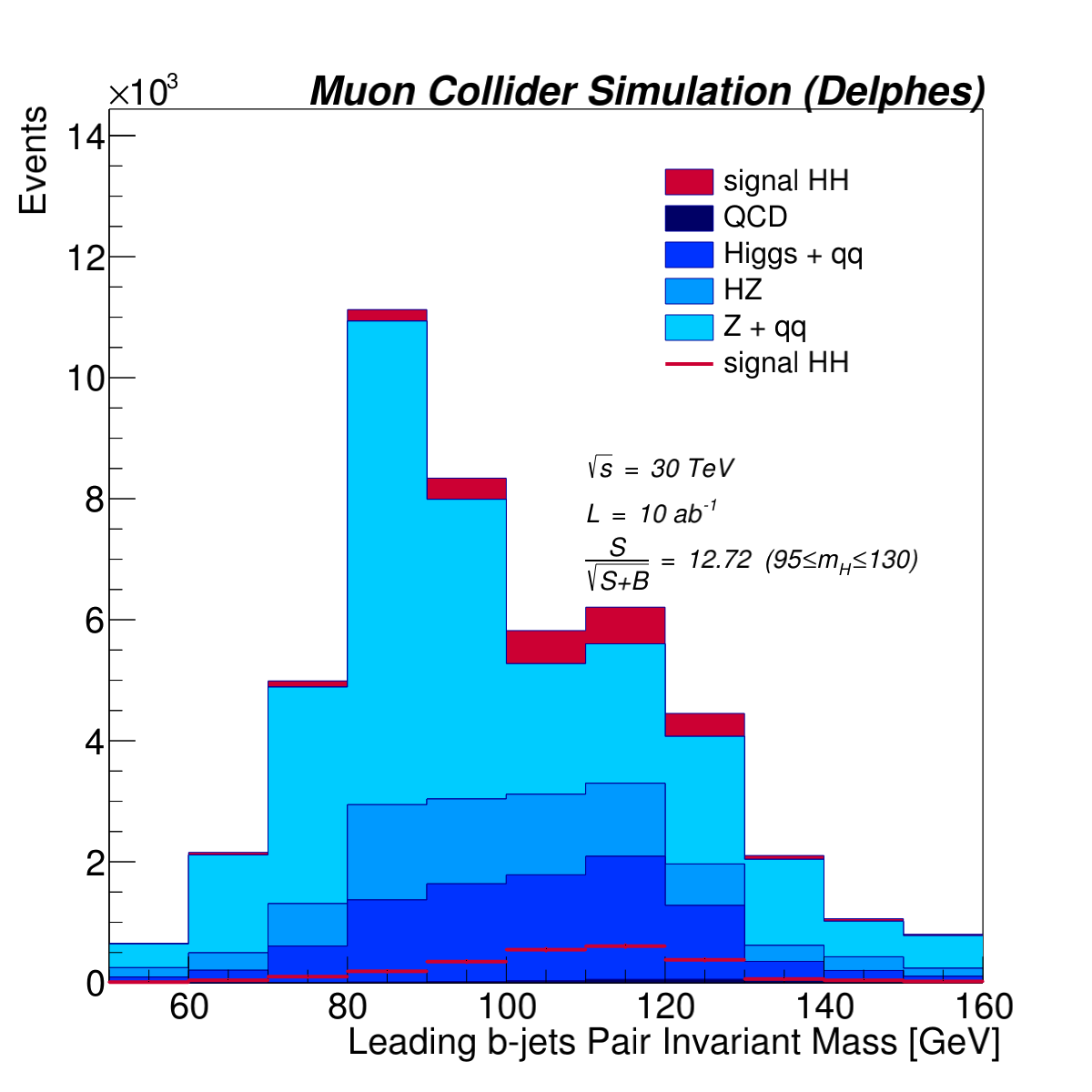}
    \includegraphics[width=3in]{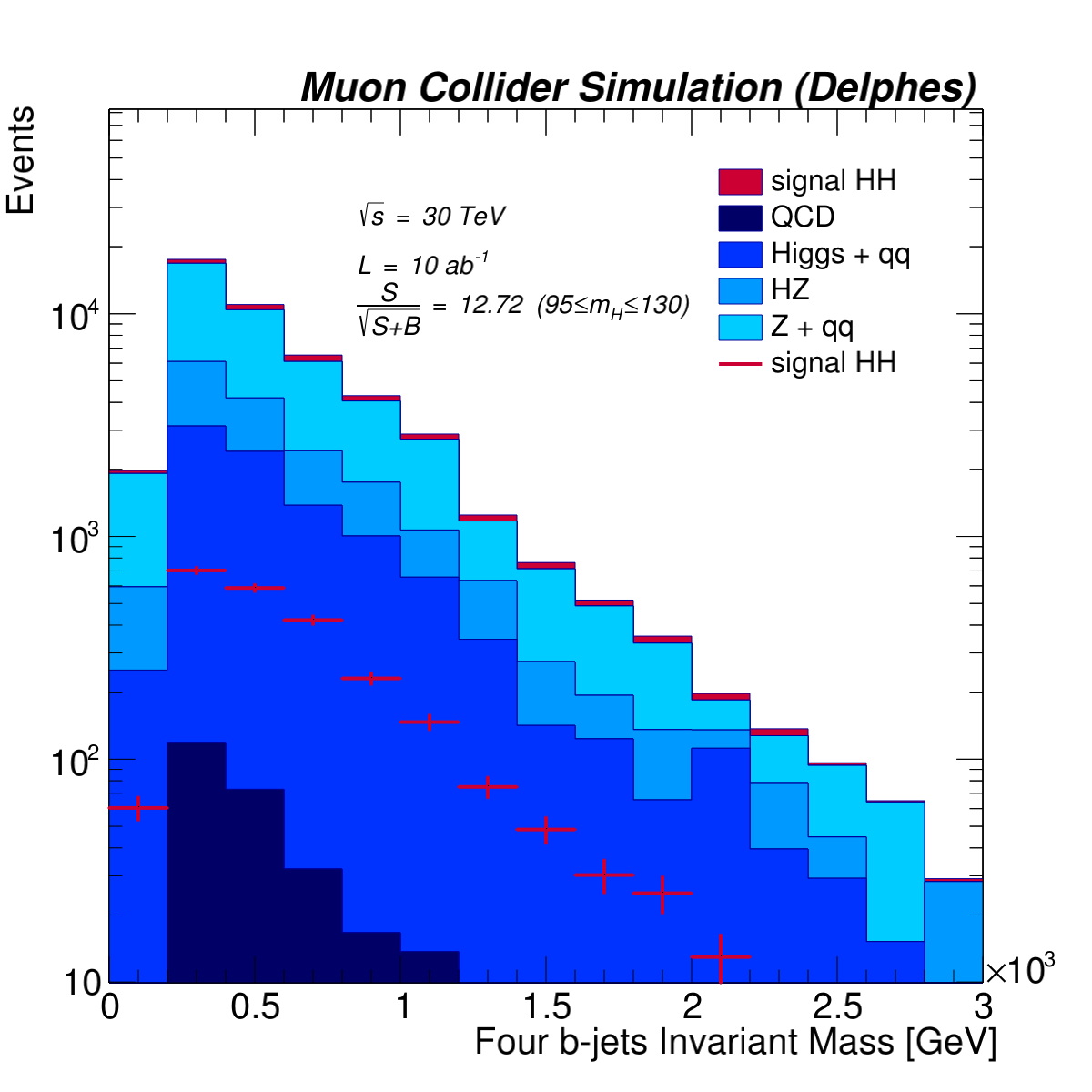}
    \caption{Event distribution of both signal and backgrounds channels is shown in the plane of the leading and sub-leading $b$-jets pair invariant mass for $\sqrt{s}$ = 30 TeV data. Signal is shown in red and the background in various shades of blue. Projections to the leading (bottom-left) and sub-leading (top-right) $b$-jets pair invariant mass are also shown. The four $b$-jet invariant mass is shown (bottom-right).}
    \label{JetPair_30TeV}
\end{figure}

All combinations of $b$-jet pairs are formed, and the pairing which minimizing 
\begin{align*}
    (M_1-M_H)^2+(M_2-M_H)^2,
\end{align*}
where $M_{H} = 125$ GeV and $M_1, M_2$ are the best two di-jets pairs, is used for further analysis. A scatter plot of the leading and sub-leading $b$-jet pair invariant mass is shown in the top-left planes of the Figures~\ref{JetPair_3TeV}, \ref{JetPair_6TeV}, \ref{JetPair_10TeV}, and \ref{JetPair_30TeV} for 3, 6, 10, and 30 TeV center of mass energies. Projects to the sub-leading and leading $b$-jet pair invariant mass are shown in the top-right and bottom-left planes of the same figures. The bottom-right plane of these figures shows the quad-$b$-jet invariant mass. Note that the calibration of the jet energies is yet to be optimized. Nevertheless, the Z and the H mass peaks are separated and can be used for extracting the signal significance. 

The signal events are plotted in red. The QCD background, i.e., events in which $b$-pairs emanate from radiated gluons, shown in dark blue, was found to be negligible. The Z-pair and Z + qq backgrounds are shown in lighter shades of blue and are seen to peak lower than the Higgs. The most-difficult-to-model background, shown in dark shade of blue, is coming from single Higgs production with $b$-pair radiation, where both leading and sub-leading $b$-pair invariant masses are peaked in the Higgs region. Using integrated luminosity settings of 1, 4, 10 and 10 ab$^{-1}$ for 3, 6, 10, and 30 TeV center of mass energies, a simple cut and count analysis is performed with tight cuts around both (uncalibrated) $b$-pair invariant mass: 
\begin{align*}
    95\text{ GeV} < m_{bb} < 130\text{ GeV},
\end{align*}
with another simple cut imposed on the four jets invariant mass:
\begin{align*}
     M_{4b} > 150\text{ GeV}.
\end{align*} The estimated significance results are shown in the figures and tabulated for all four collider settings in the Table~\ref{tab:Delphes_HH}.

As is to be expected, the single Higgs production accompanied by $b$-pair radiation from the final state is the biggest challenge to handle. The jet resolutions are such that the Z and H peaks are overlapping, resulting in difficult signal extraction. Further improvement in the analysis is possible using angular variables and more sophisticated machine learning algorithms. 

\begin{table}[ht!]
    \centering
    \begin{tabular}{|c|c|}
        \hline
        &\\
        $\sqrt{s}~(\int{d{\cal L}})$ & Estimated signal significance\\
        &\\\hline
        &\\
        3 TeV (1 ab$^{-1}$) & 2.629\\
        &\\\hline
        &\\
        6 TeV (4 ab$^{-1}$) & 6.287\\
        &\\\hline
        &\\
        10 TeV (10 ab$^{-1}$) & 10.22\\
        &\\\hline
        &\\
        30 TeV (10 ab$^{-1}$) & 12.72\\
        &\\\hline
    \end{tabular}
    \caption{Significance for the extraction of di-Higgs to four $b$ quarks events for muon colliders operating at various centers of mass and integrated luminosity.}
    \label{tab:Delphes_HH}
\end{table}

\clearpage
\subsection{The $\nu \bar{\nu} b \bar{b} \gamma\gamma$ channel} 
\begin{figure}[ht!]
    \centering
    \includegraphics[width=3in]{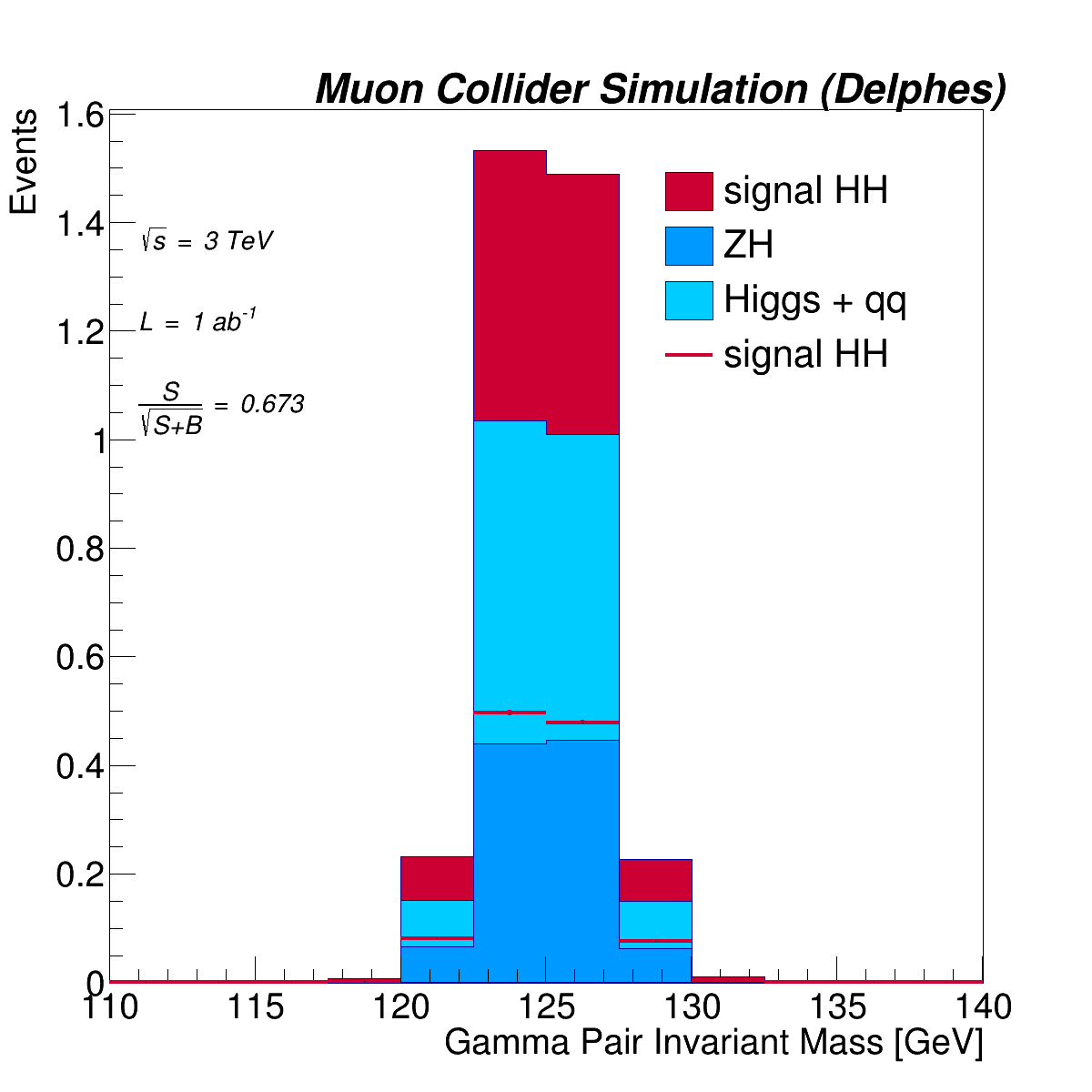}
    \includegraphics[width=3in]{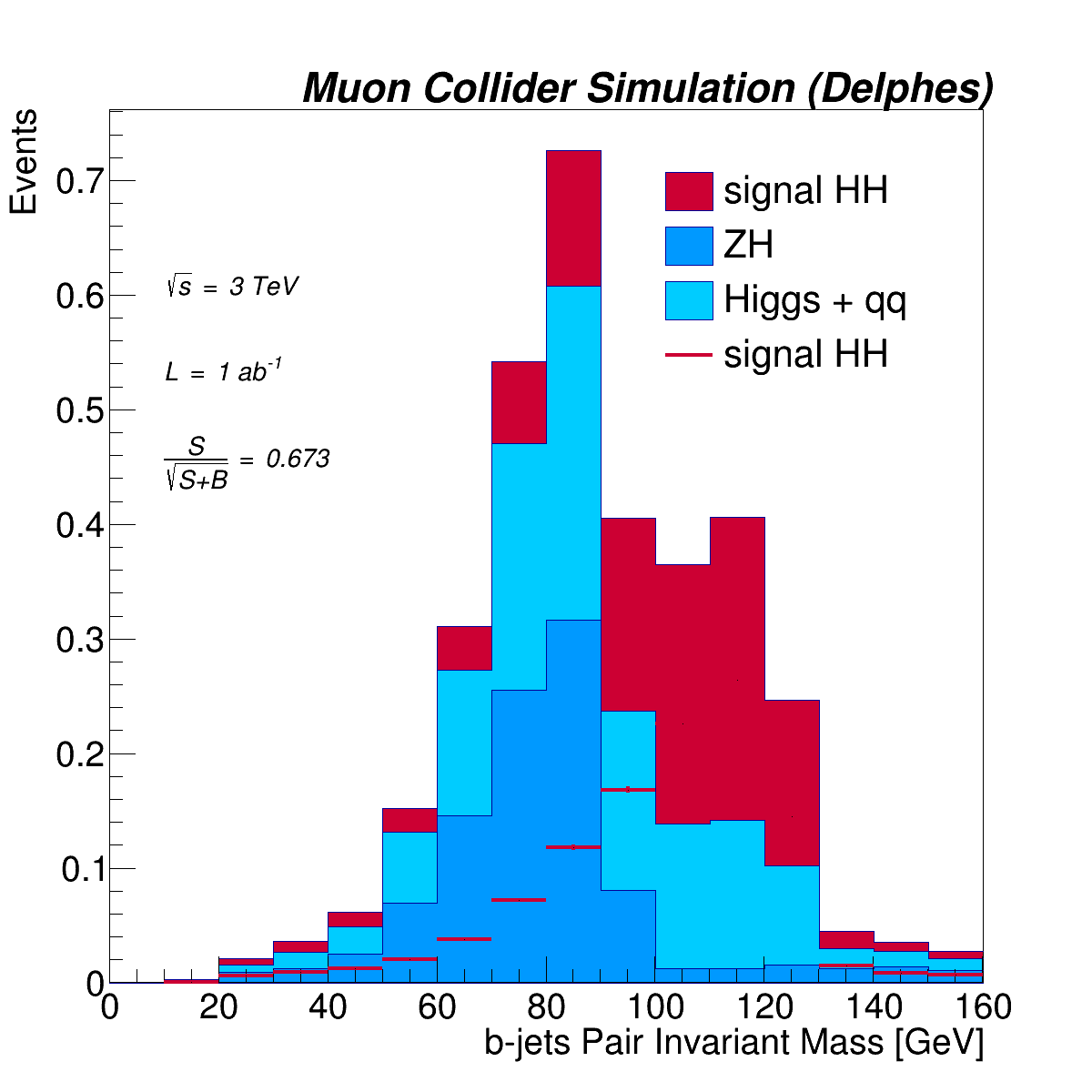}\\
    \includegraphics[width=3in]{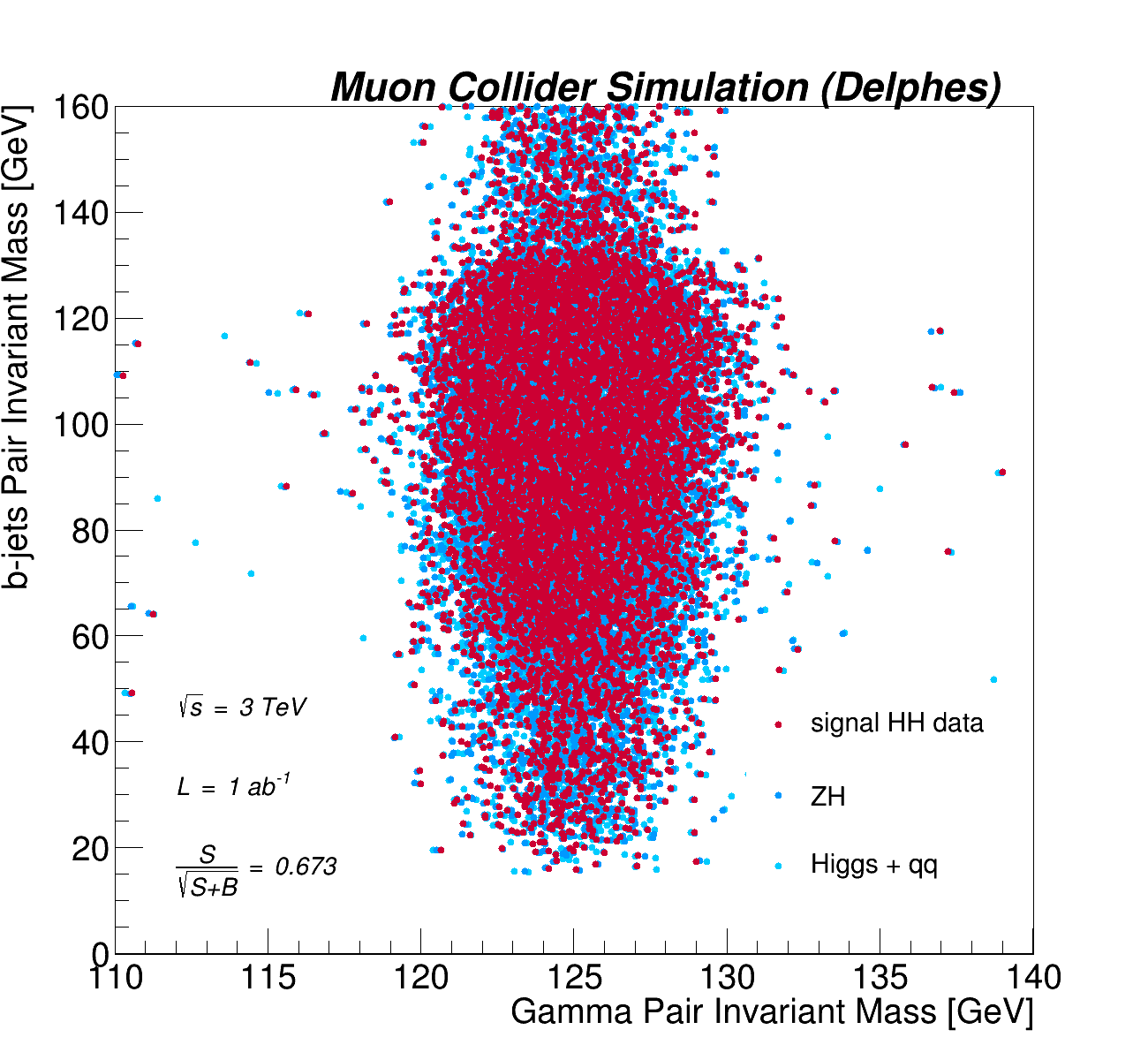}
    \caption{Event distribution of both signal and backgrounds channels is shown in the plane of the $b$-jet pair and the $\gamma\gamma$ pair invariant mass for $\sqrt{s}$ = 3 TeV data (bottom). Signal is shown in red and the background in various shades of blue. Projections to the $\gamma$ pair (top-left) and the $b$-jet pair (top-right) invariant mass are also shown.}
    \label{JetPair_bbaa_3TeV}
\end{figure}

\begin{figure}[ht!]
    \centering
    \includegraphics[width=3in]{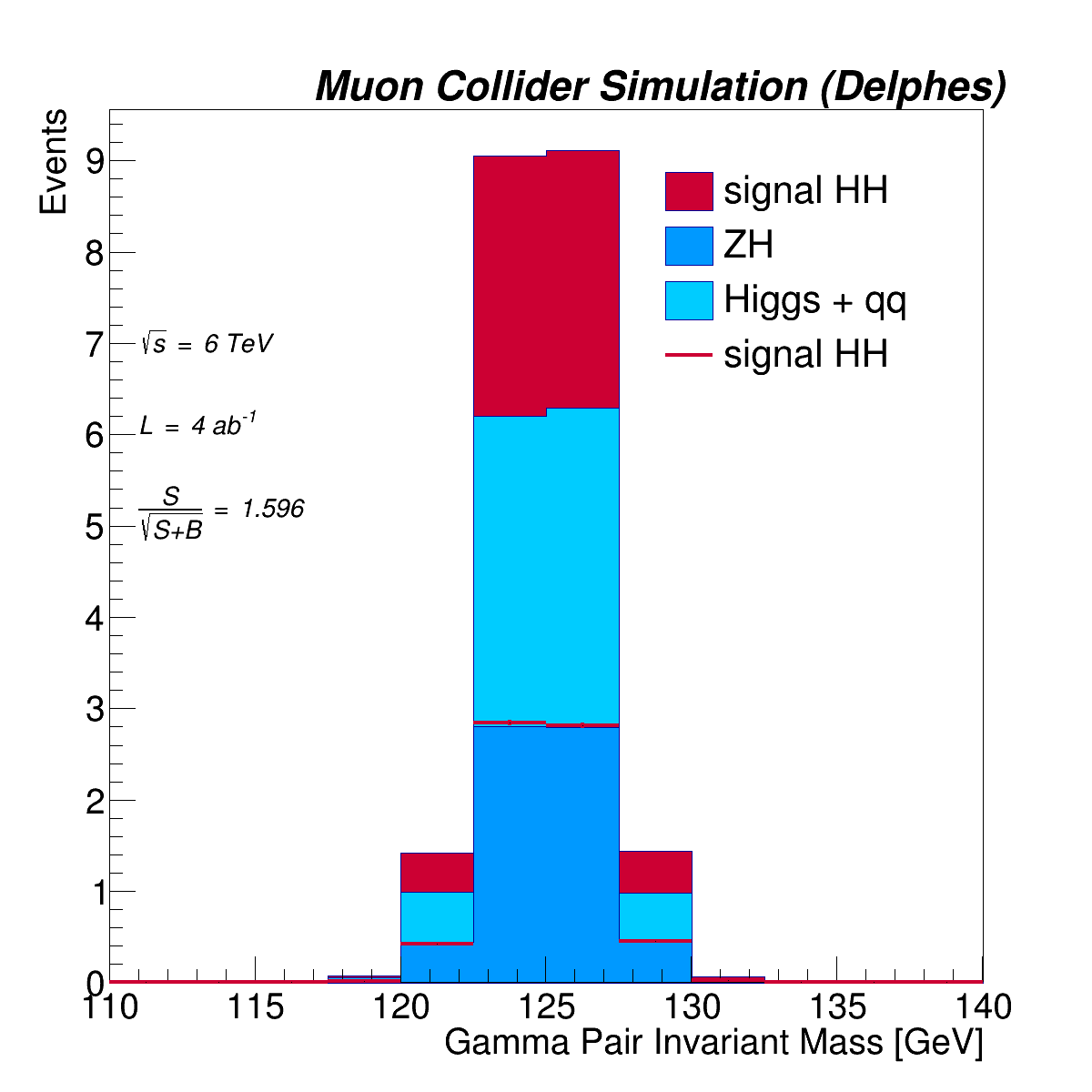}
    \includegraphics[width=3in]{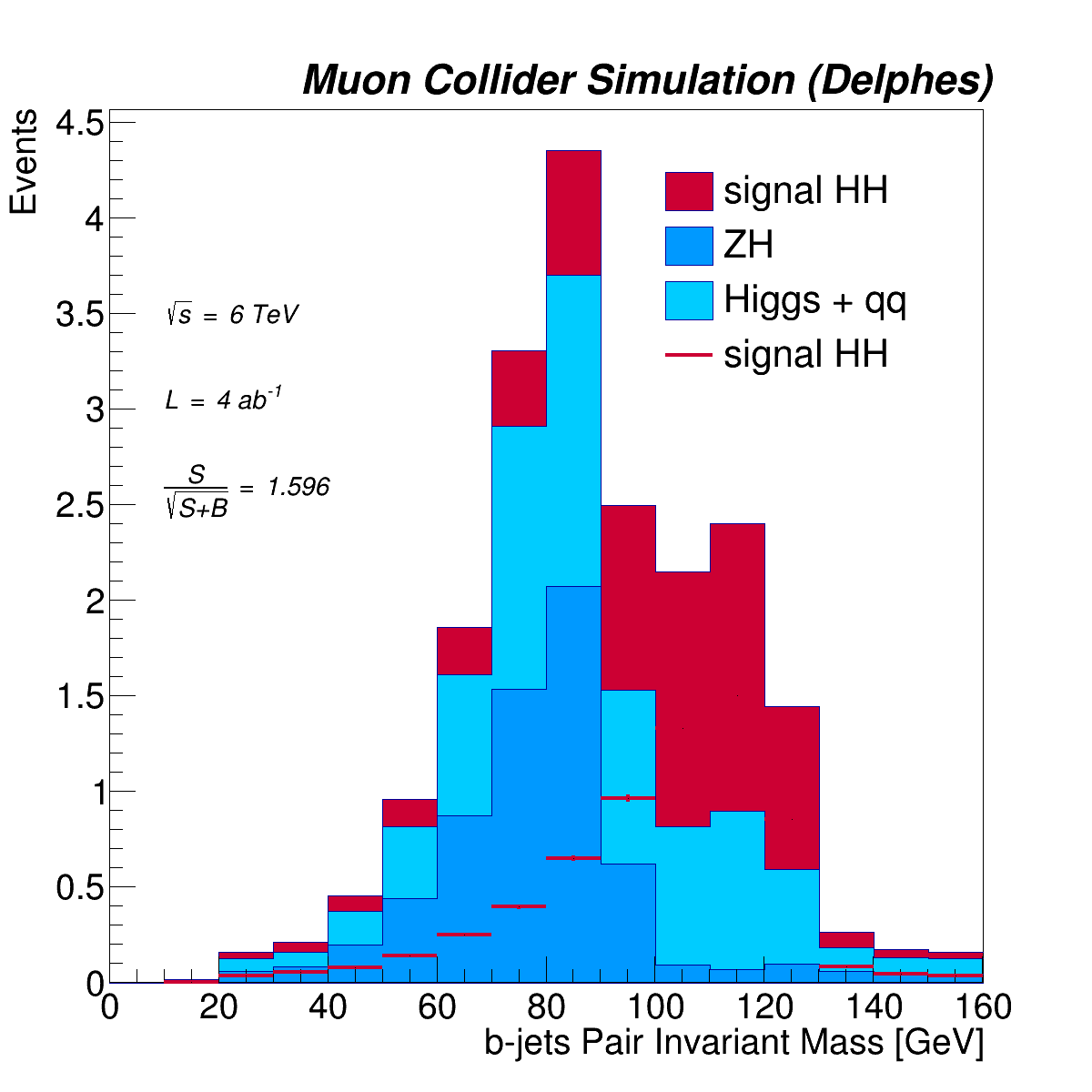}\\
    \includegraphics[width=3in]{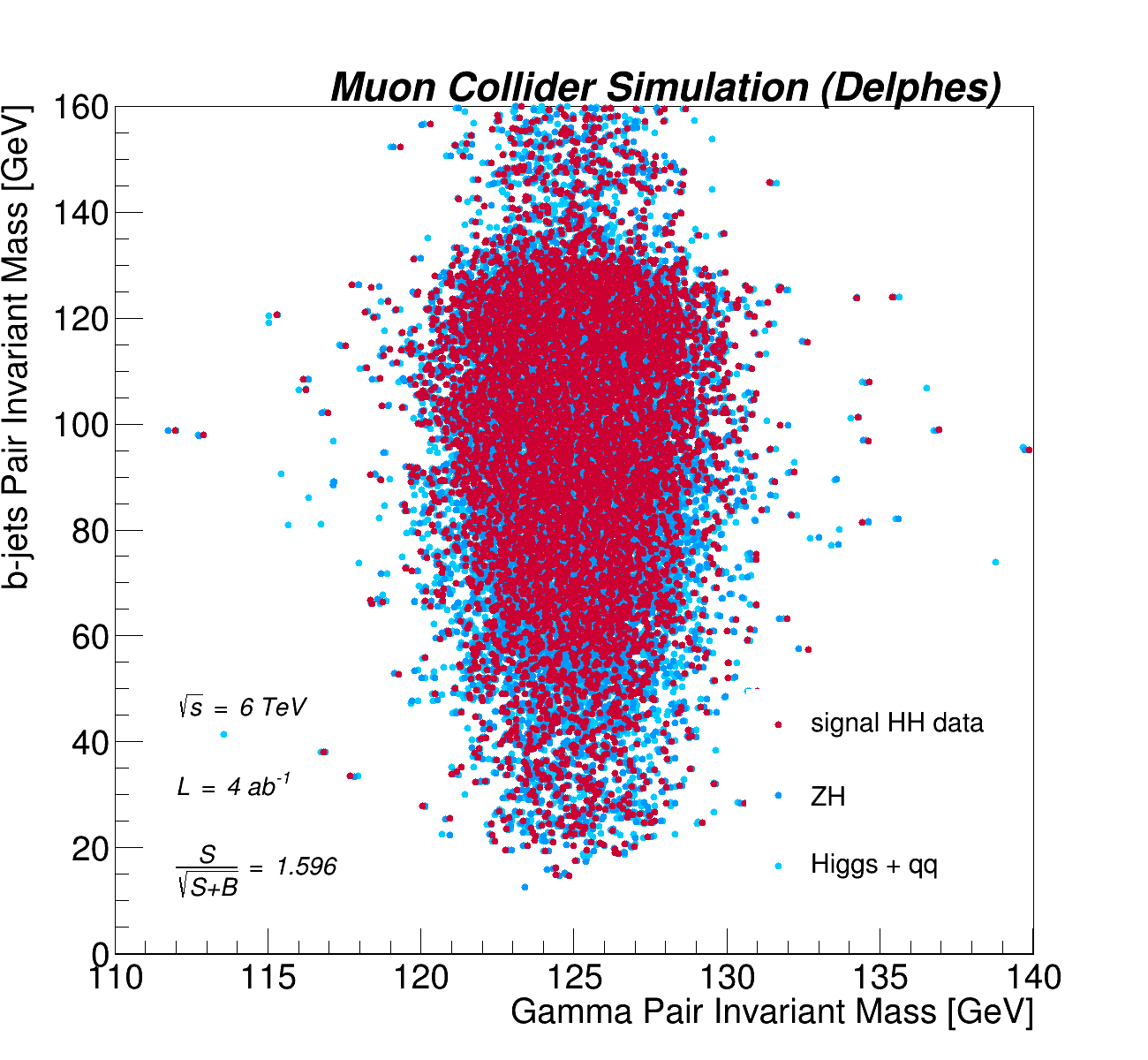}
    \caption{Event distribution of both signal and backgrounds channels is shown in the plane of the $b$-jet pair and the $\gamma\gamma$ pair invariant mass for $\sqrt{s}$ = 6 TeV data (bottom). Signal is shown in red and the background in various shades of blue. Projections to the $\gamma$ pair (top-left) and the $b$-jet pair (top-right) invariant mass are also shown.}
    \label{JetPair_bbaa_6TeV}
\end{figure}

\begin{figure}[ht!]
    \centering
    \includegraphics[width=3in]{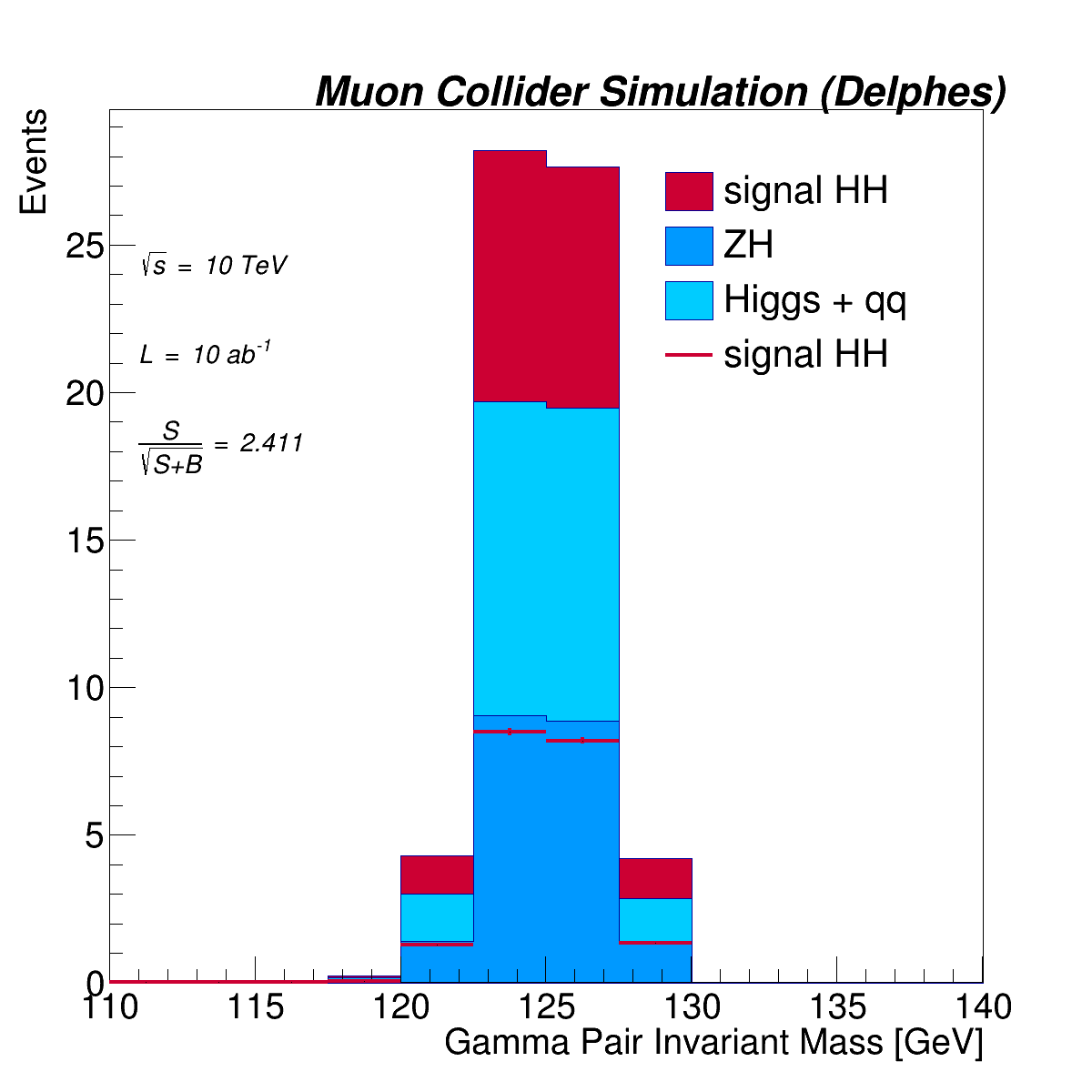}
    \includegraphics[width=3in]{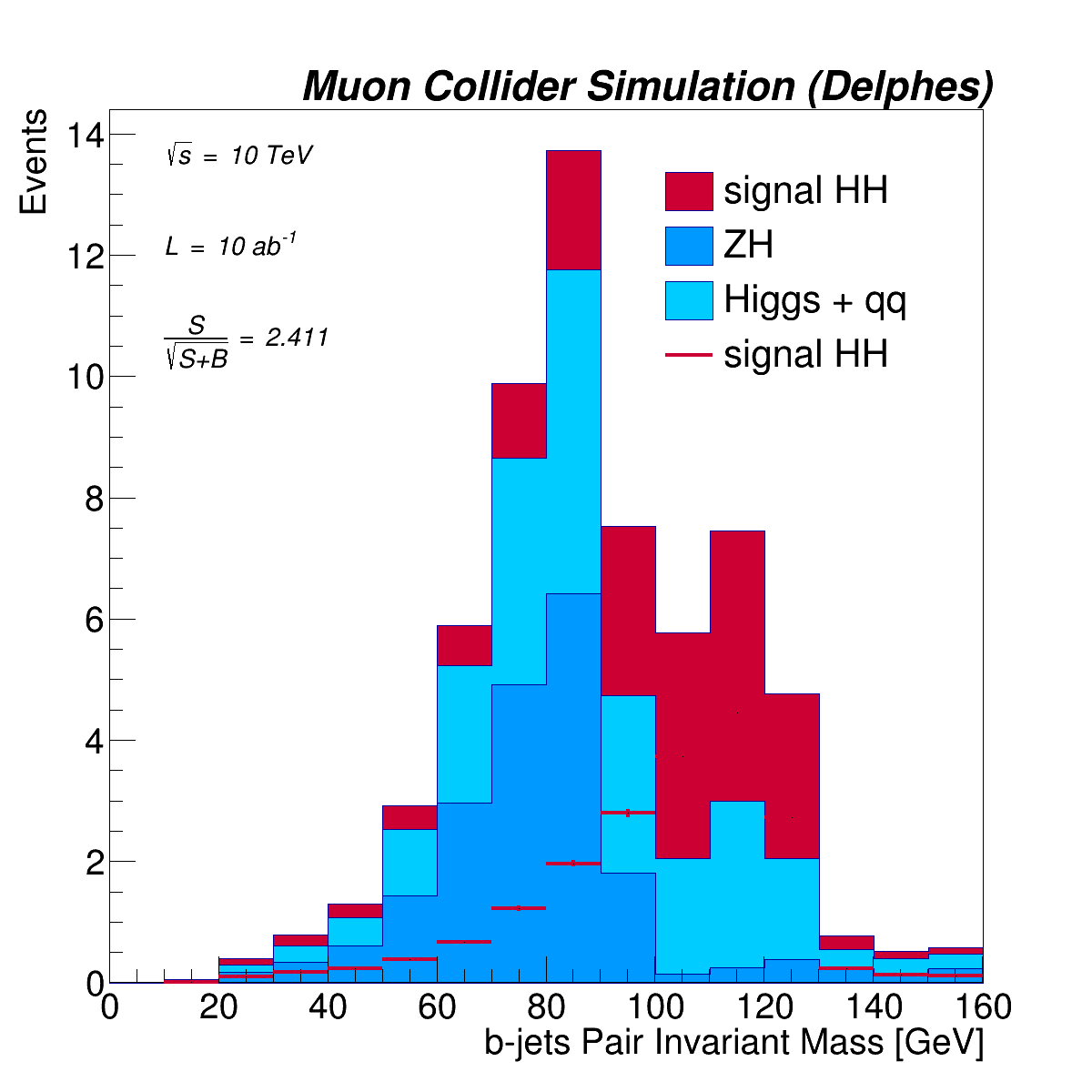}\\
    \includegraphics[width=3in]{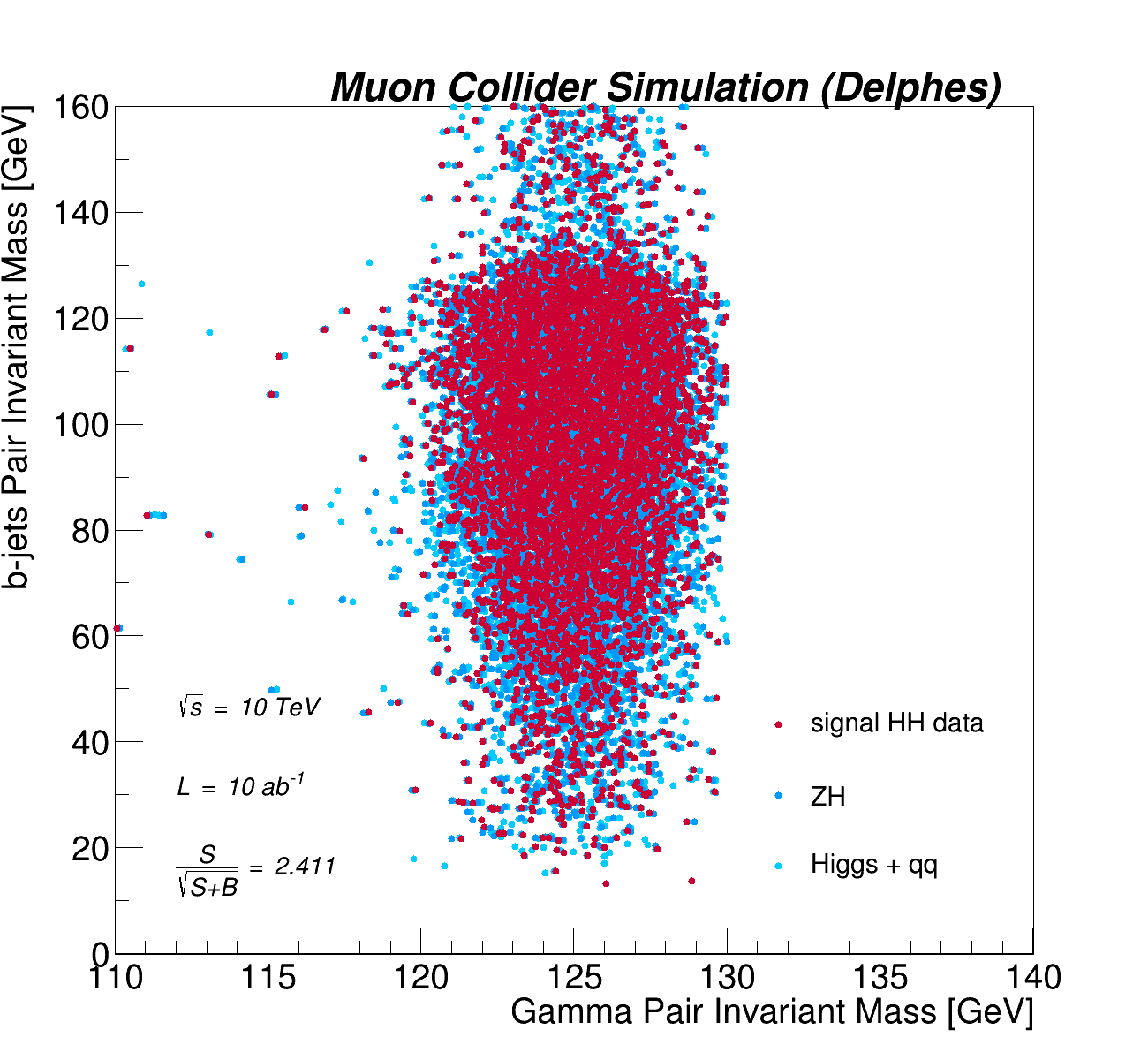}
    \caption{Event distribution of both signal and backgrounds channels is shown in the plane of the $b$-jet pair and the $\gamma\gamma$ pair invariant mass for $\sqrt{s}$ = 10 TeV data (bottom). Signal is shown in red and the background in various shades of blue. Projections to the $\gamma$ pair (top-left) and the $b$-jet pair (top-right) invariant mass are also shown.}
    \label{JetPair_bbaa_10TeV}
\end{figure}

\begin{figure}[ht!]
    \centering
    \includegraphics[width=3in]{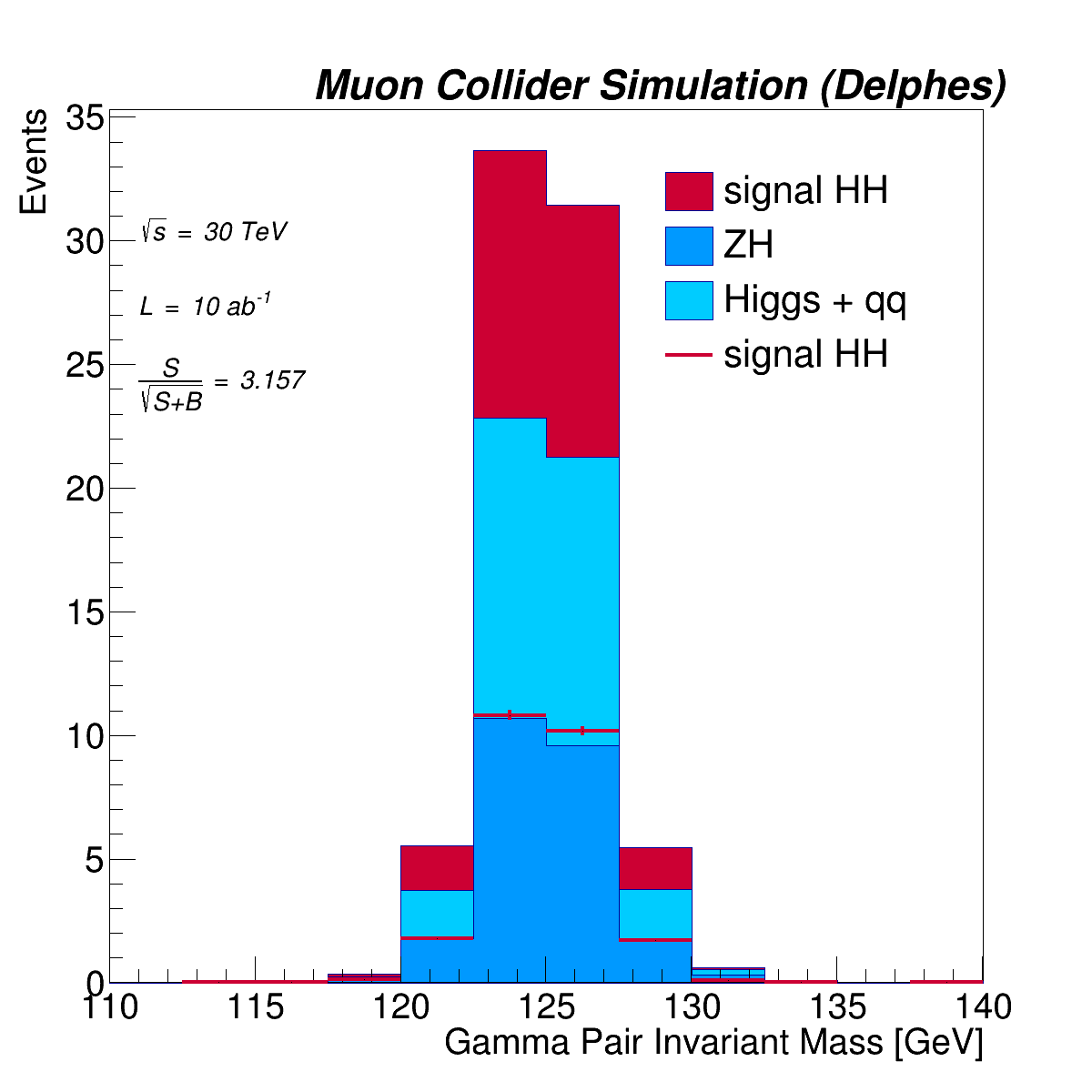}
    \includegraphics[width=3in]{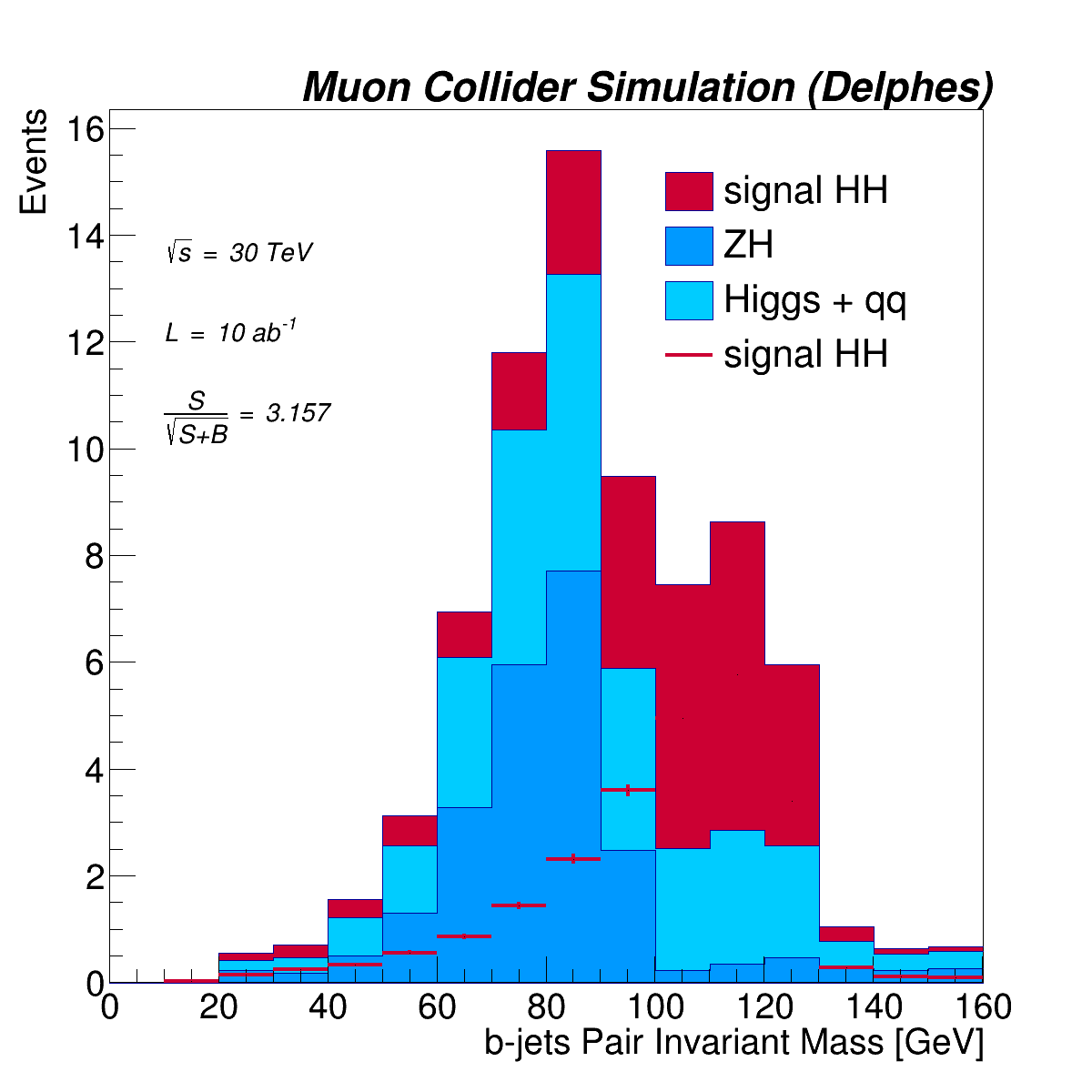}\\
    \includegraphics[width=3in]{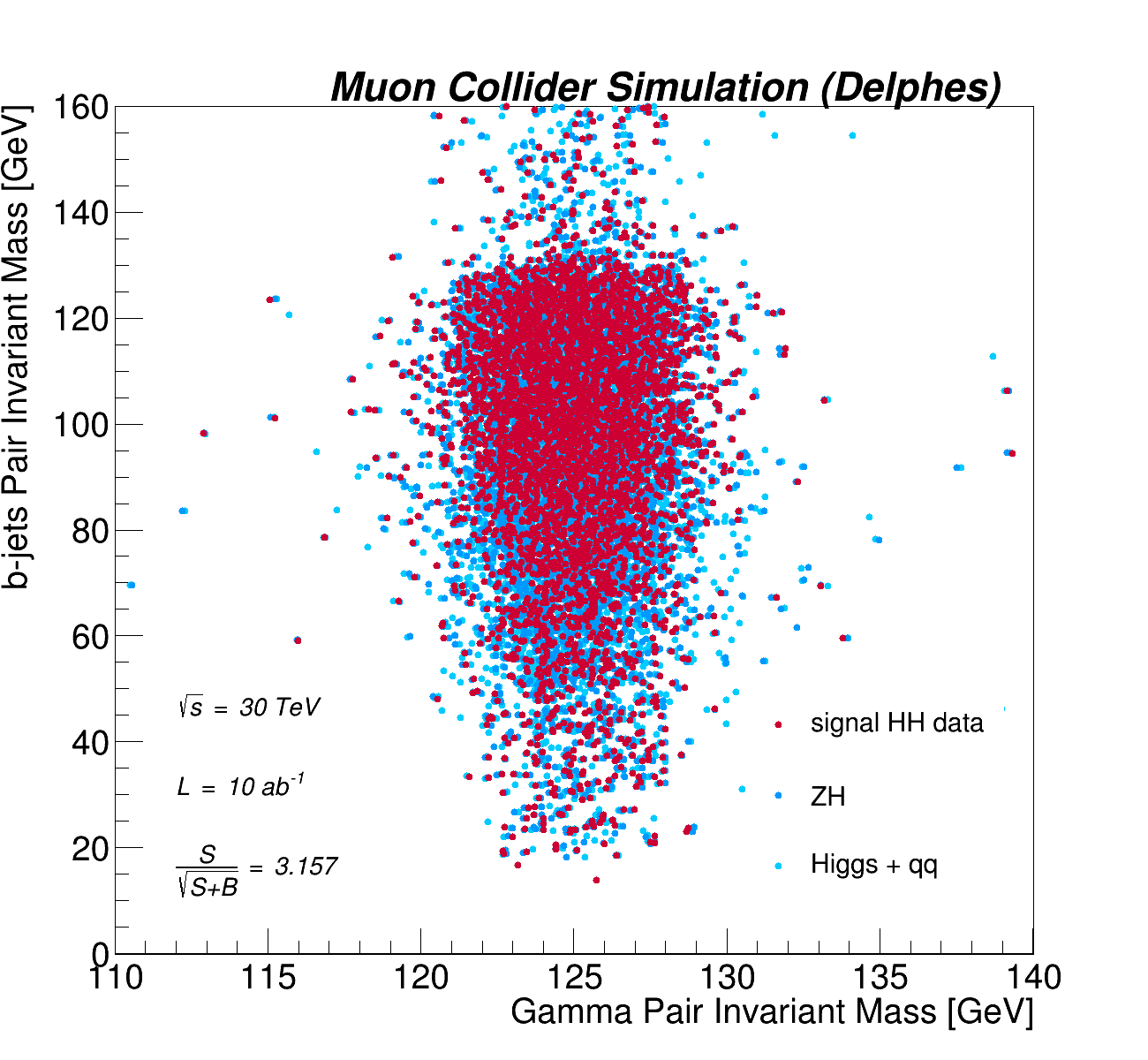}
    \caption{Event distribution of both signal and backgrounds channels is shown in the plane of the $b$-jet pair and the $\gamma\gamma$ pair invariant mass for $\sqrt{s}$ = 30 TeV data (bottom). Signal is shown in red and the background in various shades of blue. Projections to the $\gamma$ pair (top-left) and the $b$-jet pair (top-right) invariant mass are also shown.}
    \label{JetPair_bbaa_30TeV}
\end{figure}

Despite the disadvantage of $h\to\gamma\gamma$ decay's small cross section, the $\nu \bar{\nu} b \bar{b} \gamma\gamma$ channel has its compensation with its cleanliness. For this channel, we have simulated both the signal, $\mu^+\mu^- \to \nu \bar{\nu} H H \to \nu \bar{\nu} b \bar{b} \gamma\gamma$, and background, 
$\mu^+\mu^- \to \nu \bar{\nu} Z H, H\to \gamma \gamma $ (ZH), 
$\mu^+\mu^- \to \nu \bar{\nu} H q \bar{q}, H\to \gamma \gamma$ (Higgs + qq) events using fast simulation. The $h\to \gamma\gamma$ decay in both signal and background channel are simulated with the Higgs Effective Field Theory (HEFT) model in MadGraph5~\cite{2015}.
The analysis starts with the requirement of two loosely $b$-tagged jets with $p_T > 20$ GeV and two photon with $p_T > 15$ GeV. For the Higgs to two photon pairs, we reconstructed the Higgs with photon pairs with the invariant mass closest to $125$ GeV. For optimization of the signal significance, a tight cut is placed on the di-photon invariant mass at
\begin{align*}
    110\text{ GeV} \leq M_{\gamma\gamma} \leq 140\text{ GeV}.
\end{align*}
Then for reconstructing the $b \bar{b}$ jets pairs, for each event, we have selected a $b \bar{b}$ jets pairs with invariant mass closest to $125$ GeV where both jets are loosely $b$-tagged.
Similarly, a scatter plot of the $b$-jet pair and the $\gamma\gamma$ pair invariant mass is shown in the bottom planes of the Figures~\ref{JetPair_bbaa_3TeV}, \ref{JetPair_bbaa_6TeV}, \ref{JetPair_bbaa_10TeV}, and \ref{JetPair_bbaa_30TeV} for 3, 6, 10, and 30 TeV center of mass energies. Projects to the $b$-jet pair and the $\gamma\gamma$ pair invariant mass are shown in the top-right and top-left planes of the same figures. The $H\to b \bar{b}$ mass peaks are separated from the background and can be used for extracting the signal significance. With the same integrated luminosity settings for different center of mass energies, a simple cut and count analysis is performed with tight cuts around the (uncalibrated) $b$-pair invariant mass:
\begin{align*}
    90 \text{ GeV} < m_{bb} < 130\text{ GeV}.
\end{align*}
The estimated significance results are shown in the figures and tabulated for all four collider settings in the Table~\ref{tab:Delphes_HH_bbaa}.
\begin{table}[ht!]
    \centering
    \begin{tabular}{|c|c|}
        \hline
        &\\
        $\sqrt{s}~(\int{d{\cal L}})$ & Estimated signal significance\\
        &\\\hline
        &\\
        3 TeV (1 ab$^{-1}$) & 0.673\\
        &\\\hline
        &\\
        6 TeV (4 ab$^{-1}$) & 1.596\\
        &\\\hline
        &\\
        10 TeV (10 ab$^{-1}$) & 2.411\\
        &\\\hline
        &\\
        30 TeV (10 ab$^{-1}$) & 3.157\\
        &\\\hline
    \end{tabular}
    \caption{Significance for the extraction of di-Higgs to $b \bar{b} \gamma \gamma$ events for muon colliders operating at various centers of mass and integrated luminosity.}
    \label{tab:Delphes_HH_bbaa}
\end{table}
\clearpage
\subsection{The $\nu \bar{\nu} b \bar{b} \tau\tau$ channel} 
The $\nu \bar{\nu} b \bar{b} \tau\tau$ channel with its three different final states also provides its unique value for di-Higgs study. For this channel, we have simulated both the signal, $\mu^+\mu^- \to \nu \bar{\nu} H H \to \nu \bar{\nu} b \bar{b} \tau \tau$, and dominant background, 
$\mu^+\mu^- \to \nu \bar{\nu} Z H, H\to \tau \tau $ (ZH($\tau \tau$)), 
$\mu^+\mu^- \to \nu \bar{\nu} H q \bar{q}, H\to \tau \tau$ (Higgs + qq), and $\mu^+\mu^- \to \nu \bar{\nu} Z H, Z\to \tau \tau $ (Z($\tau \tau$)H) events using fast simulation. 
Because of the neutrinos in the background, simple approximation methods like collinear approximation does not perform well. Therefore, in this study, we simply reconstructed the di-$\tau$ invariant mass by its visible mass. For this reason, we neglect the full leptonic final states with poor reconstruction of the di-$\tau$ mass. Analysis on full hadronic and semi-leptonic final states are present in following sub-subsections.

\subsubsection{The $\nu \bar{\nu} b \bar{b} \tau_{\text{had}} \tau_{\text{had}}$ channel}
\begin{figure}[ht!]
    \centering
    \includegraphics[width=3in]{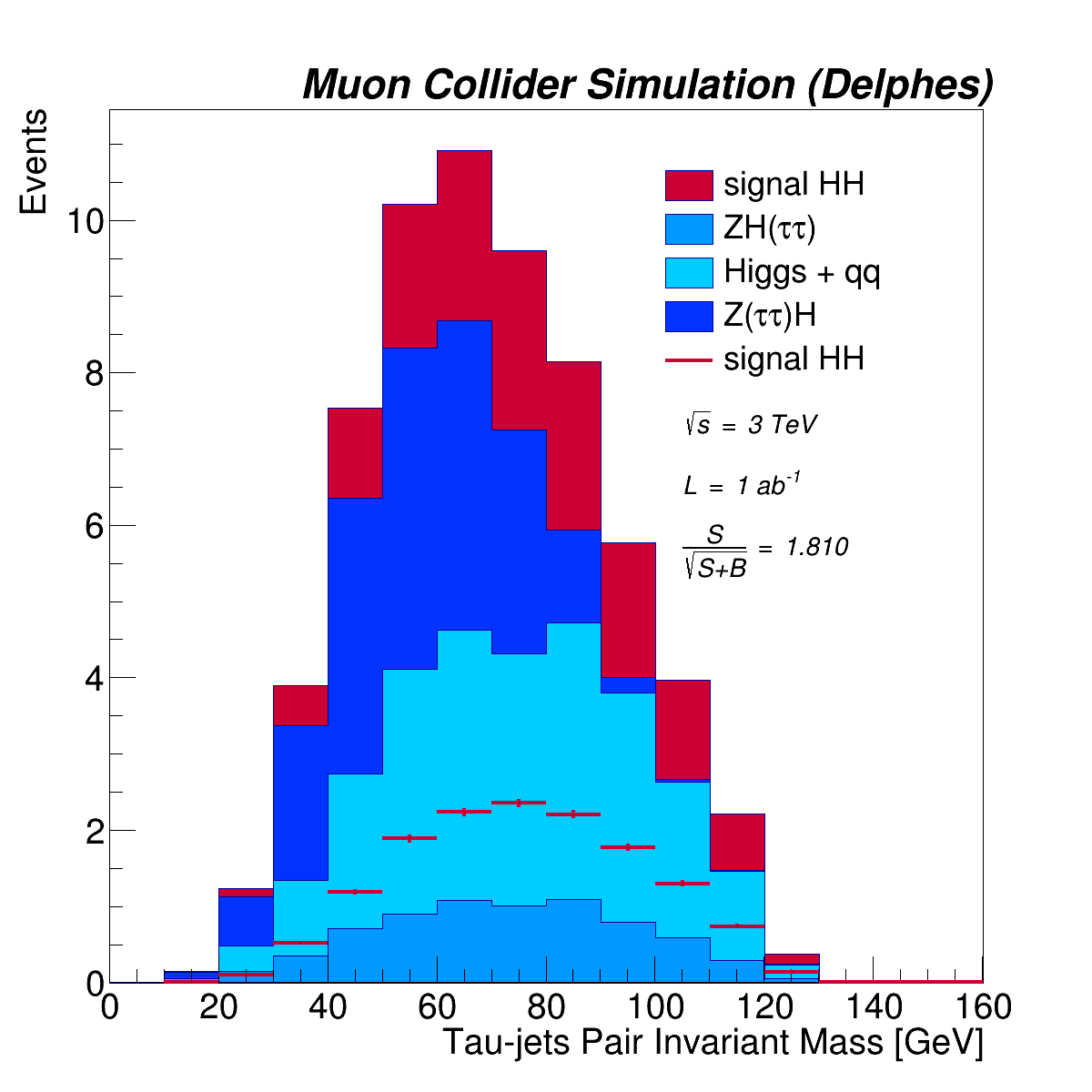}
    \includegraphics[width=3in]{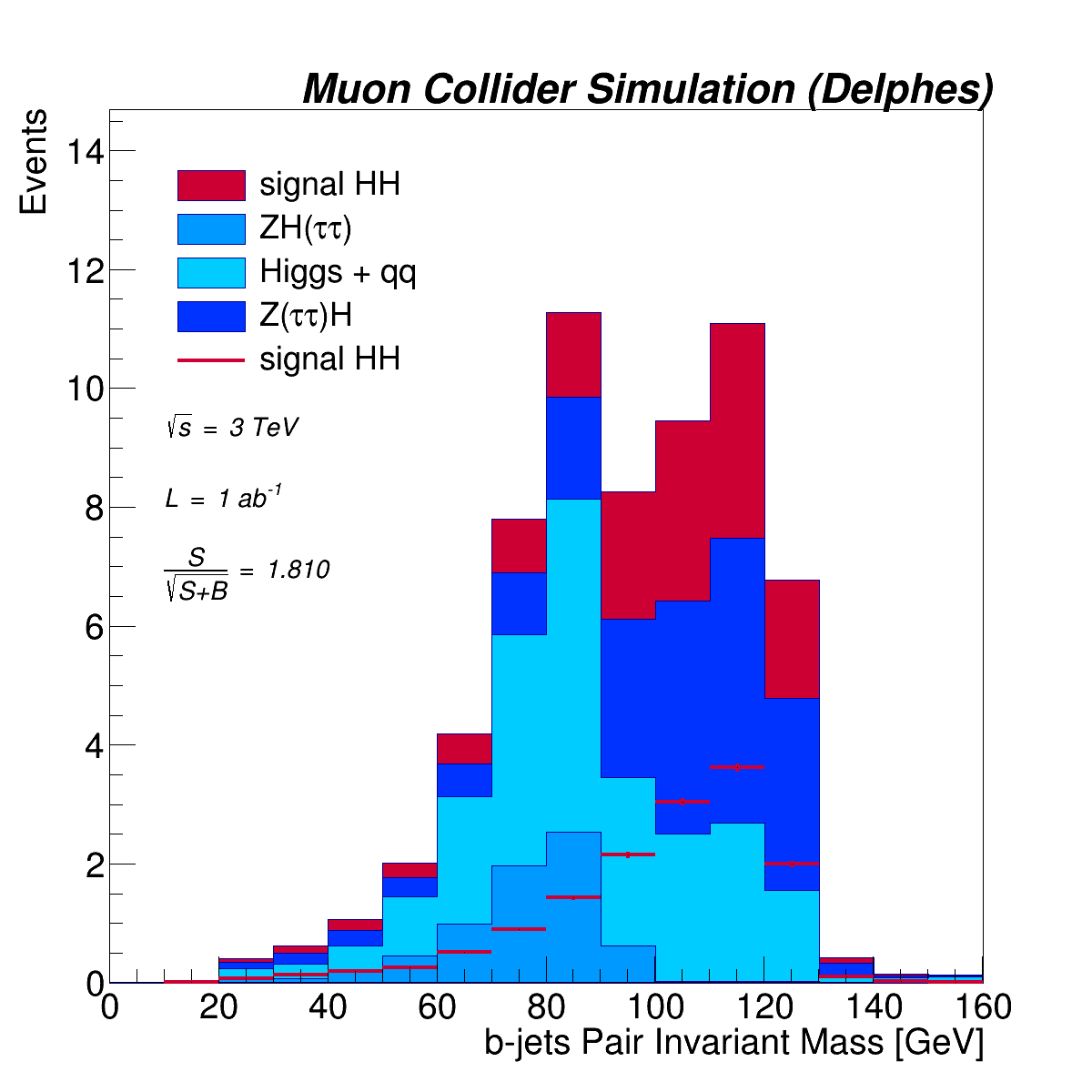}\\
    \includegraphics[width=3in]{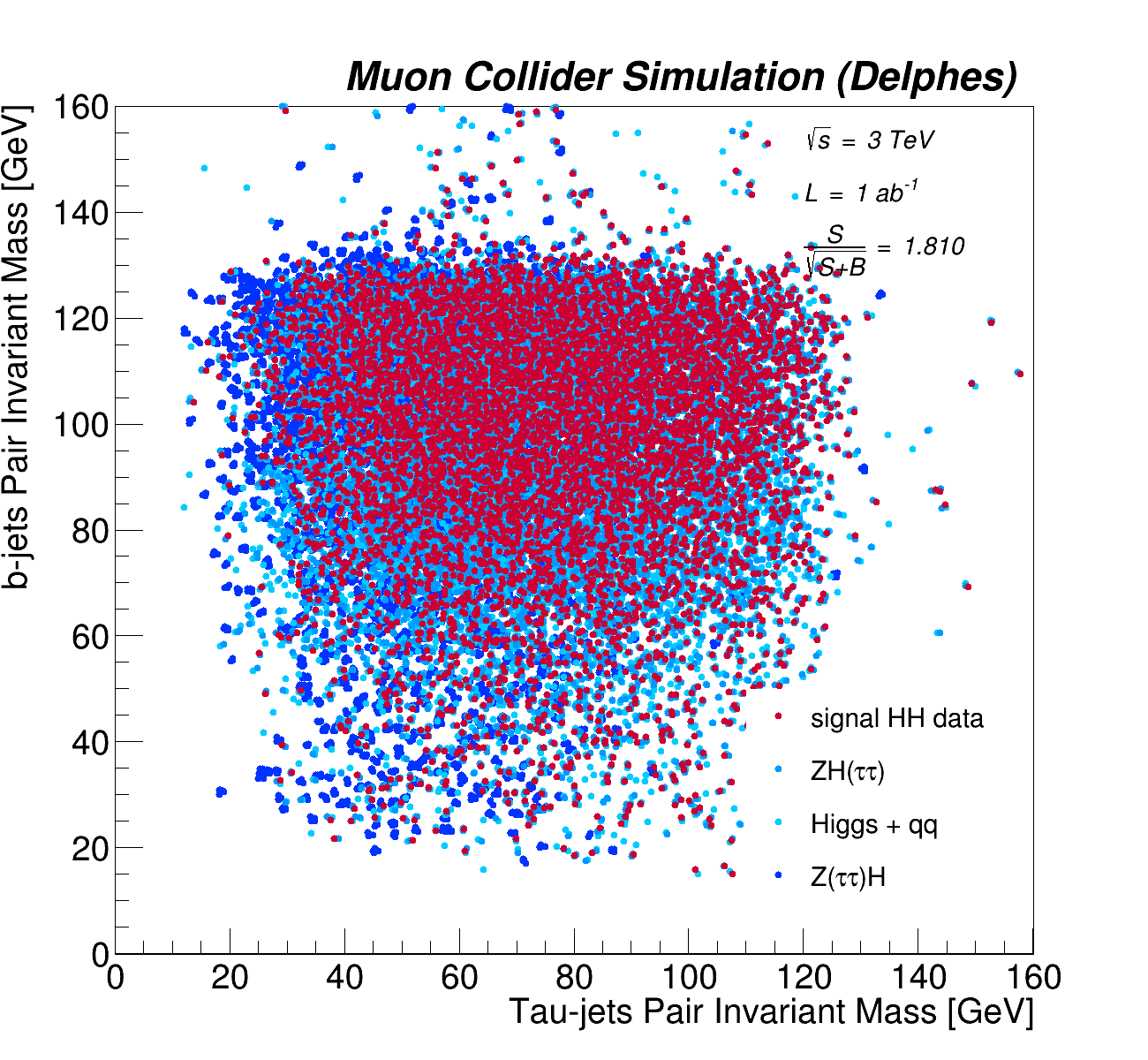}
    \caption{Event distribution of both signal and backgrounds channels is shown in the plane of the $b$-jet pair and the $\tau_{\text{had}}\tau_{\text{had}}$ pair invariant mass for $\sqrt{s}$ = 3 TeV data (bottom). Signal is shown in red and the background in various shades of blue. Projections to the $\tau_{\text{had}}\tau_{\text{had}}$ pair (top-left) and the $b$-jet pair (top-right) invariant mass are also shown.}
    \label{JetPair_tau_had_3TeV}
\end{figure}

\begin{figure}[ht!]
    \centering
    \includegraphics[width=3in]{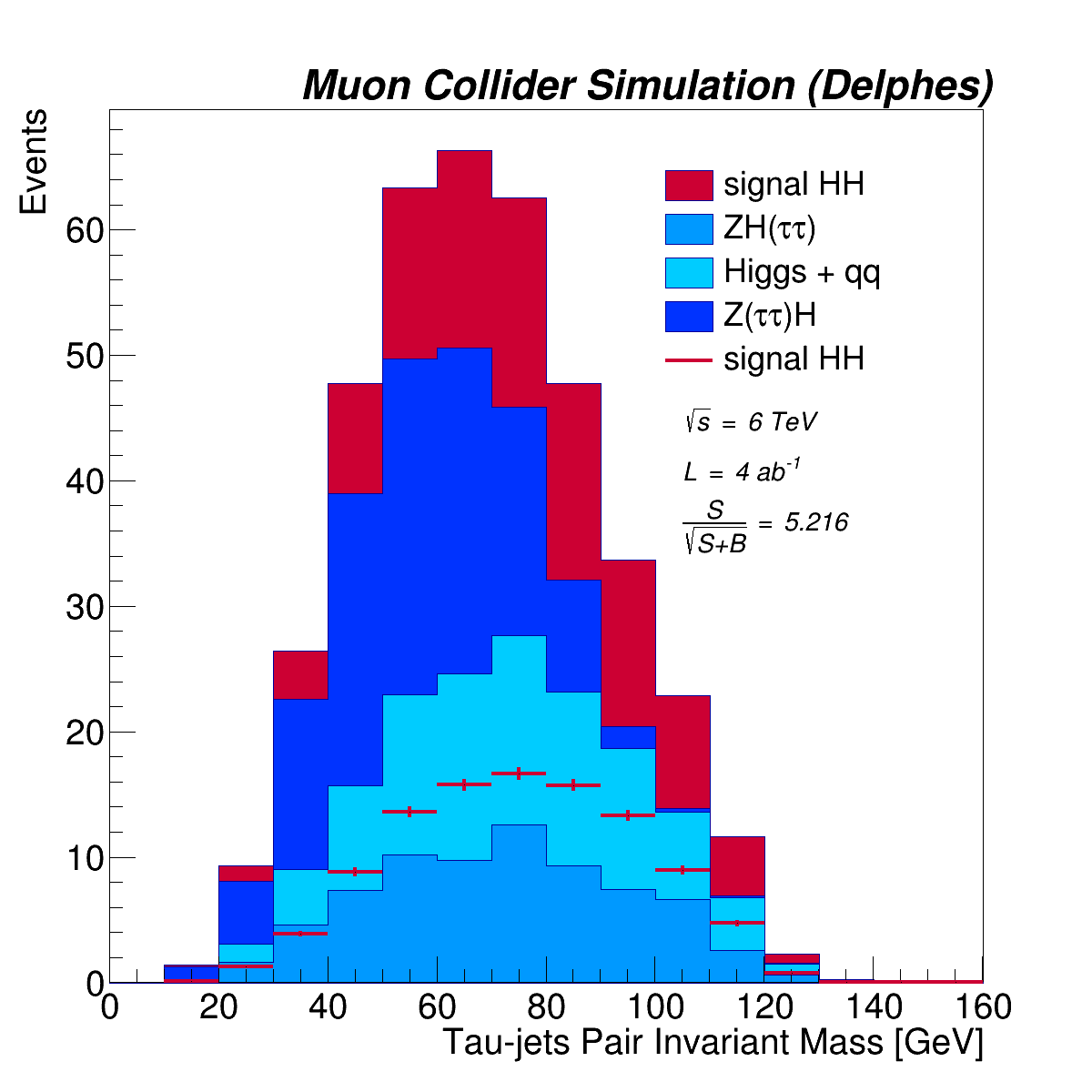}
    \includegraphics[width=3in]{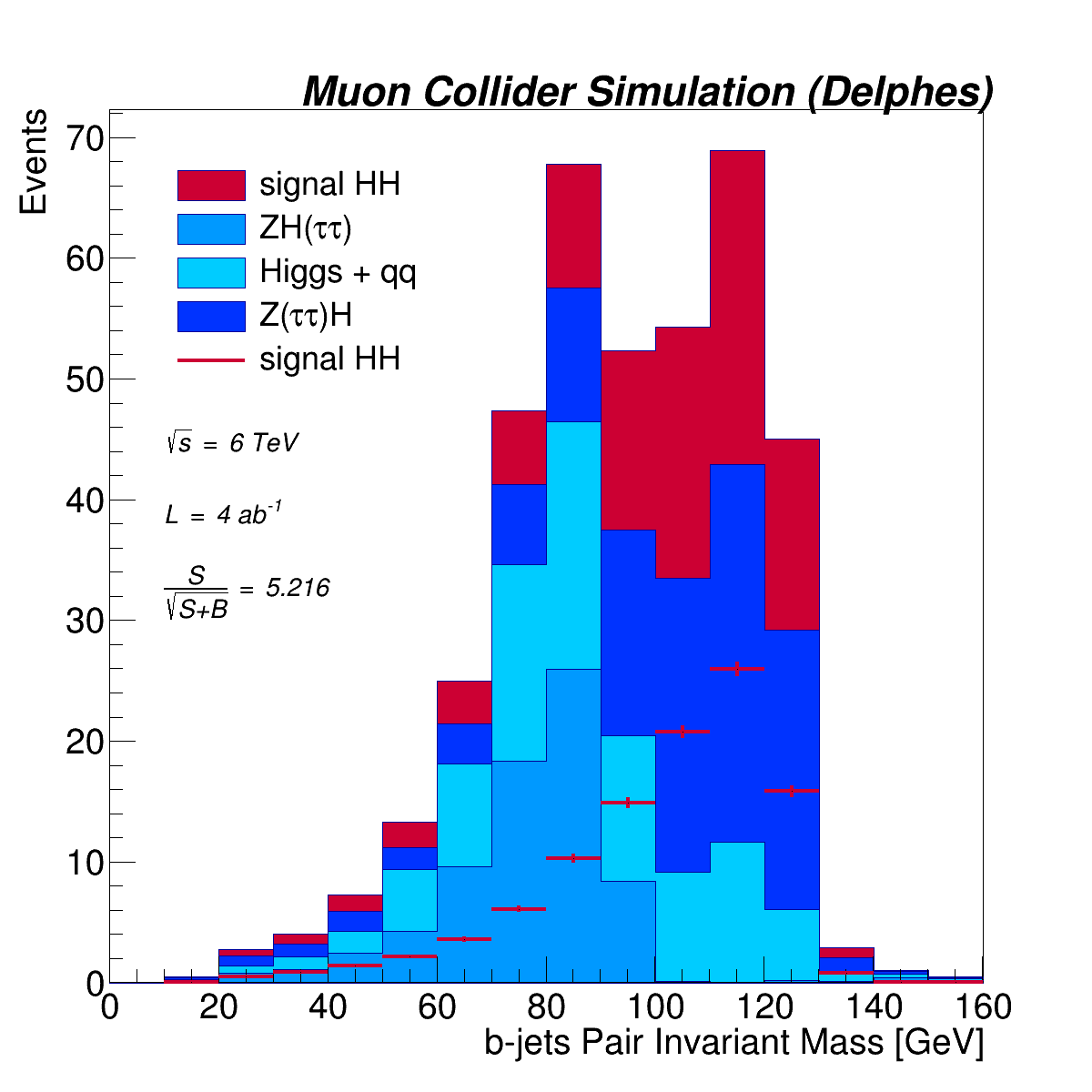}\\
    \includegraphics[width=3in]{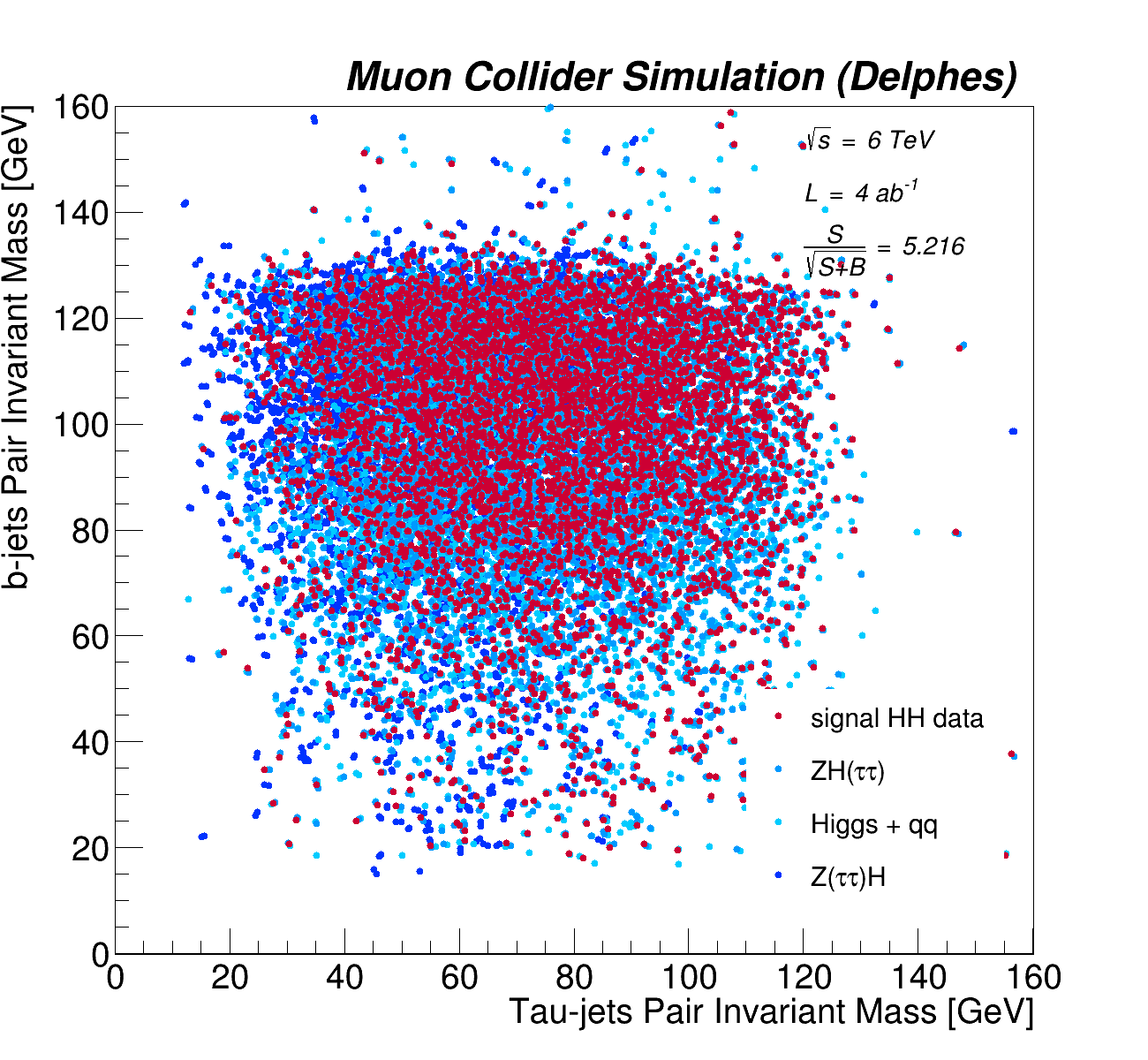}
    \caption{Event distribution of both signal and backgrounds channels is shown in the plane of the $b$-jet pair and the $\tau_{\text{had}}\tau_{\text{had}}$ pair invariant mass for $\sqrt{s}$ = 6 TeV data (bottom). Signal is shown in red and the background in various shades of blue. Projections to the $\tau_{\text{had}}\tau_{\text{had}}$ pair (top-left) and the $b$-jet pair (top-right) invariant mass are also shown.}
    \label{JetPair_tau_had_6TeV}
\end{figure}

\begin{figure}[ht!]
    \centering
    \includegraphics[width=3in]{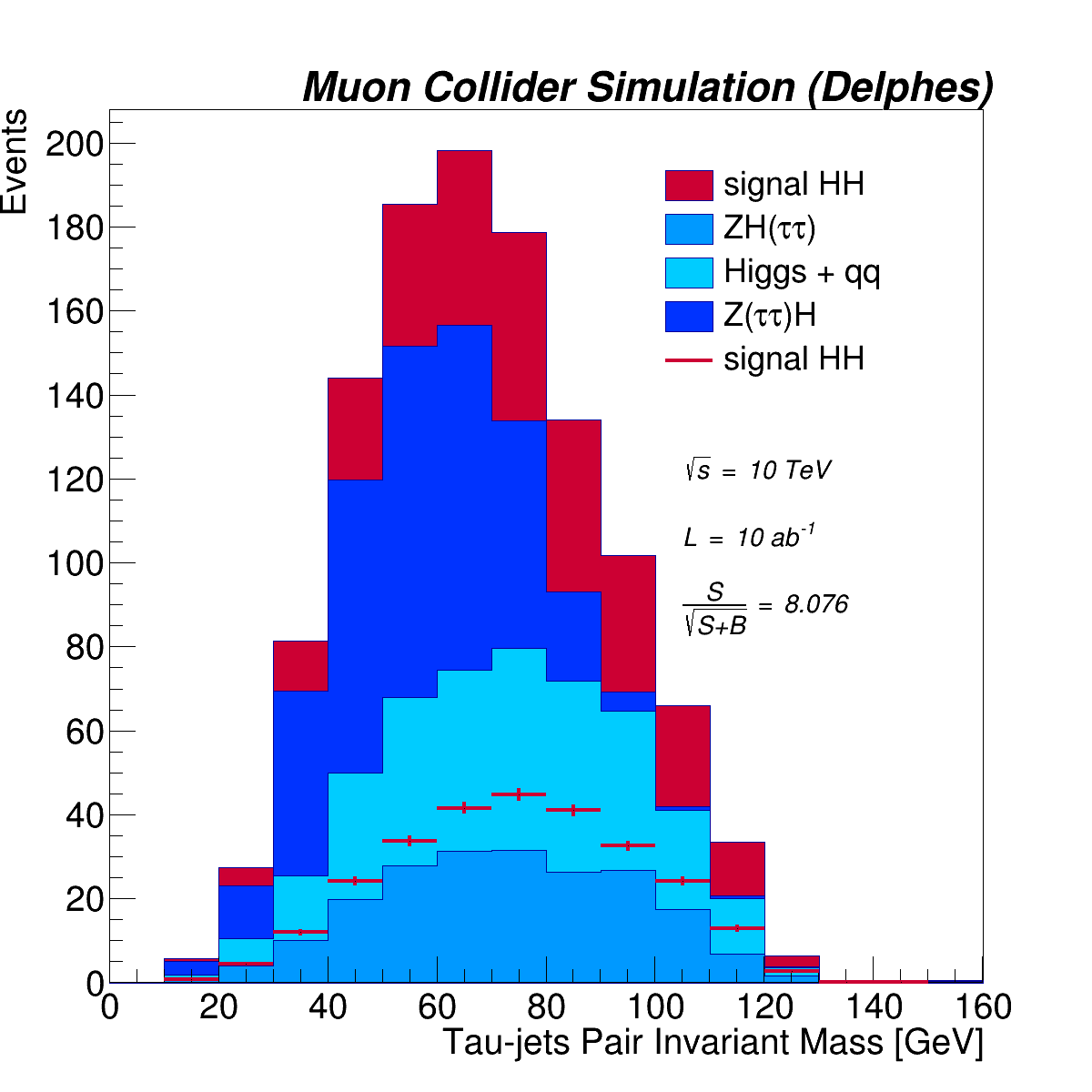}
    \includegraphics[width=3in]{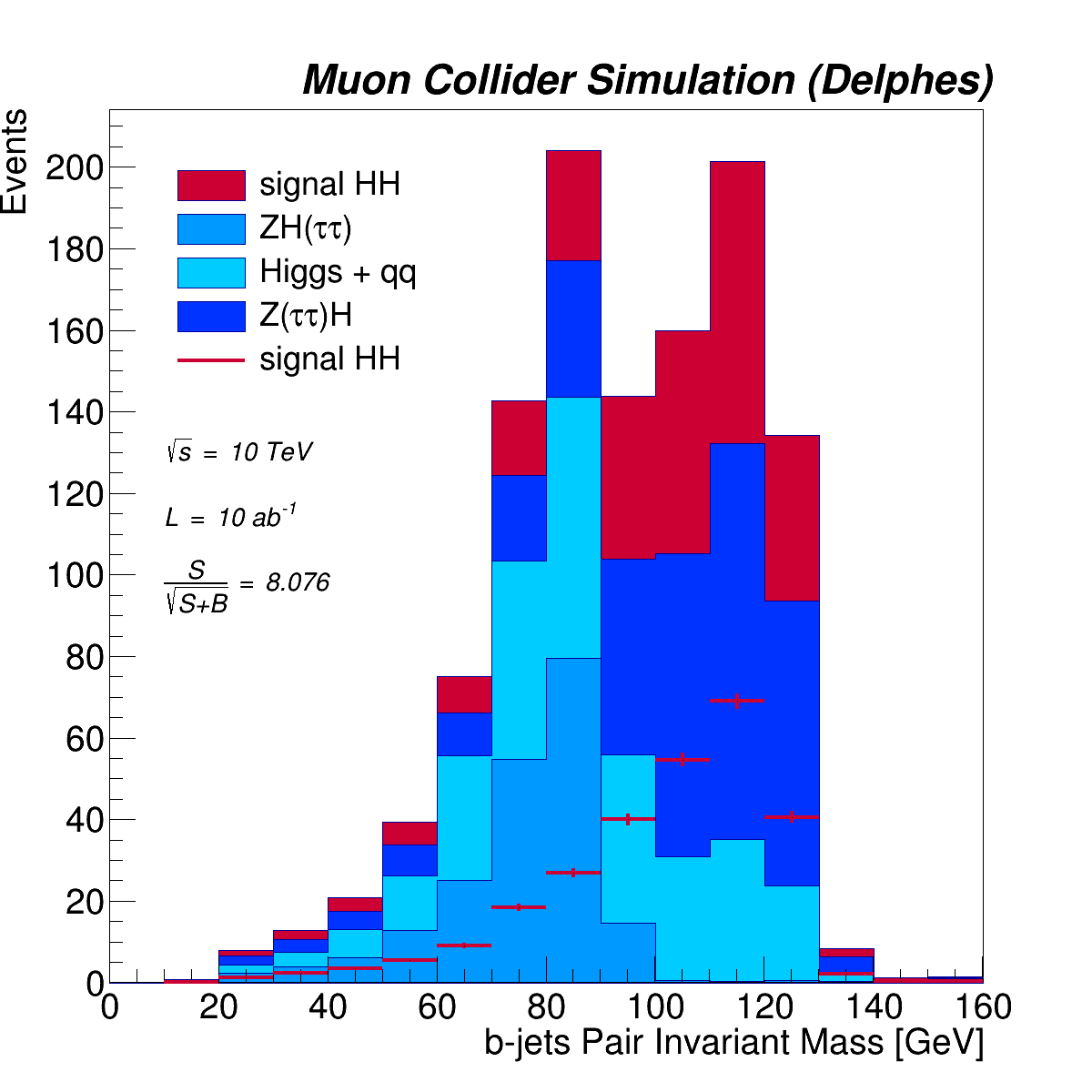}\\
    \includegraphics[width=3in]{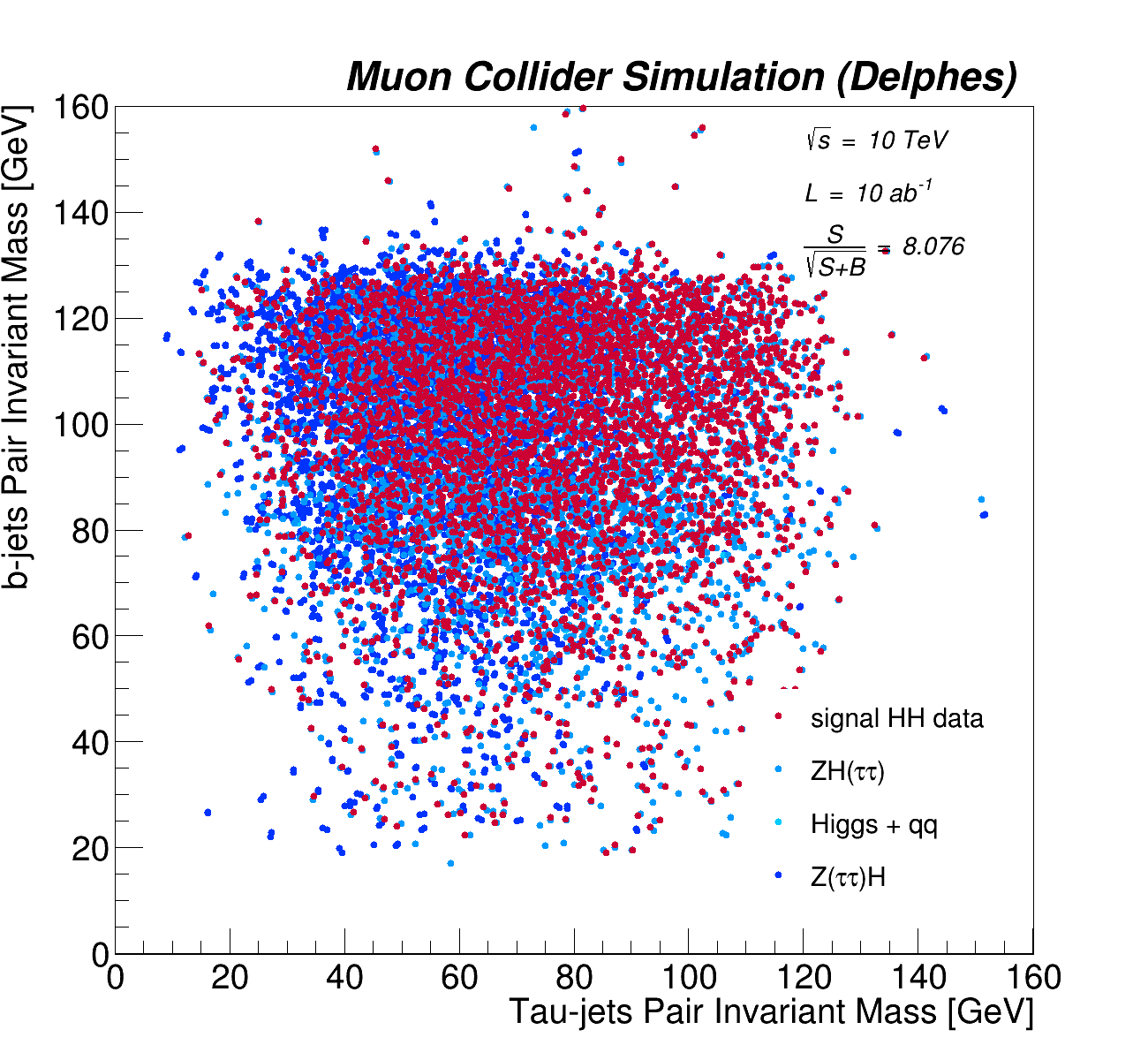}
    \caption{Event distribution of both signal and backgrounds channels is shown in the plane of the $b$-jet pair and the $\tau_{\text{had}}\tau_{\text{had}}$ pair invariant mass for $\sqrt{s}$ = 10 TeV data (bottom). Signal is shown in red and the background in various shades of blue. Projections to the $\tau_{\text{had}}\tau_{\text{had}}$ pair (top-left) and the $b$-jet pair (top-right) invariant mass are also shown.}
    \label{JetPair_tau_had_10TeV}
\end{figure}

\begin{figure}[ht!]
    \centering
    \includegraphics[width=3in]{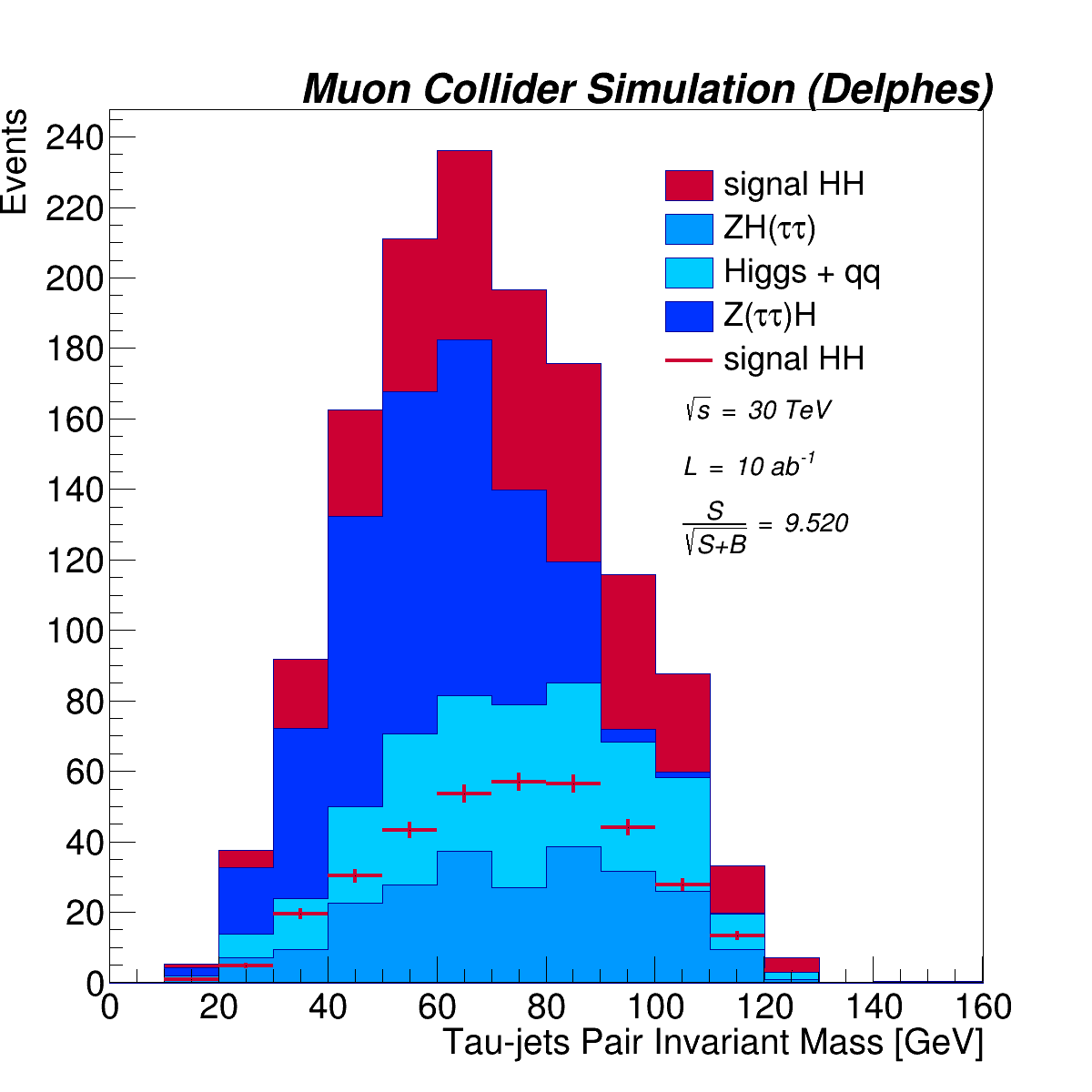}
    \includegraphics[width=3in]{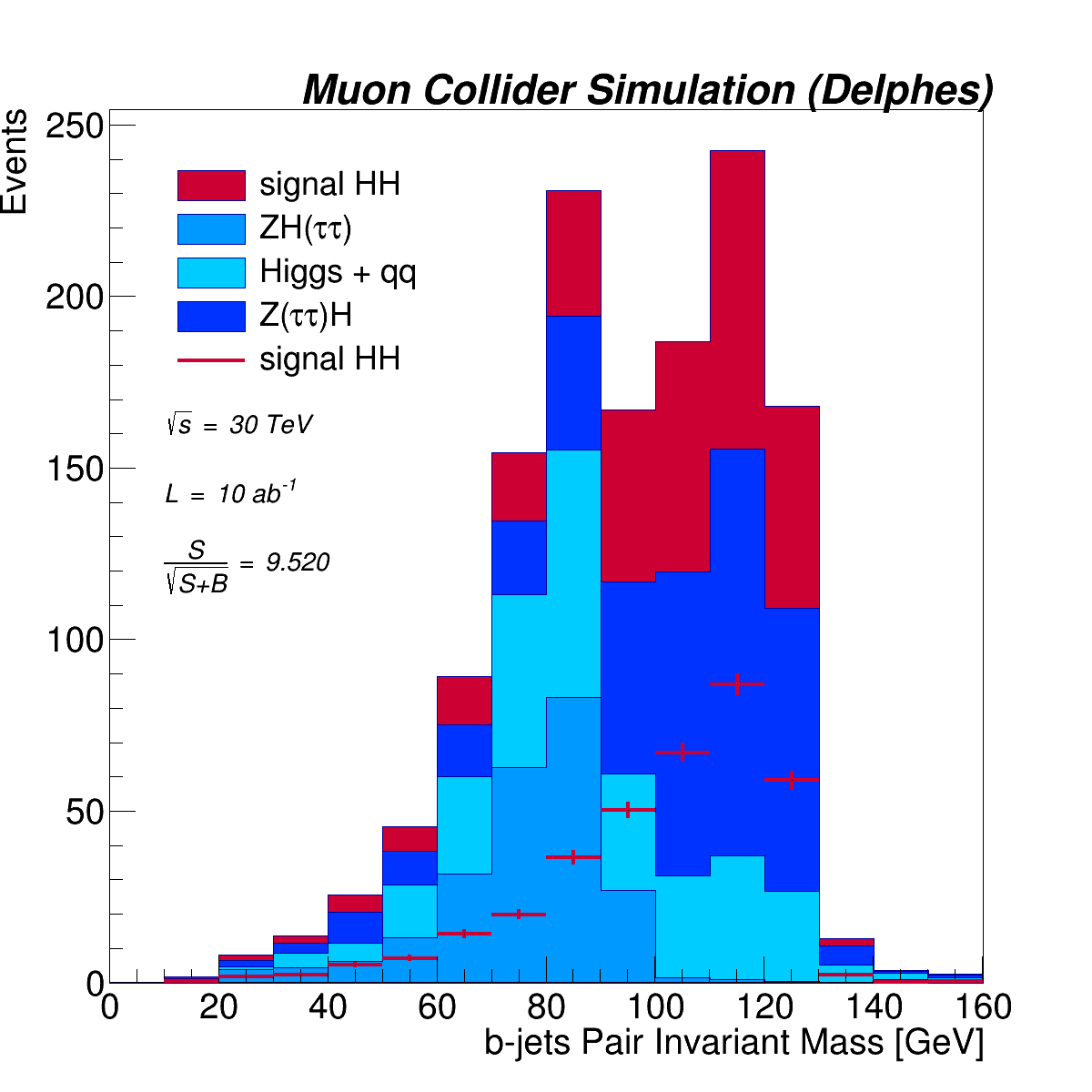}\\
    \includegraphics[width=3in]{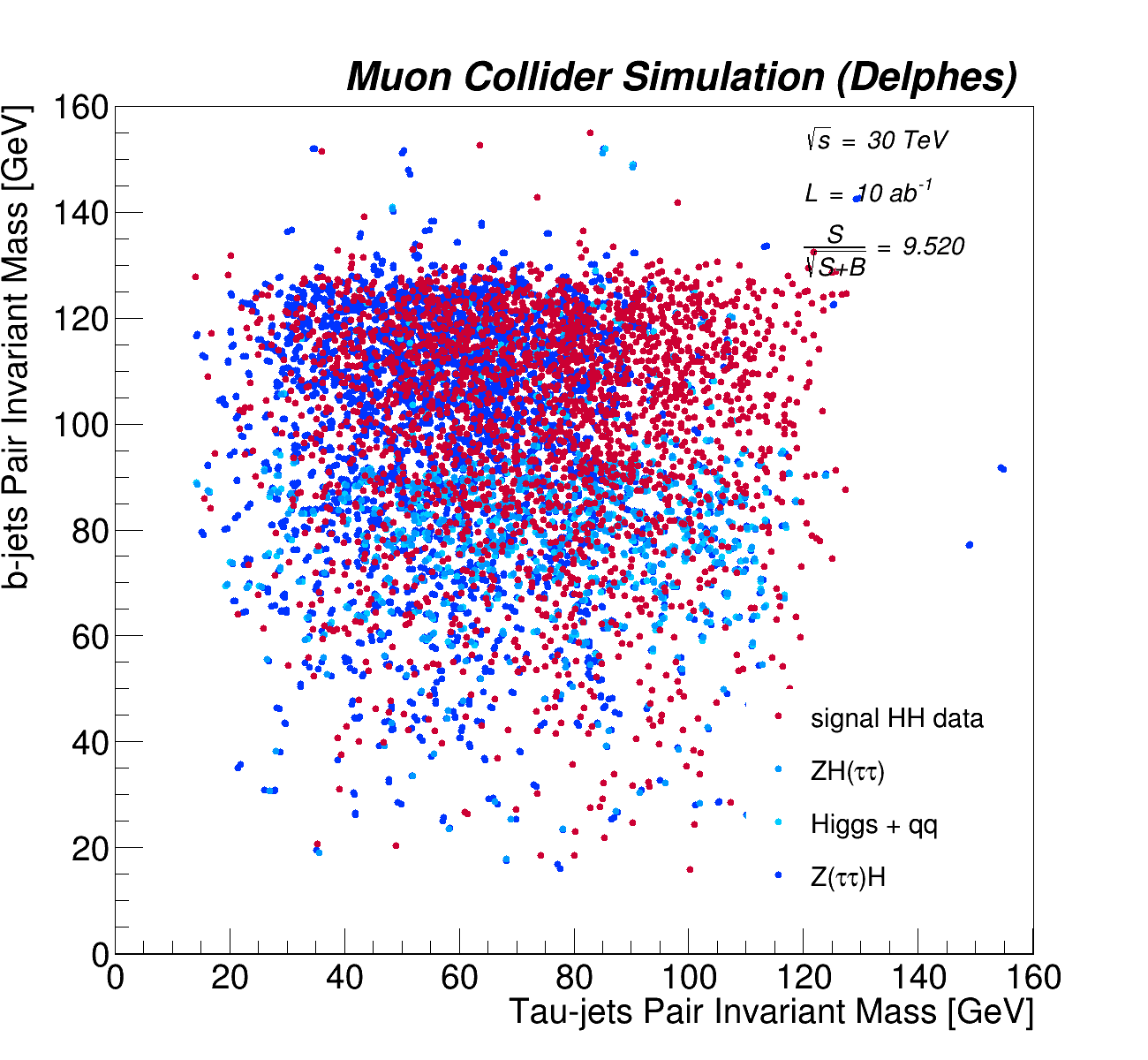}
    \caption{Event distribution of both signal and backgrounds channels is shown in the plane of the $b$-jet pair and the $\tau_{\text{had}}\tau_{\text{had}}$ pair invariant mass for $\sqrt{s}$ = 30 TeV data (bottom). Signal is shown in red and the background in various shades of blue. Projections to the $\tau_{\text{had}}\tau_{\text{had}}$ pair (top-left) and the $b$-jet pair (top-right) invariant mass are also shown.}
    \label{JetPair_tau_had_30TeV}
\end{figure}

In this channel, the Higgs decayed into close di-$\tau$ pairs might be highly boosted, which results in $\tau$ pair that are close to each other. Hence, we used the anti-$k_T$ jet algorithm with jet cone size $R=0.2$ and requiring $p_{T_{\min}}>15$ GeV. Then, we reconstructed the Higgs to di-$\tau$ pairs by selecting all possible $R=0.2$ jet pairs which both jets are $\tau$-tagged but not $b$-tagged, in order to avoid selecting $b$ jet fake $\tau$, and charge product $Q=-1$. Further, we reconstructed the Higgs with the one with the invariant mass closest to $125$ GeV. To reconstruct the $b \bar{b}$ jets pairs, we selected a $b \bar{b}$ jets pairs with invariant mass closest to $125$ GeV where both jets are loosely $b$-tagged.
Similarly, a scatter plot of the $b$-jet pair and the di-$\tau$ pair invariant mass is shown in the bottom planes of the Figures~\ref{JetPair_tau_had_3TeV}, \ref{JetPair_tau_had_6TeV}, \ref{JetPair_tau_had_10TeV}, and \ref{JetPair_tau_had_30TeV} for 3, 6, 10, and 30 TeV center of mass energies. Projects to the $b$-jet pair and the $\tau\tau$ pair invariant mass are shown in the top-right and top-left planes of the same figures. Because of the poor reconstruction of the di-$\tau$ mass, the $H$ and $Z$ mass peak in the di-$\tau$ pair invariant mass histogram is overlapping. Hence simple cut method's performance to separate the $Z(\tau \tau)H$ from the signal is very limited. With the same integrated luminosity settings for different center of mass energies, the estimated significance results with no cut are shown in the figures and tabulated for all four collider settings in the Table~\ref{tab:Delphes_HH_tau_had}.
\begin{table}[ht!]
    \centering
    \begin{tabular}{|c|c|}
        \hline
        &\\
        $\sqrt{s}~(\int{d{\cal L}})$ & Estimated signal significance\\
        &\\\hline
        &\\
        3 TeV (1 ab$^{-1}$) & 1.810\\
        &\\\hline
        &\\
        6 TeV (4 ab$^{-1}$) & 5.216\\
        &\\\hline
        &\\
        10 TeV (10 ab$^{-1}$) & 8.076\\
        &\\\hline
        &\\
        30 TeV (10 ab$^{-1}$) & 9.520\\
        &\\\hline
    \end{tabular}
    \caption{Significance for the extraction of di-Higgs to $b \bar{b} \tau_{\text{had}} \tau_{\text{had}}$ events for muon colliders operating at various centers of mass and integrated luminosity.}
    \label{tab:Delphes_HH_tau_had}
\end{table}

\clearpage
\subsubsection{The $b \bar{b} \tau_{\text{lep}} \tau_{\text{had}}$ channel}
\begin{figure}[ht!]
    \centering
    \includegraphics[width=3in]{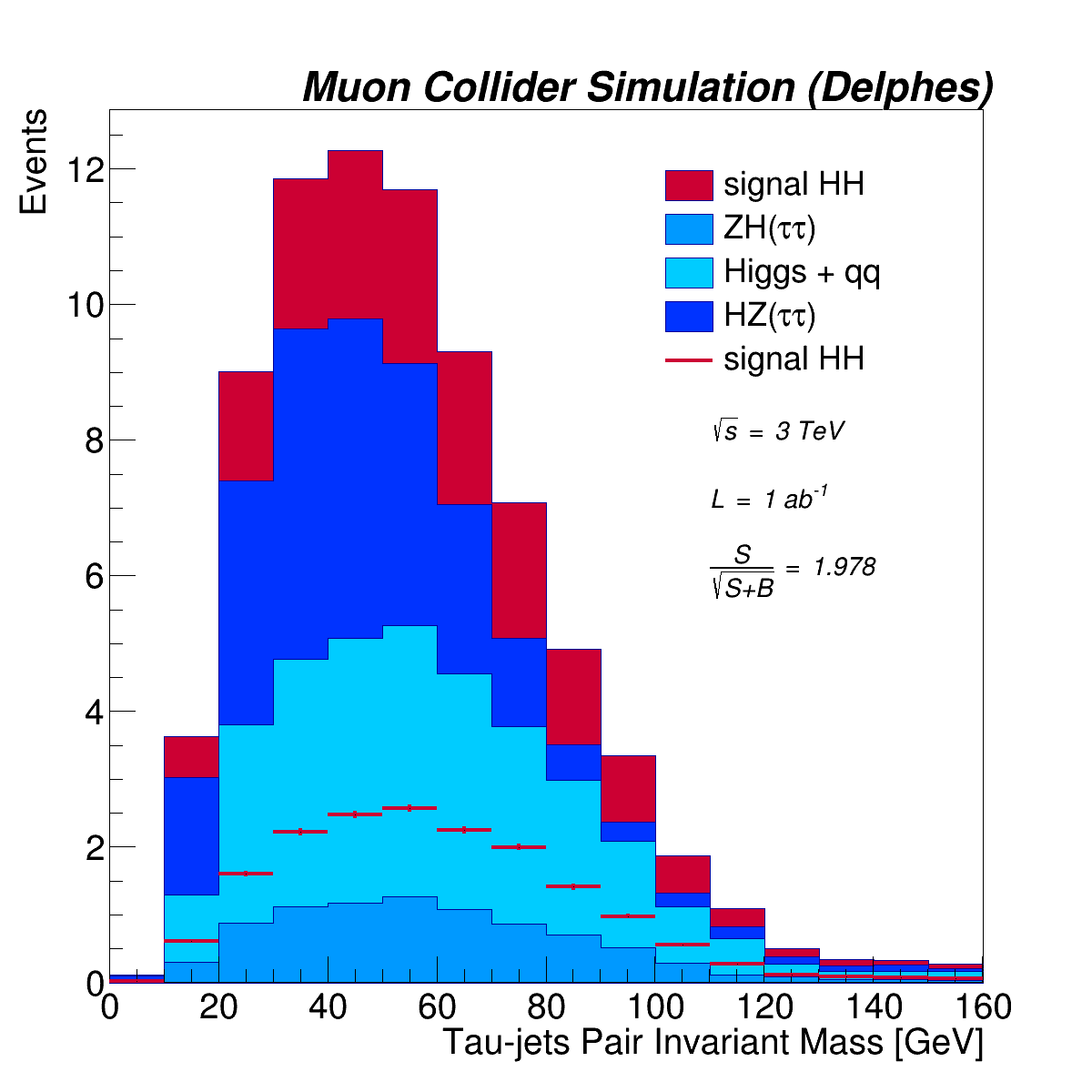}
    \includegraphics[width=3in]{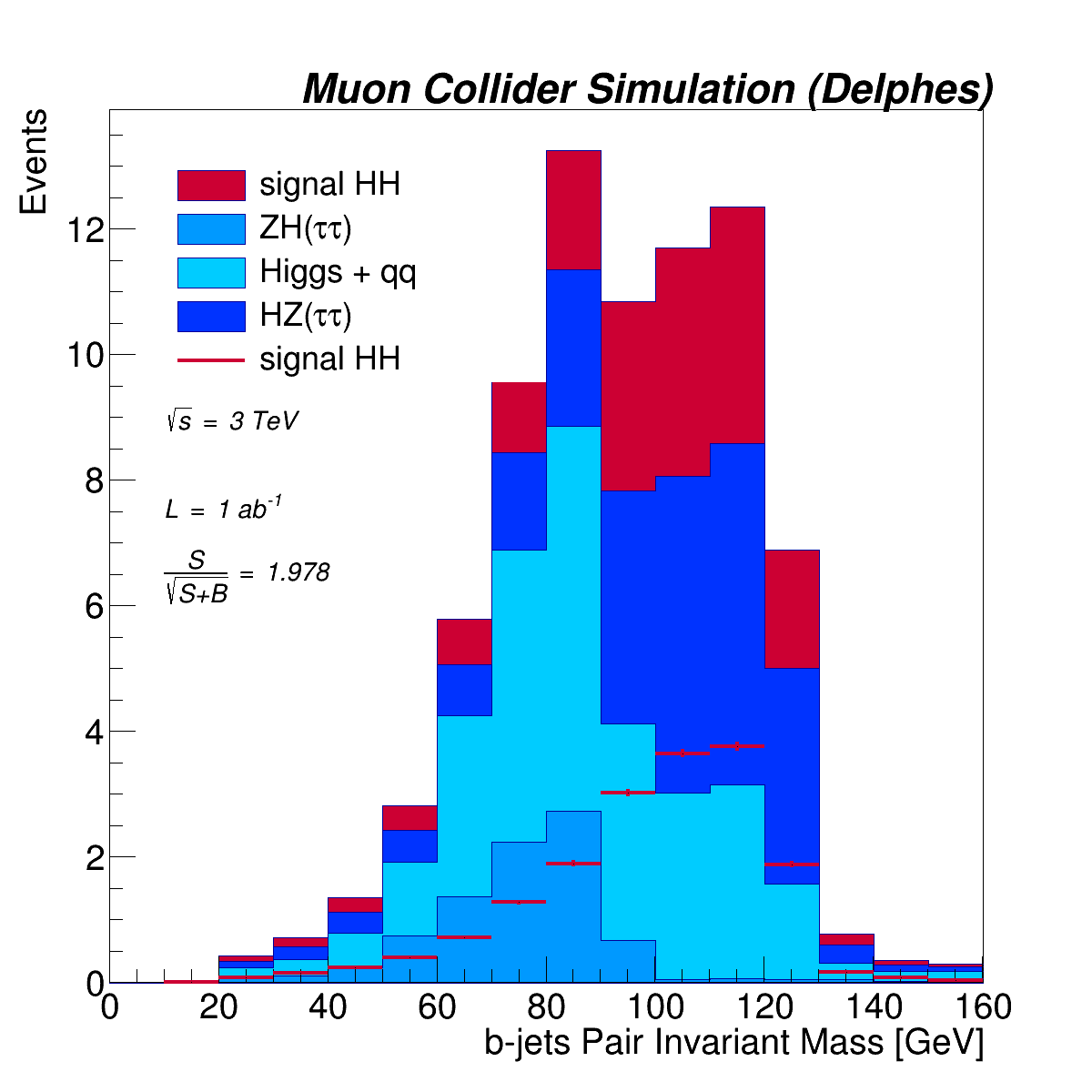}\\
    \includegraphics[width=3in]{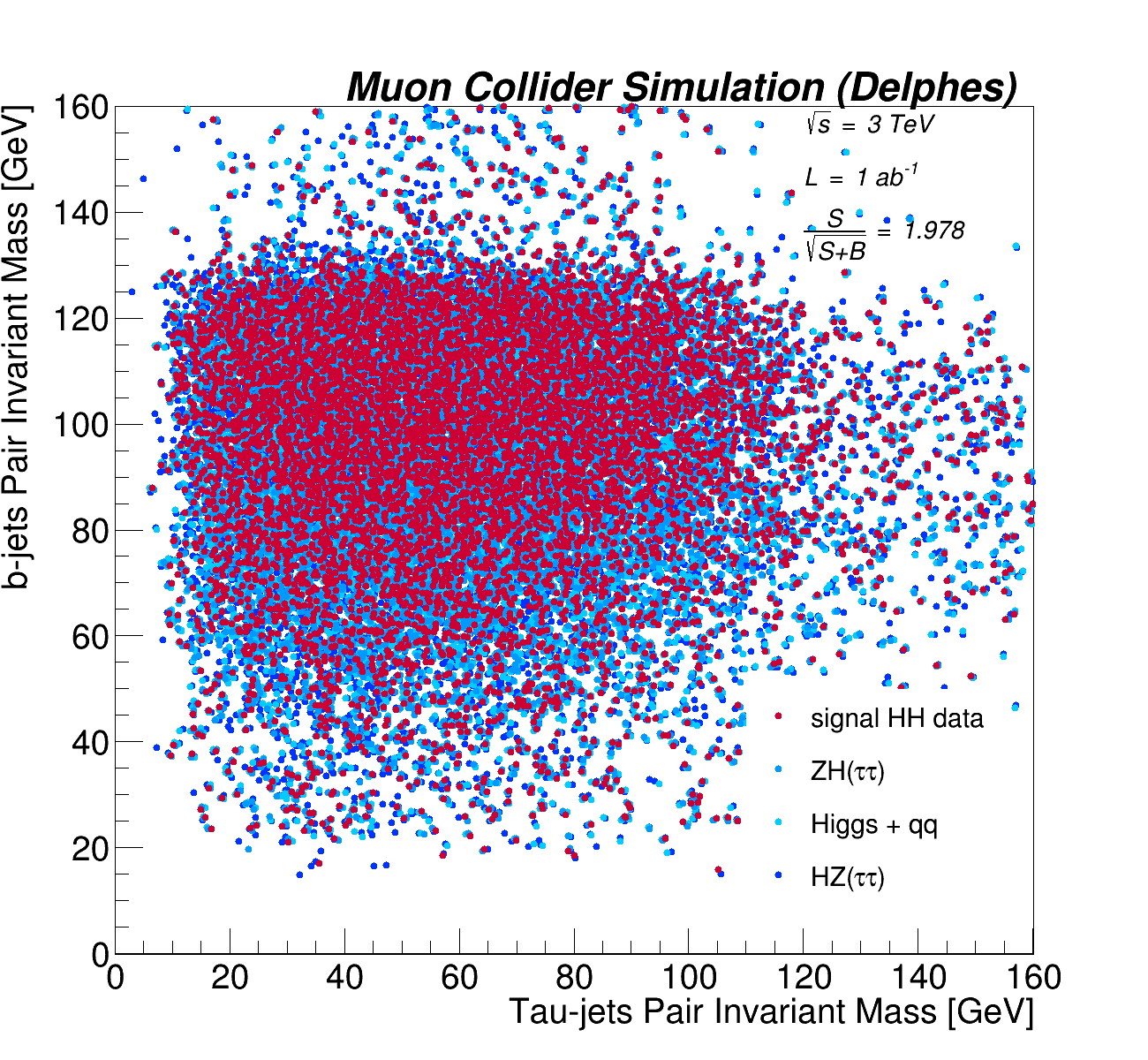}
    \caption{Event distribution of both signal and backgrounds channels is shown in the plane of the $b$-jet pair and the $\tau_{\text{lep}}\tau_{\text{had}}$ pair invariant mass for $\sqrt{s}$ = 3 TeV data (bottom). Signal is shown in red and the background in various shades of blue. Projections to the $\tau_{\text{lep}}\tau_{\text{had}}$ pair (top-left) and the $b$-jet pair (top-right) invariant mass are also shown.}
    \label{JetPair_tau_mix_3TeV}
\end{figure}

\begin{figure}[ht!]
    \centering
    \includegraphics[width=3in]{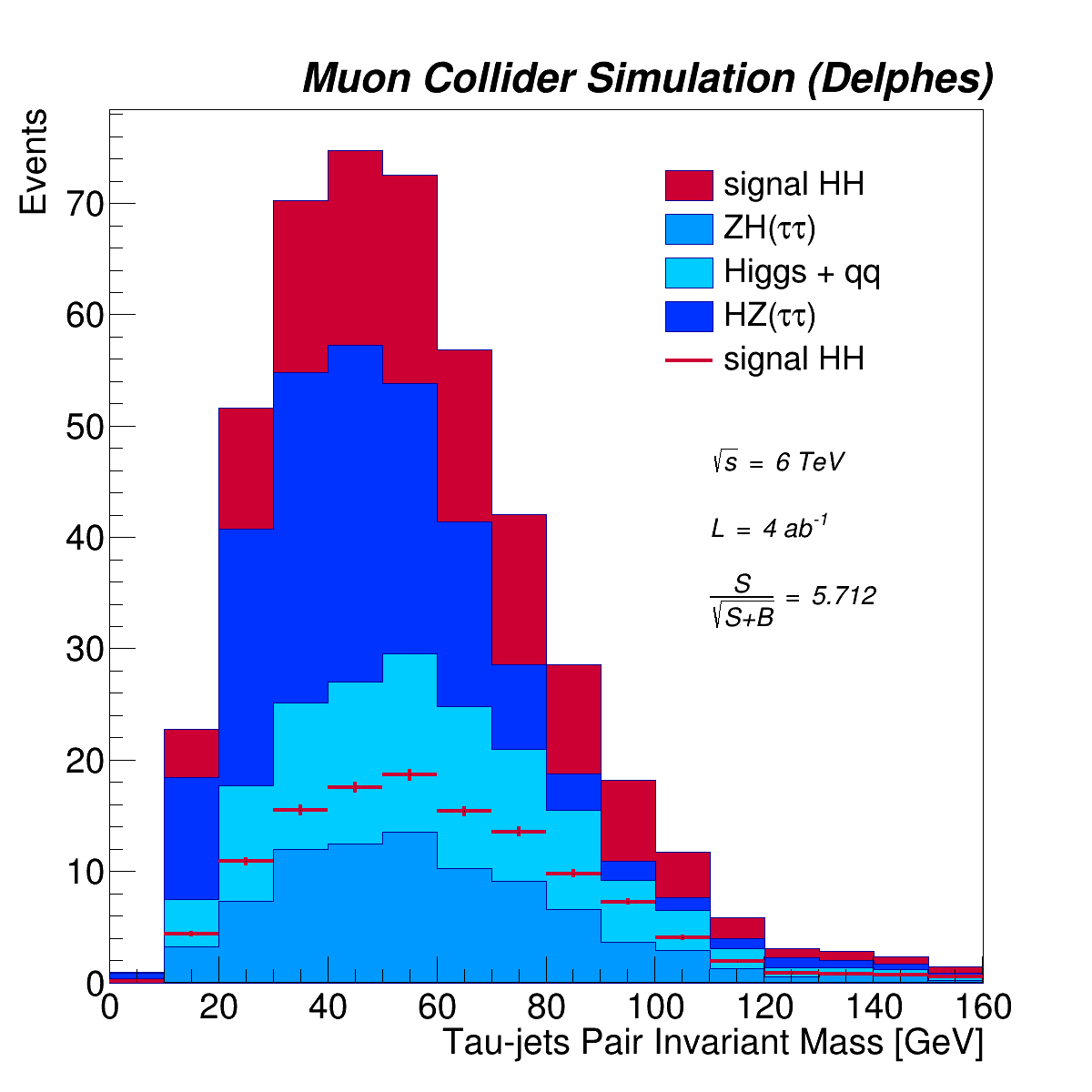}
    \includegraphics[width=3in]{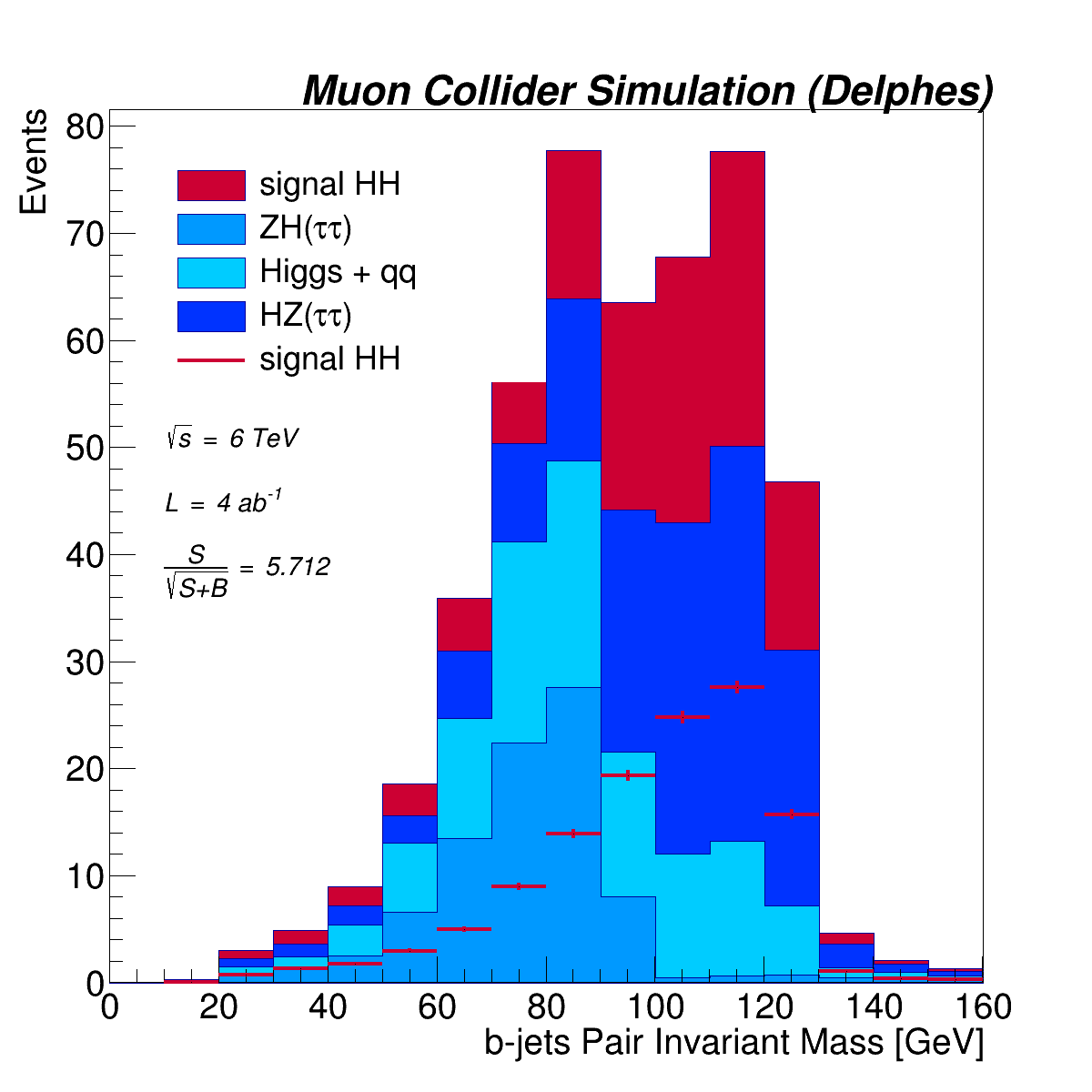}\\
    \includegraphics[width=3in]{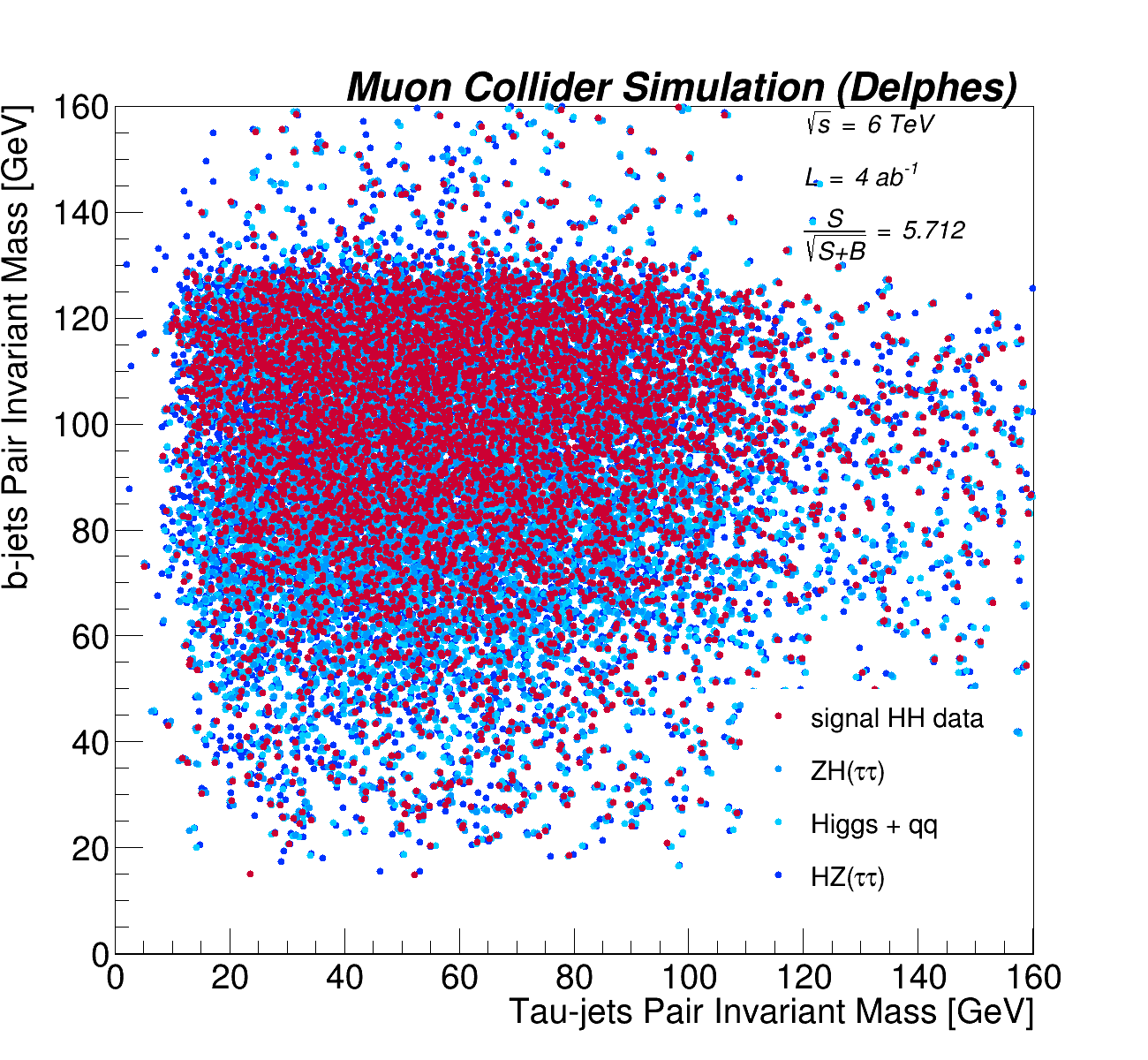}
    \caption{Event distribution of both signal and backgrounds channels is shown in the plane of the $b$-jet pair and the $\tau_{\text{lep}}\tau_{\text{had}}$ pair invariant mass for $\sqrt{s}$ = 6 TeV data (bottom). Signal is shown in red and the background in various shades of blue. Projections to the $\tau_{\text{lep}}\tau_{\text{had}}$ pair (top-left) and the $b$-jet pair (top-right) invariant mass are also shown.}
    \label{JetPair_tau_mix_6TeV}
\end{figure}

\begin{figure}[ht!]
    \centering
    \includegraphics[width=3in]{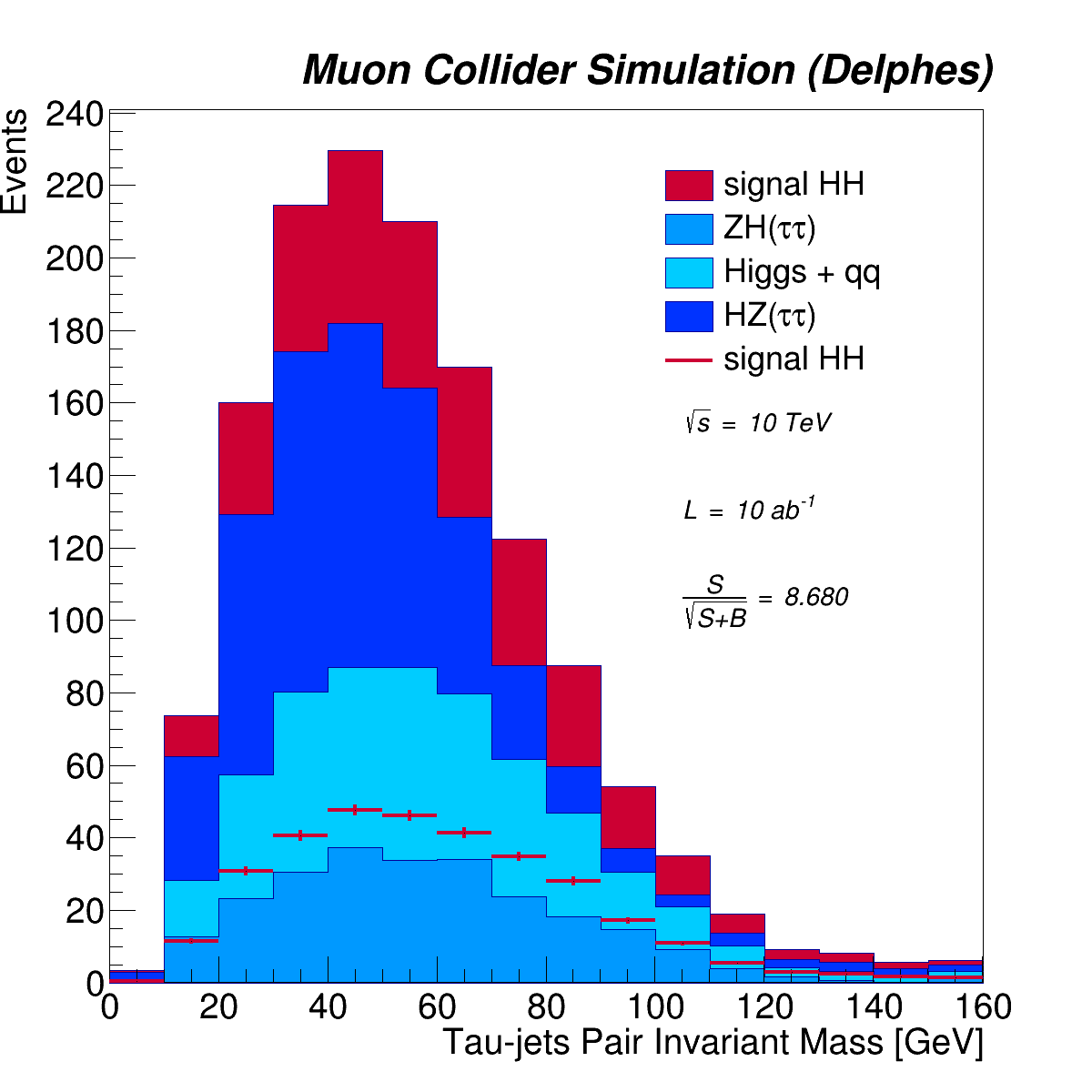}
    \includegraphics[width=3in]{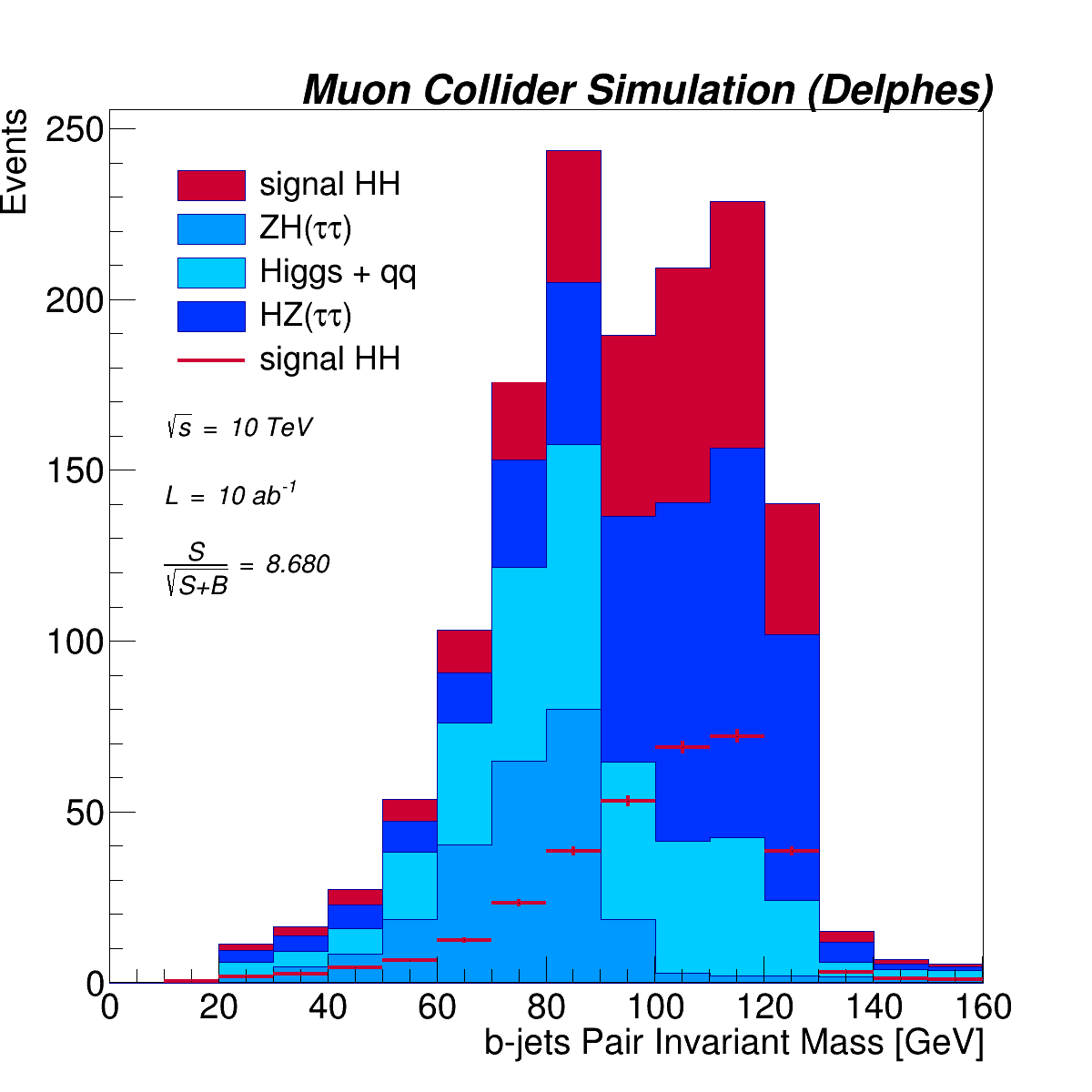}\\
    \includegraphics[width=3in]{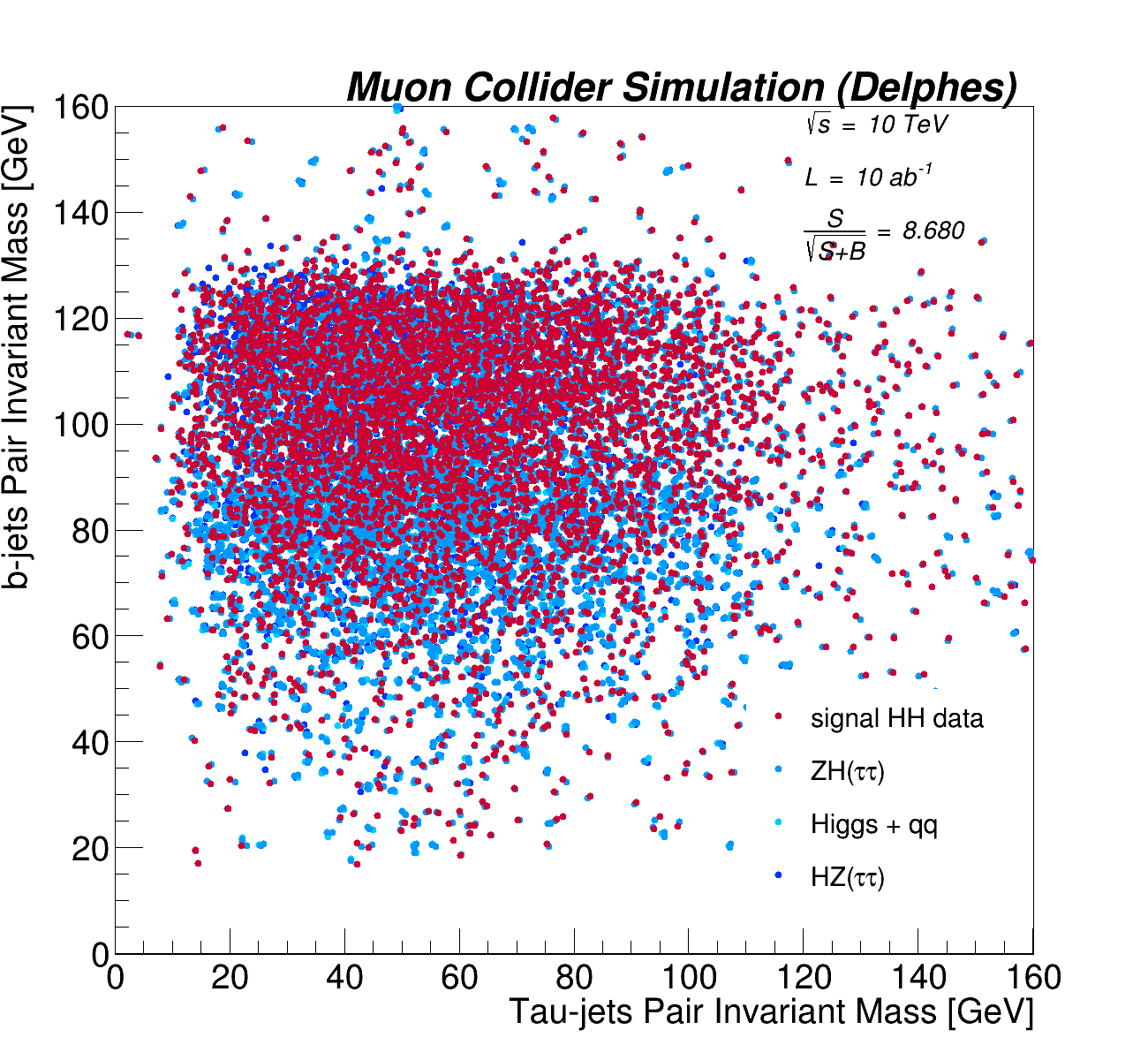}
    \caption{Event distribution of both signal and backgrounds channels is shown in the plane of the $b$-jet pair and the $\tau_{\text{lep}}\tau_{\text{had}}$ pair invariant mass for $\sqrt{s}$ = 10 TeV data (bottom). Signal is shown in red and the background in various shades of blue. Projections to the $\tau_{\text{lep}}\tau_{\text{had}}$ pair (top-left) and the $b$-jet pair (top-right) invariant mass are also shown.}
    \label{JetPair_tau_mix_10TeV}
\end{figure}

\begin{figure}[ht!]
    \centering
    \includegraphics[width=3in]{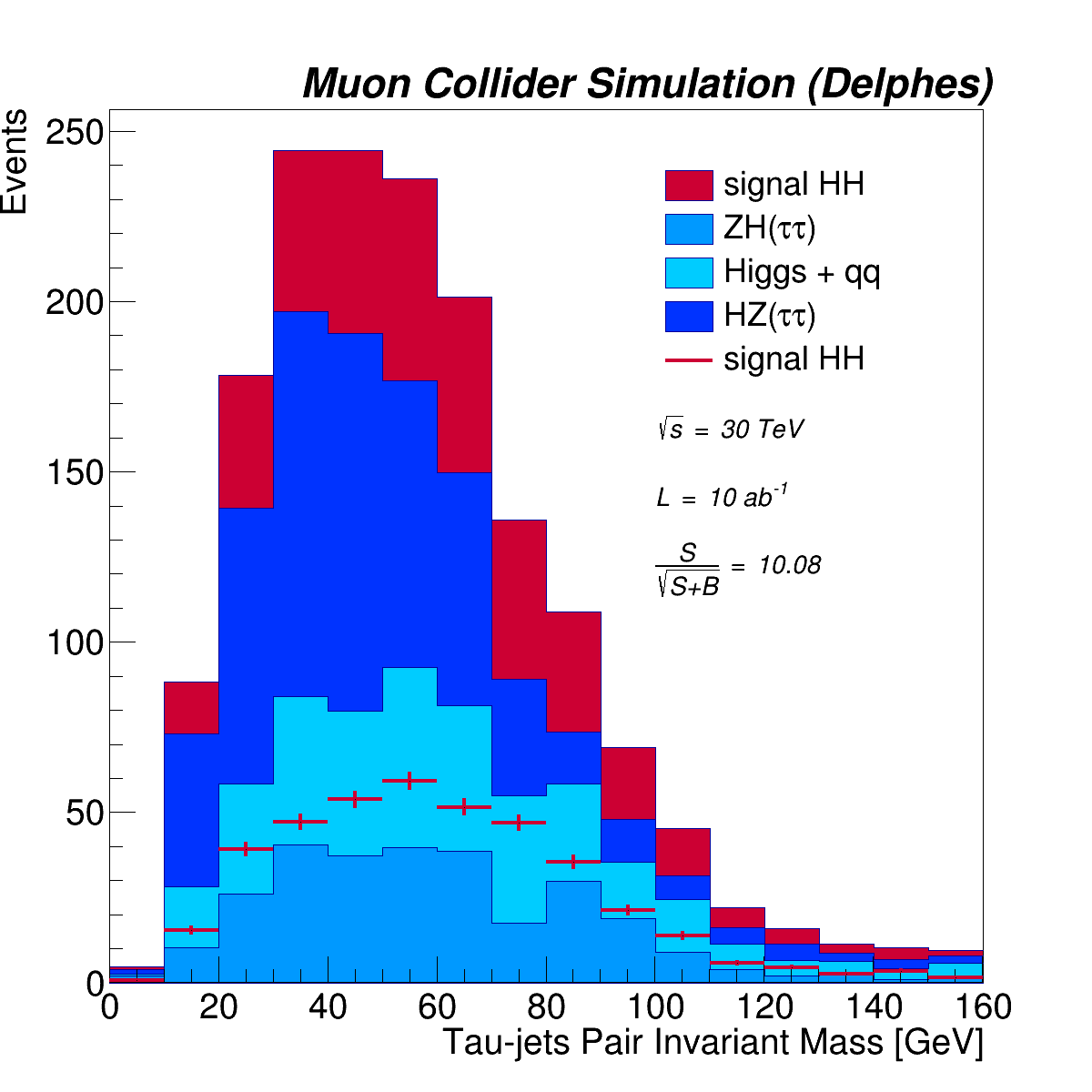}
    \includegraphics[width=3in]{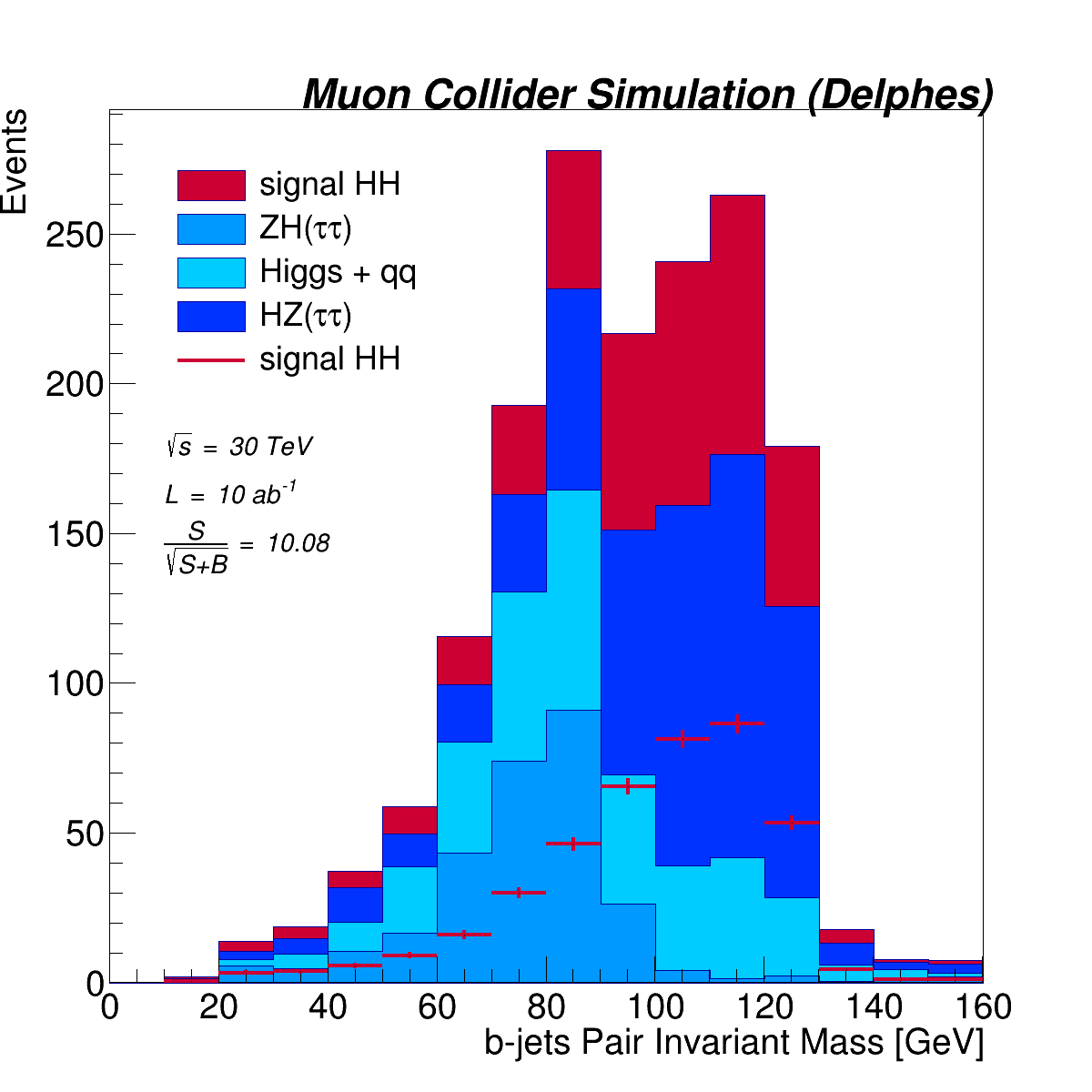}\\
    \includegraphics[width=3in]{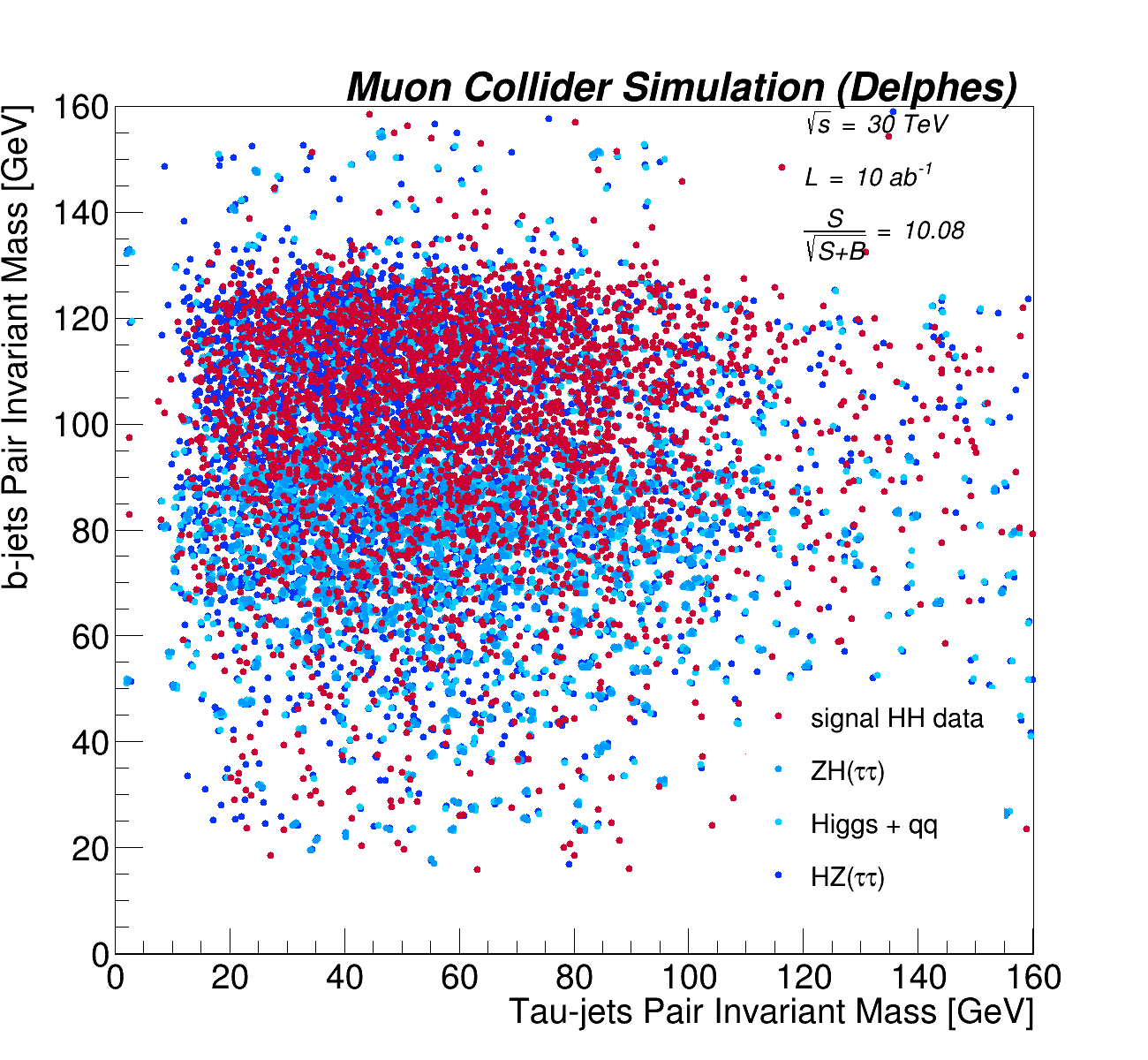}
    \caption{Event distribution of both signal and backgrounds channels is shown in the plane of the $b$-jet pair and the $\tau_{\text{lep}}\tau_{\text{had}}$ pair invariant mass for $\sqrt{s}$ = 30 TeV data (bottom). Signal is shown in red and the background in various shades of blue. Projections to the $\tau_{\text{lep}}\tau_{\text{had}}$ pair (top-left) and the $b$-jet pair (top-right) invariant mass are also shown.}
    \label{JetPair_tau_mix_30TeV}
\end{figure}

In this channel, we used the anti-$k_T$ jet algorithm with jet cone size $R=0.5$ and requiring $p_{T_{\min}}>20$ GeV for all jets used in this channel. We reconstructed the Higgs to di-$\tau$ pairs by pairing jet which are $\tau$-tagged and not $b$-tagged with  a lepton carrying opposite charge with the jet. Then, we reconstructed the Higgs with the one with the invariant mass closest to $125$ GeV. To reconstruct the $b \bar{b}$ jets pairs, we selected a $b \bar{b}$ jets pairs with invariant mass closest to $125$ GeV which both jets are loosely $b$-tagged.
Similarly, a scatter plot of the $b$-jet pair and the di-$\tau$ pair invariant mass is shown in the bottom planes of the Figures~\ref{JetPair_tau_mix_3TeV}, \ref{JetPair_tau_mix_6TeV}, \ref{JetPair_tau_mix_10TeV}, and \ref{JetPair_tau_mix_30TeV} for 3, 6, 10, and 30 TeV center of mass energies. Projects to the $b$-jet pair and the $\tau\tau$ pair invariant mass are shown in the top-right and top-left planes of the same figures. Similar to the full hadronic case, the $H$ and $Z$ mass peak in the di-$\tau$ pair invariant mass histogram is also overlapping. Hence simple cut performs poorly on this channel. With the same integrated luminosity settings for different center of mass energies, the estimated significance results with no cut are shown in the figures and tabulated for all four collider settings in the Table~\ref{tab:Delphes_HH_tau_mix}.
\begin{table}[ht!]
    \centering
    \begin{tabular}{|c|c|}
        \hline
        &\\
        $\sqrt{s}~(\int{d{\cal L}})$ & Estimated signal significance\\
        &\\\hline
        &\\
        3 TeV (1 ab$^{-1}$) & 1.978\\
        &\\\hline
        &\\
        6 TeV (4 ab$^{-1}$) & 5.712\\
        &\\\hline
        &\\
        10 TeV (10 ab$^{-1}$) & 8.680\\
        &\\\hline
        &\\
        30 TeV (10 ab$^{-1}$) & 10.08\\
        &\\\hline
    \end{tabular}
    \caption{Significance for the extraction of di-Higgs to $b \bar{b} \tau_{\text{lep}} \tau_{\text{had}}$ events for muon colliders operating at various centers of mass and integrated luminosity.}
    \label{tab:Delphes_HH_tau_mix}
\end{table}

\clearpage
\subsection{Combined results}
We combined the different decay modes and obtained results for the feasibility of measuring the Standard Model Higgs pair production at the Muon Colliders using fast simulation, with the assumption that BIB mitigation will be perfect. With the most branching ratio, four $b$ channel can use advanced statistical analysis techniques and angular variables for signal extraction to improve further. For comparison, the $b\bar{b}\gamma\gamma$ channel has the cleanest background. But it requires much more amount of data for reaching the $5\sigma$ standard. For $b\bar{b}\tau\tau$, we expect $\tau$-tagging algorithm developments for the Muon Collider to improve the sensitivity of this channel.

We use the Stouffer's method~\cite{Stouffer} for combination and estimated the combined signal significance of all three channel we studied. The combined estimated signal significance are tabulated for all four collider settings in the Table~\ref{tab:Delphes_HH_combine}.
\begin{table}[ht!]
    \centering
    \begin{tabular}{|l|c|c|c|c|}
        \hline
        &&&&\\
        Final States / $\sqrt{s} (\int{d{\cal L}})$ & 3 TeV (1 ab$^{-1}$) & 6 TeV (4 ab$^{-1}$) & 10 TeV( 10 ab$^{-1}$) & 30 TeV (10 ab$^{-1}$)\\ 
        &&&&\\\hline
        &&&&\\
        $b\bar{b} b \bar{b}$ & 2.629 & 6.287 & 10.22 & 12.72\\
        &&&&\\
         $b \bar{b} \gamma \gamma$ & 0.673 & 1.596 & 2.411 & 3.157\\
        &&&&\\ $b \bar{b} \tau_{\text{had}} \tau_{\text{had}}$ & 1.810 & 5.216 & 8.076 & 9.520\\
        &&&&\\ $b \bar{b} \tau_{\text{lep}} \tau_{\text{had}}$ & 1.978 & 5.712 & 8.680 & 10.08\\
        &&&&\\ Combined & 3.545 & 9.406 & 13.19 & 17.74\\
        &&&&\\\hline
    \end{tabular}
    \caption{Significance for the extraction of di-Higgs events combining all studied channel for muon colliders operating at various centers of mass and integrated luminosity.}
    \label{tab:Delphes_HH_combine}
\end{table}
\clearpage

\section{Conclusion}

At a Muon Collider, the beam-induced background due to muon decay is the biggest challenge to jet reconstruction performance. In this report, we discussed preliminary studies of $b$-jet-pair invariant mass reconstruction in HH events, which degrades substantially in the presence of BIB. Improvements in BIB mitigation are necessary to provide the desired jet reconstruction performance to begin physics performance studies. Significant improvements are needed both for speeding up the simulation to develop a robust BIB mitigation strategy. 

Exploring the physics reach at the Muon Collider using full simulation program requires substantial effort. Therefore, fast Monte Carlo (Delphes) based studies were conducted, with the assumption that beam-induced background mitigation will be perfect. MadGraph5, Pythia8 and Delphes were used to produce both di-Higgs to  $b\bar{b}b\bar{b}$, $b\bar{b}\gamma\gamma$, and $b\bar{b}\tau\tau$ signals and their respective dominant backgrounds. We have provided estimates of significance that can be reached for various centers of mass energies and luminosities using a simple cut-and-count strategy for each channel and combined. $b\bar{b}\gamma\gamma$ channel serves as the cleanest channel among all three, while signal extraction in the rest two is limited by the jet resolution and di-tau reconstruction. Besides BIB mitigation, further in-depth study of improving the jet resolution and the tau reconstruction in present of neutrinos in the final states at a multi-TeV muon collider is necessary for improvements in measuring the di-Higgs production at the Muon Collider.
\section{Acknowledgements}

This work is supported by the US Department of Energy (Award number DE-SC0017647) and the University of Wisconsin.

\printbibliography

\end{document}